\documentclass[reprint,amsmath,amssymb,aps,prl,superscriptaddress]{revtex4-2}

\usepackage{graphicx}
\usepackage{bm}
\usepackage{amsfonts}
\usepackage{hyperref}
\usepackage{xcolor}
\usepackage{physics}
\usepackage{empheq}
\usepackage{enumitem}
\usepackage{orcidlink}

\newcommand\mbf[1]{\mathbf{#1}}

\newcommand{\argmin}{\mathop{\mathrm{argmin}}\limits}
\newcommand{\E}{\mathbb{E}}

\newcommand{\diag}{\text{diag}}
\begin{document}

\title{Attaining Fundamental Limits of Multiparameter Incoherent Optical Imaging\\ Using Joint-Detection Quantum Measurements}

\author{Nico Deshler\,\orcidlink{0000-0001-8657-3237}}
\affiliation{Wyant College of Optical Sciences, University of Arizona, Tucson, AZ, USA}
\author{Aakash Warke}
\affiliation{Clarendon Laboratory, University of Oxford, Parks Road, OX1 3PU, Oxford, United Kingdom}
\author{Michael R. Grace\,\orcidlink{0000-0002-4699-4931}}
\affiliation{Physical Sciences and Systems, Raytheon BBN, Cambridge, MA 02138, USA}
\author{Amit Ashok\,\orcidlink{0000-0002-1108-2481}}
\affiliation{Wyant College of Optical Sciences, University of Arizona, Tucson, AZ, USA}
\author{Saikat Guha\,\orcidlink{0000-0002-2581-4380}}
\affiliation{Department of Electrical and Computer Engineering, University of Maryland, College Park, MD, USA}

\begin{abstract}
     Resolving extended incoherent objects below the diffraction limit poses an application-rich imaging challenge whose solution may enable a new generation of observational instruments and capabilities. In this work, we invoke a practical model for general imaging by approximating an arbitrary extended incoherent object as a finite grid of thermal point emitters parameterized by their brightnesses. We derive the quantum Fisher information matrix (QFIM) for simultaneous brightness estimation and show that the symmetric logarithmic derivatives weakly commute, indicating that the Helstrom bound furnishes the ultimate quantum limit on the estimation error for incoherent imaging. Furthermore, for deeply sub-diffraction scenes, we find numerical evidence of a gap between the Nagaoka-Hayashi (NH) bound and the Helstrom bound. This gap reveals that {\em separable measurements}, though more experimentally accessible, are insufficient to reach the quantum limit, and points to the prospective advantage of {\em joint measurements} acting on multiple state copies. Additionally, we show that spatial mode-demultiplexing (SPADE) often saturates the NH bound solidifying its status as a near-optimal separable measurement strategy that significantly outperforms direct imaging. Finally, we articulate two joint detection receivers implemented with {\em bona fide} quantum resources that asymptotically achieve the Helstrom bound.
\end{abstract}

\maketitle
{\em Introduction}--- The past decade has seen a steep surge of scientific interest in quantum-inspired incoherent superresolution imaging. This burgeoning field, pioneered by Tsang and collaborators in their seminal publication~\cite{Tsang:2016_TwoSource}, applies mathematical tools from quantum information theory to determine fundamental performance limits endowed by physics for broad classes of imaging tasks ranging from hypothesis testing to parameter estimation (see recent review \cite{lvovsky:2026}). In the context of {\em general imaging} where one wishes to recover the full spatial intensity profile of an unknown sub-diffraction object characterized by many (possibly infinite) parameters, SPADE has been shown to outperform conventional direct imaging both in theory~\cite{Tsang:2017,Tsang:2018_SPADE_semiclassical,Tsang:2019_ClassicalSemiparametric} and in numerical/laboratory experiments~\cite{Bisketzi:2019,Pushkina:2021,Bearne:2021,Tan:2023,Tan:2023_experiment,Lee:2023,Duplinskiy:2025,Lian:2026}. The relative advantage of SPADE over direct imaging is typically most pronounced when the extent of the object is smaller than the Rayleigh limit $w\sim \lambda/D$ of the imaging system characterized by the operating wavelength $\lambda$ and the aperture diameter $D$. 

Furthermore, by developing a theory of quantum semiparametric estimation, Tsang rigorously showed that the classical  Cram\'er-Rao bound (CRB) for SPADE scales identically to the quantum CRB, hereafter referred to as the Helstrom bound, when estimating individual moments (single-parameter) of an incoherent subdiffraction scene~\cite{Tsang:2019_Qlimits_SubdiffractionImaging_Part1,Tsang:2021_Qlimits_SubdiffractionImaging_Part2}. However, it remains unknown whether separable measurements like SPADE can generally achieve the multiparameter Helstrom bound for {\em simultaneous estimation of many object parameters} given the possibility of parameter incompatibility~\cite{Ragy:2016,Yang:2019,Albarelli:2020,Nurdin:2024}. Indeed, achieving the multiparameter Helstrom bound for imaging may instead require joint measurements that act collectively on many state copies~\cite{Guta:2009,Demkowicz-Dobrzanski:2020,Tsang:2026_HolevoBound_ManyBody,zhou:2026_MixedStatePurification}.

\begin{figure*}
\centering
\includegraphics[width=1.8\columnwidth]{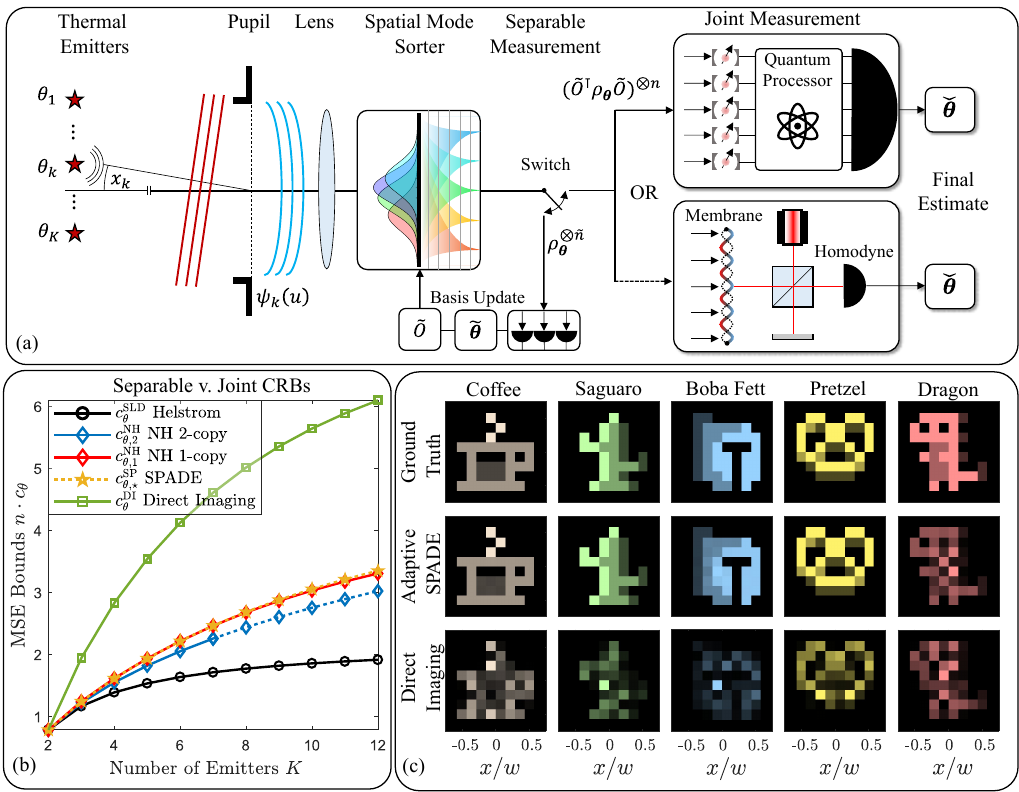}
\caption{(a) Schematic of the physical model involving $K$ incoherent thermal emitters with known positions $x_k$ and unknown relative brightnesses $\theta_k$. The receiver consists of a separable SPADE measurement designed to process $\tilde{n}$ copies of the state in order to pre-estimate the parameters $\tilde{\bm{\theta}}$. Subsequently, the remaining $n$ copies are (nearly) diagonalized by a spatial mode sorter using an approximation of the Williamson basis $\tilde{O}^{\intercal}\rho_{\bm{\theta}}\tilde{O}$. These are then routed to a joint measurement module (implemented using either transduction into quantum memories or interaction with an optomechanical membrane) to estimate residual deviations from a diagonal state and achieve the Helstrom bound for the brightness parameters. (b) Comparison between quantum bounds (Helstrom and NH) and classical CRBs (SPADE and Direct Imaging) for estimating the brightnesses of a 1D chain of equally bright and equally-spaced emitters $x_{k+1}-x_{k}= w/2$ imaged through a circular pupil. The CRB of SPADE (under an optimized basis) saturates the 1-copy NH bound. (c) (false-color) Simulated reconstructions comparing the brightness estimation performance of an adaptive SPADE protocol and direct imaging. The target scenes are $8\times8$ sub-diffraction images represented by 64 unequally bright mutually incoherent point sources placed on a grid. Both direct imaging and adaptive SPADE (running 50 updates per scene) were allotted a total photon budget of $N=100/\lambda_{\min}$ where $\lambda_{\min}$ is the minimum eigenvalue of the single-photon density operator associated with a particular scene.}
\label{fig: setup}
\end{figure*}

This raises an important question in the pursuit of quantum-optimal measurement protocols for general imaging: Is there a gap between fundamental precision limits attainable with separable versus joint measurement strategies? In this Letter, we answer this question in the affirmative. Specifically, we consider simultaneous estimation of the brightnesses (mean photon numbers) of individual thermal point sources in a $K$-emitter ensemble at known locations. This problem formulation is a stepping stone towards general imaging, since estimating the brightness of a dense grid of emitters is approximately tantamount to estimating a pixelated representation of a bounded extended object. Importantly, this finite dimensional parameterization allows us to compute the single-copy NH bound on the mean-squared error achievable with separable measurements, which in general is tighter than the Helstrom bound.

Our main results are summarized as follows: {(I)} Using a Gaussian state model for thermal emitters \cite{Weedbrook:2012}, we derive simple closed-form expressions for the QFIM and the symmetric logarithmic derivatives (SLDs) of the emitter brightnesses. We show that the SLDs weakly commute indicating that the Helstrom bound furnishes the ultimate quantum limit for simultaneous brightness estimation. {(II)} Under the weak-source approximation, we compare the NH and Helstrom bounds for different emitter geometries, identifying a gap between the two that trends with the number of emitters and their spacing. This provides numerical evidence for the necessity of joint measurement strategies in attaining the ultimate quantum limit. {(III)} We show that an optimized SPADE basis often saturates the NH bound even at deeply sub-diffraction scales, indicating that SPADE constitutes a near-ideal separable measurement. {(IV)} We describe two distinct joint detection receivers that asymptotically achieve the Helstrom bound. The first uses a generalization of the Duan-Kimble protocol~\cite{DuanKimble:2004} to load unary photonic states onto quantum memories followed by quantum computation and measurement. The second, based on recent work~\cite{Tsang:2026_HolevoBound_ManyBody}, performs a homodyne measurement on optomechanical phononic ancillas whose dynamics are governed by a specific nonlinear interaction Hamiltonian related to the SLDs and QFIM derived herein.
\\
\\
{\em Physical model}--- For pedagogical purposes our presentation is simplified to one-dimension, though the generalization to two-dimensions is straightforward. Consider $K$ quasi-monochromatic, mutually incoherent thermal emitters located at known positions $\{x_k\}_{k=1}^{K}$ in the far field of an imaging system (see Fig.~\ref{fig: setup}(a)) and let $\bar{\mbf{n}} =[\bar{n}_1,\ldots,\bar{n}_{K}]^{\intercal}\in \mathbb{R}^{K}_{+}$ be the mean photon numbers of the orthogonal plane-wave modes excited by each source impinging on the pupil. We assume that these parameters are strictly positive $\bar{n}_k>0$ such that the state is full rank. Additionally, we will assume that the pupil function of the imaging system is inversion-symmetric $\psi(-u) = \psi(u)$ with normalization $(\psi|\psi) =\int du |\psi(u)|^2 =1$. 

Upon propagation through the pupil, the plane-wave modes are spatially truncated to a set of non-orthogonal modes $\psi_k(u) = \frac{1}{\sqrt{2\pi}}e^{ix_k u}\psi(u)$ whose overlaps are collected in the Gram matrix $G_{jk} =(\psi_j|\psi_k) \in \mathbb{R}$. From the Gram matrix, we define a convenient orthonormal mode basis for $\overline{\rm span}\{\psi_k\}$ given by the parameter-independent modes $\{g_{k}(u)=\sum_{j=1}^{K}[G^{-1/2}]_{kj}\psi_{j}(u)\}$ with annihilation operators $\hat{a}_{k}$. The mode quadratures $\hat{q}_{k} = (\hat{a}_k +\hat{a}_{k}^{\dagger})/\sqrt{2}$ and $\hat{p}_{k} = (\hat{a}_{k}-\hat{a}_{k}^{\dagger})/i\sqrt{2}$ satisfy a canonical commutation relation $[\hat{q}_j,\hat{p}_{k}]=i\delta_{jk}$ and are collected into a vector $\hat{\mathbf{x}} = [\hat{q}_{1},\ldots,\hat{q}_{K},\hat{p}_{1}\ldots,\hat{p}_K]^{\intercal}$. Letting $\hat{\rho}_{\bar{\mbf{n}}}$ be the Gaussian state received by the imaging system, its  displacement vector $\bar{\mbf{x}} = \langle\hat{\mbf{x}} \rangle$ and covariance matrix $\sigma_{jk} = \frac{1}{2}\langle \{\hat{x}_{j}-\bar{x}_{j},\hat{x}_{k} -\bar{x}_{k}\}\rangle$ are (see Supplement Sec. \ref{ss: Gaussian State Model}),
\begin{equation}
\bar{\mathbf{x}} = \mathbf{0},\qquad \sigma = I_{2}\otimes(\sqrt{G}\bar{N}\sqrt{G} + \tfrac{1}{2}I).
\label{eq:input_gaussian_moments}
\end{equation}
where $\bar{N} \equiv \text{Diag}(\bar{\mbf{n}})$. By Williamson's theorem~\cite{Houde:2024,Weedbrook:2012}, the covariance matrix can be diagonalized as $\sigma = SDS^{\intercal}$ where $S$ is a symplectic matrix and $D = I_{2}\otimes \text{Diag}(\nu_1,\ldots,\nu_{K})$ is a diagonal matrix of symplectic eigenvalues satisfying $\nu_{k}\geq \tfrac{1}{2}$. It is straightforward to show that $S =I_{2}\otimes O$ and $D = I_{2}\otimes (\bar{M} + \tfrac{1}{2}I)$ where $O$ and $\bar{M}$ are obtained from the spectral decomposition,
\begin{equation}
\sqrt{G}\bar{N}\sqrt{G} = O\bar{M} O^{\intercal}.
\label{eq:spectral_decomposition}
\end{equation}
The orthogonal matrix $O\in\mathbb{O}(K)\equiv \{O\in\mathbb{R}^{K\times K} : O^{\intercal}O = I = OO^{\intercal}\}$ defines a set of uncorrelated spatial modes of the field $\hat{c}_{k} = \sum_{j=1}^{K}O_{jk}\hat{a}_{j}$ which we hereafter refer to as the {\em Williamson modes}. The matrix $\bar{M} \equiv \text{Diag}(\bar{\mbf{m}})$ contains the mean photon numbers of the Williamson modes. Importantly, both $O$ and $\bar{\mathbf{m}}$ implicitly depend on the parameters $\bar{\mbf{n}}$ through Eq. \ref{eq:spectral_decomposition}. In this new mode basis, the state of the optical field conveniently factors into a tensor product of thermal states,
\begin{equation}
\hat{\rho}_{\bar{\mbf{n}}} = \bigotimes_{k=1}^{K}\frac{1}{(\bar{m}_k+1)}\bigg(\frac{\bar{m}_{k}}{\bar{m}_{k}+1}\bigg)^{\hat{m}_{k}}
\label{eq: thermal_state}
\end{equation}
where $\hat{m}_{k} = \hat{c}_{k}^{\dagger}\hat{c}_k$ are number operators.

For Gaussian states, the SLDs and QFIM of the emitter brightnesses $\bar{\mbf{n}}$ can be calculated entirely from derivatives of the displacement vector $\bar{\mbf{x}}$ and the covariance matrix $\sigma$~\cite{Safranek:2019,Albarelli:2026_Efficient_SLD_CRB_GaussianStates,Sorelli:2024,Liu:2020_QFIM}. We find that the SLD of the parameter $\bar{n}_{a}$ is
\begin{equation}
\hat{L}_{a} = \sum_{j,k=1}^{K} \frac{\Psi_{ja}\Psi_{ka}}{M_{jk}}(\hat{c}_j^{\dagger}\hat{c}_k - \bar{m}_k\delta_{jk})
\label{eq:sld_La}
\end{equation}
where $\Psi \equiv O^{\intercal}\sqrt{G}$ is the representation of the modes $\psi_k$ in the Williamson basis and $M_{jk} \equiv \bar{m}_{j}\bar{m}_{k} + \frac{1}{2}(\bar{m}_{j} + \bar{m}_{k})$. Meanwhile, we find the QFIM to be
\begin{equation}
[Q_{\bar{\mbf{n}}}]_{ab} = \sum_{j,k=1}^{K}\frac{\Psi_{ja}\Psi_{ka}\Psi_{jb}\Psi_{kb}}{M_{jk}}.
    \label{eq: qfim}
\end{equation}
Although tractable versions of Eq. \ref{eq: qfim} have been previously proposed under the so-called weak source approximation \cite{Fiderer:2021}, here Eq. \ref{eq: qfim} is explicit and applies to general thermal states. Additionally, we note that the $(j=k)$ terms in the sums of Eqs \ref{eq:sld_La} and \ref{eq: qfim} correspond to the results of Ref.~\cite{Nair:2015_PhotonCountingOptimality_ThermalNumber} for the estimation of mean photon numbers of independent thermal states. The $(j\neq k)$ terms arise because of the parameter-dependence of the Williamson modes themselves. Using Eqs. \ref{eq: thermal_state} and \ref{eq:sld_La}, it is straightforward to show that the SLDs for the brightness parameters weakly commute:
\begin{equation}
\Tr\hat{\rho}_{\bar{\mbf{n}}}[\hat{L}_{a},\hat{L}_{b}]=0.  \label{eq:weak_commutativity}
\end{equation}
Importantly, Eq. \ref{eq:weak_commutativity} holds under all emitter positions and brightnesses. Moreover, it implies that the Helstrom bound is asymptotically attainable, and thus provides the ultimate quantum limit for the mean-squared error of any locally unbiased   brightness estimator.

{\em Hierarchy of Cram\'er-Rao Bounds}--- To better orient our results within the landscape of multiparameter estimation, we briefly review the hierarchy of scalar CRBs and their implications for optimal measurements (see Refs. ~\cite{Hayashi:2023_tightCRB,Imai:2026_boundhierarchy,Demkowicz-Dobrzanski:2020} for more comprehensive accounts). In general, the task at hand is to estimate some collection of real parameters $\bm{\theta}\in\Theta\subseteq\mathbb{R}^{K}$ from the outcome of a measurement $\Pi$ acting on $n$ identical state copies $\hat{\rho}^{\otimes n}_{\bm{\theta}}$. Without loss of generality, let $\Pi = \{\hat{\Pi}_{x\in\mathcal{X}}\geq 0:\sum_{x\in\mathcal{X}}\hat{\Pi}_{x}=\hat{I}\}$ be a positive operator-valued measure (POVM) on the joint Hilbert space $\mathcal{H}^{\otimes n}$, and let $\check{\bm{\theta}}(x)$ be a locally unbiased estimator. To evaluate the quality of the estimator-measurement pair it is standard to use the (weighted) mean squared error (MSE),
\begin{equation}
    c_{\bm{\theta}}^{\rm MSE}(\check{\bm{\theta}},\Pi) =\sum_{x\in\mathcal{X}} p_{\bm{\theta}}(x|\Pi) [\check{\bm{\theta}}(x) - \bm{\theta}]^{\intercal}W [\check{\bm{\theta}}(x) - \bm{\theta}]
\end{equation}
where $W>0$ is a user-defined positive-definite weight matrix and $p_{\bm{\theta}}(x|\Pi) = \Tr\hat{\rho}_{\bm{\theta}}\hat{\Pi}_{x}$ is the classical statistical model induced by the measurement $\Pi$. It is well known from classical statistics that the MSE for unbiased estimators is subject to the CRB $
c^{\rm MSE}_{\bm{\theta}}(\check{\bm{\theta}},\Pi) \geq c_{\bm{\theta}}^{\rm CR}(\Pi) \equiv \tr[WF_{\bm{\theta}}^{-1}(\Pi)]
$, where $F_{\bm{\theta}}(\Pi)$ is the classical Fisher information matrix (CFIM) of the model $p_{\bm{\theta}}(x|\Pi)$. In the quantum setting, we have the freedom to choose our measurement, which begs the question: what is the smallest $c_{\bm{\theta}}^{\rm CR}(\Pi)$ that one can attain? It is a celebrated result from quantum estimation theory that the CRB for {\em any} measurement $\Pi$ on $\mathcal{H}^{\otimes n}$ is beholden to the following (non-exhaustive) chain of inequalities: 
\begin{equation}
c_{\bm{\theta}}^{\rm CR}(\Pi) \geq c_{\bm{\theta},n}^{\rm NH} \geq \tfrac{1}{n} c_{\bm{\theta}}^{\rm H} \geq  \tfrac{1}{n}c_{\bm{\theta}}^{\rm SLD}.
\label{eq: scalar_quantum_bounds}
\end{equation}
Here, $c_{\bm{\theta},n}^{\rm NH}$ is the NH bound, $c_{\bm{\theta}}^{\rm H}$ is the Holevo bound, and $c^{\rm SLD}_{\bm{\theta}} \equiv \tr[WQ^{-1}_{\bm{\theta}}]$ is the Helstrom bound (see definitions in Supplement Sec. \ref{sub: CRB Definitions}). The Holevo and Helstrom bounds are {\em additive} $c(\hat{\rho}^{\otimes n}_{\bm{\theta}})= \frac{1}{n}c(\hat{\rho}_{\bm{\theta}})$ while the NH bound is {\em subadditive} $c(\hat{\rho}^{\otimes n}_{\bm{\theta}})\leq \frac{1}{n}c(\hat{\rho}_{\bm{\theta}})$ and has thus been adorned with an additional copy number subscript `$n$' as a reminder. The subadditivity of the NH bound provides an avenue for exploring how the advantage of joint measurements improves when applied to progressively larger numbers of state copies. The Holevo bound defines the ultimate quantum bound for multiparameter estimation -- it is guaranteed to be asymptotically attainable in the limit of infinite copies $\lim_{n\rightarrow\infty}\inf_{\Pi} nc^{\rm CR}_{\bm{\theta}}(\Pi) = c_{\bm{\theta}}^{H}$. In special estimation problems where the SLDs weakly commute (as in Eq. \ref{eq:weak_commutativity}), the Holevo bound and the Helstrom bound coincide $c_{\bm{\theta}}^{\rm H} = c_{\bm{\theta}}^{\rm SLD}$. Therefore, in the context of brightness estimation, the Helstrom bound is asymptotically attainable motivating us to restrict our attention to $c_{\bm{\theta}}^{\rm SLD}$ and $c_{\bm{\theta},1}^{\rm NH}$. By examining the discrepancy between these two bounds we can identify the circumstances under which separable measurements are strictly suboptimal for brightness estimation in view of the quantum limit~\cite{Tsang:2026_NH}.

{\em Separable v. Joint Measurements}--- To compare the quantum bounds alongside CRBs for well-known measurements like SPADE and direct imaging, it is necessary to move to a finite dimensional Hilbert space whereupon the NH bound is computable. To this end, we take the weak-source approximation $\bar{n}_0 \equiv \mbf{1}^{\intercal}\bar{\mbf{n}}\ll 1$ such that the mean photon number per temporal mode is very small. Upon post-selecting on the single-photon component, the parametric family we consider is,
\begin{equation}
\hat{\rho}_{\bm{\theta}} \equiv \sum_{k=1}^{K} \theta_{k}\dyad{\psi_{k}},\qquad \bm{\theta}\in\Theta
\label{eq:weak_source_model}
\end{equation}
where $\ket{\psi_k}$ is a single-photon state of the mode $\psi_k$ and $\bm{\theta}=\bar{\mbf{n}}/\bar{n}_0 \in \Theta \equiv \{\bm{\theta}\in \mathbb{R}^K :\theta_{k}>0, \mbf{1}^{\intercal}\bm{\theta} = 1\}$ is a probability vector of {\em relative brightnesses} living in the probability simplex $\Theta$. Eq. \ref{eq:weak_source_model} is the foundational model underpinning nearly all quantum-inspired imaging publications in recent years and is central to the formulation of Poisson states \cite{Tsang:2021_PoissonStates} in quantum optics.

Evaluating the Helstrom bound for $\bm{\theta}$ is almost trivial given the QFIM of Eq. \ref{eq: qfim} but requires rigorously accounting for the simplex constraint (see Supplement Sec. \ref{ss: QFIM}). Hereafter taking $W=I$, we find the QFIM for the relative brightness parameters under the single-photon model $\hat{\rho}_{\bm{\theta}}$ to be,
\begin{equation}
Q_{\bm{\theta}} \equiv \Gamma \tilde{Q}_{\bm{\theta}}\Gamma,\qquad [\tilde{Q}_{\bm{\theta}}]_{ab} = 2\sum_{jk} \frac{\Psi_{ja}\Psi_{ka}\Psi_{jb}\Psi_{kb}}{\lambda_{j}+\lambda_{k}}
\end{equation}
where $\Gamma =I - \frac{1}{K}\mbf{1}\mbf{1}^{\intercal}$ is a projector onto the tangent space of the probability simplex and $\lambda_{k} =\bar{m}_{k}/\bar{n}_0$ are the eigenvalues of $\hat{\rho}_{\bm{\theta}}$. Note that the projection makes $Q_{\bm{\theta}}$ rank-deficient such that the Helstrom bound is given by
\begin{equation}
c_{\bm{\theta}}^{\rm SLD} = \tr Q_{\bm{\theta}}^{+}
\label{eq:helstrom_bound}
\end{equation}
where $A^{+}$ is the Moore-Penrose pseudo-inverse of the matrix $A$. 

The NH bound can be numerically evaluated using semidefinite programming \cite{Conlon:2021_NHBound}. For computational purposes we choose an explicit coordinate system for the tangent space of the probability simplex such that relative brightnesses are re-parameterized as $\bm{\theta} = \frac{1}{K} \mbf{1} + V \bm{\tau}$ in terms of the unconstrained variables $\bm{\tau} \in \mathbb{R}^{K-1}$ ~\footnote{It remains an interesting open problem whether a coordinate-free formulation of the NH bound, similar to that of $Q_{\bm{\theta}}$ and $c_{\bm{\theta}}^{\rm SLD}$, exists in the context of constrained parameters (see Supplementary Materials for more discussion).}. Here, the matrix $V\in\mathbb{R}^{K\times (K-1)}$ satisfies  $V^{\intercal}V = I_{K-1}$ and $\mbf{1}^{\intercal}V = \mbf{0}$ such that its columns define orthonormal coordinate vectors on the tangent space. Since any choice of orthonormal coordinates $V$ (or $V'$) induces the same projector $\Gamma = VV^\intercal=V'V'^{\intercal}$, the NH bound is coordinate-independent (see Supplement Sec. \ref{sub: CRB Invariance})

We now consider the Cram\'er-Rao bound associated with SPADE and direct imaging measurements. Since the pupil is assumed to be inversion symmetric (leading to a real-valued amplitude spread function $\tilde{\psi}(x)\equiv \mathcal{F}[\psi(u)]\in\mathbb{R}$) the state $\hat{\rho}_{\bm{\theta}}$ and its derivatives are supported on the real span of $\{\ket{\psi_{k}}\}_{k=1}^{K}$. Therefore, the space of possible SPADE measurements is isomorphic to the group of orthogonal matrices $\mathbb{O}(K)$. A particular SPADE measurement, defined by $R\in\mathbb{O}(K)$, induces a classical channel with probability transfer matrix $P_{jk}= [R^{\intercal}\sqrt{G}]_{jk}^{2}$. Assuming the same SPADE measurement acts independently on each copy of $\hat{\rho}_{\bm{\theta}}$, the number of photons observed in each spatial mode is a multinomial random variable $\mbf{x}\in\mathbb{Z}^{K}_{+}$,
\begin{equation}
\mbf{x}|R \sim \text{Mu}(n,\mbf{q}),\qquad  \mbf{q} = P\bm{\theta} 
\label{eq: mutlinom}
\end{equation}
The CRB and CFIM of this model are (see derivation in Supplement Sec. \ref{ss: Separable Measurements}),
\begin{subequations}
\begin{align}
c_{\bm{\theta}}^{\rm SP}(R) &= \tr[( F_{\bm{\theta}}^{\rm SP})^{+}],\qquad F_{\bm{\theta}}^{\rm SP} = n \Gamma\tilde{F}_{\bm{\theta}}^{\rm SP}\Gamma \\
&\qquad \qquad \tilde{F}_{\bm{\theta}}^{\rm SP} \equiv P^{\intercal}D_{1/\mbf{q}}P
\end{align}
\end{subequations}
where $D_{1/\mbf{q}}\equiv \text{Diag}(1/q_1,\ldots,1/q_{K})$ and $\tilde{F}_{\bm{\theta}}$ denotes the unconstrained Fisher information matrix. Minimizing the CRB over the orthogonal group supplies the optimal SPADE measurement,
\begin{equation}
R^{\star}_{\bm{\theta}} = \argmin_{R \in \mathbb{O}(K)} c_{\bm{\theta}}^{\rm SP}(R).
\label{eq:spade_opt}
\end{equation}
The best CRB achievable with SPADE is denoted $c_{\bm{\theta},\star}^{\rm SP} \equiv c_{\bm{\theta}}^{\rm SP}(R^{\star})$. For direct imaging, the CRB and CFIM are given by
\begin{subequations}
\begin{align}
 c_{\bm{\theta}}^{\rm DI} &= \tr [(F_{\bm{\theta}}^{\rm DI})^{+}],\qquad F^{\rm DI}_{\bm{\theta}} = n\Gamma\tilde{F}_{\bm{\theta}}^{\rm DI}\Gamma
 \\ 
&\quad [\tilde{F}_{\bm{\theta}}^{\rm DI}]_{ab} \equiv \int_{-\infty}^{\infty} \frac{p_{a}(x)p_{b}(x)}{\sum_{k=1}^{K}\theta_{k}p_{k}(x)}dx
\end{align}
\end{subequations}
where $p_{k}=p(x-x_k)$ is a shifted version of the intensity point spread function $p(x) = \tilde{\psi}^2(x)$.

{\em Discussion}---Figure~\ref{fig: setup}(b) compares the Helstrom and NH bounds for a line of $K$ equally-spaced and equally-bright point sources with sub-diffraction nearest-neighbor separation $w/2$. The gap between the Helstrom bound and the 1-copy NH bound is zero when there are only two sources, and grows monotonically in $K$. This suggests that the relative advantage of joint measurements increases as more brightness parameters are introduced. As a point of comparison, we also computed the 2-copy NH bound which is uniformly lower than the 1-copy NH bound~\footnote{For emitter numbers $K>7$, computing the 2-copy NH bound became numerically intractable. Therefore, in Fig. \ref{fig: setup}(b) we extrapolate to $K=8,\ldots,12$ using a power-law fit to the NH bound at $K=2,\ldots,7$.}. Hence, even `small' joint measurements acting on just two state copies at a time stand to offer an advantage over separable measurements acting copy-by-copy on the same number of total resources $n$. Interestingly, SPADE is shown to exactly saturate the 1-copy NH bound for all emitter numbers, certifying its optimality as a separable measurement in this exploratory case. 

\begin{figure}
    \centering
    \includegraphics[width=\linewidth]{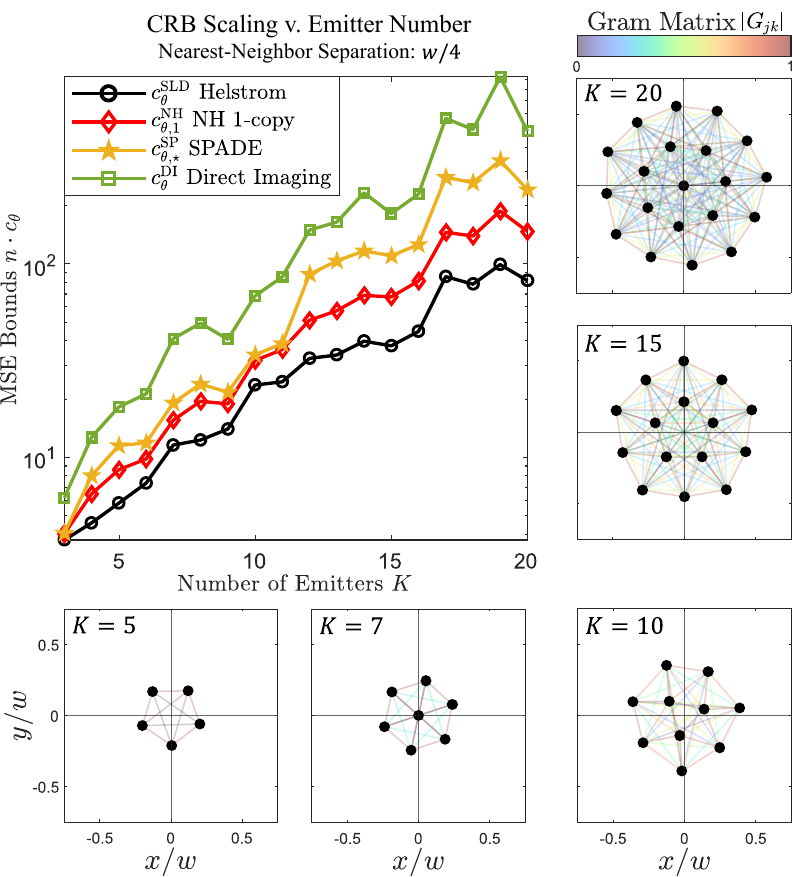}
    \caption{Scalar CRBs for brightness estimation under different emitter geometries involving $K$ sources of equal brightness. The nearest-neighbor separation for cluster was set to $w/4$. The NH bound is approximately $2\times$ greater than the Helstrom bound for all geometries considered. SPADE consistently outperforms direct imaging while saturating the single-copy NH bound under particular geometries. Condition numbers for the Gram matrices of each scene ranged between $\sim 10^{1}-10^{8}$ for $K=3,\ldots,20$. The colored lines connecting each emitter encode the magnitude of the gram matrix element $|G_{ij}|$ measuring the overlap between spatial modes excited by the $j^{\rm th}$ and $k^{\rm th}$ emitter.}
    \label{fig: K-sweep}
\end{figure}

In Fig. \ref{fig: K-sweep} we further explore the gap between joint and separable bounds for more complicated geometric arrangements of emitters with tighter spacing (nearest neighbor separation set to $w/4$) and clustering. Here the gap between the 1-copy NH bound and the Helstrom remains present, however, it does not necessarily increase monotonically with emitter number $K$. We therefore conclude that the relative performance enhancement afforded by joint measurements is highly dependent on the emitter geometry itself. 

\begin{figure}
    \centering
    \includegraphics[width=\linewidth]{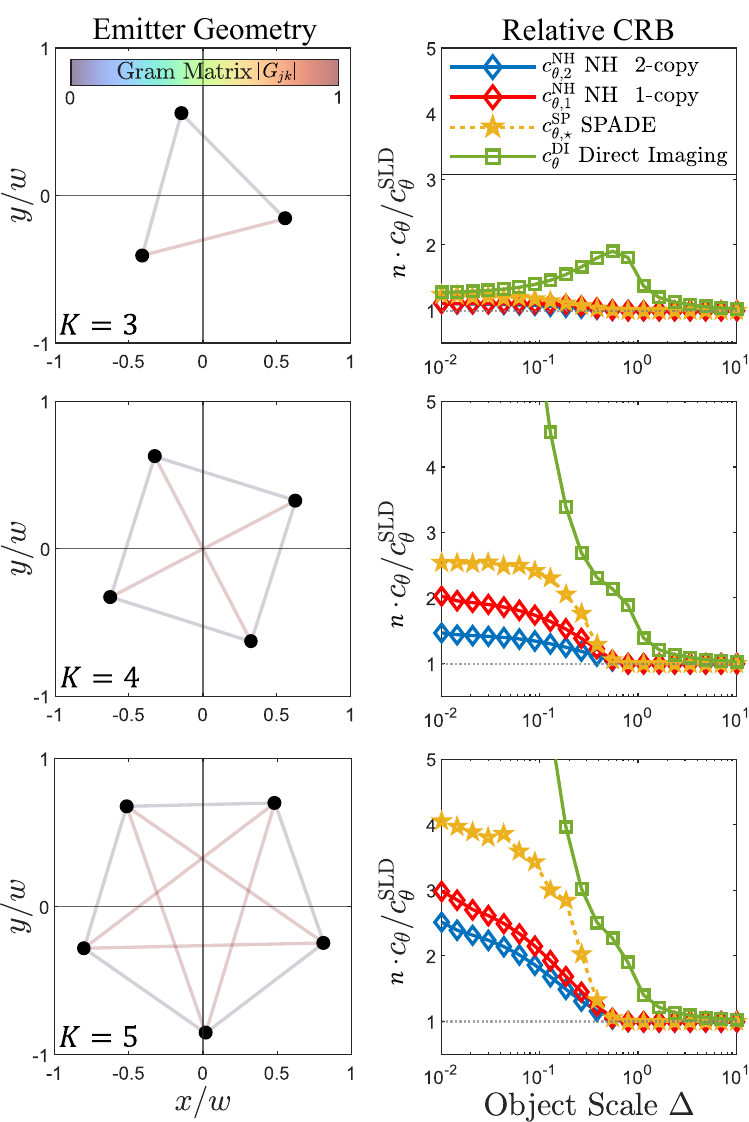}
    \caption{Relative CRBs normalized by the Helstrom bound as a function of the object scale (scene dilation factor $\Delta$) for three different emitter geometries. In the resolved regime ($\Delta \gg 1$) all bounds converge to the Helstrom bound independent of the emitter geometry. Here the Gram matrix is approximately diagonal such that the brightness of each emitter is decoupled into orthogonal modes. When the size of the object falls below the diffraction limit ($\Delta \ll 1$), we see that the ratio between CRBs exceeds $1$, highlighting a regime where joint detection measurements provide a genuine enhancement.}
    \label{fig: Dilation Sweep}
\end{figure}

To probe how the gap varies with the size of the object, Fig. \ref{fig: Dilation Sweep} illustrates the relative CRBs normalized to the Helstrom bound as a function of a dimensionless object dilation factor $\Delta$ for three example geometries consisting of $K=3,4,5$ equally bright emitters. When $\Delta = 1$, the nearest neighbor separation of the emitters is equal to the diffraction limit of the imaging system. When the emitters are well-resolved ($\Delta \gg 1$) all bounds intuitively converge to the Helstrom bound indicating that joint measurements offer no advantage over separable ones. This corresponds to a situation where the emitters are well-separated enough that the modes they excite are essentially orthogonal (i.e. the Gram matrix approaches identity). Then the estimation of relative brightness simplifies to the estimation of mean photon numbers for a collection of independent thermal states, for which photon counting in each mode (a separable measurement) is known to be optimal~\cite{Nair:2015_PhotonCountingOptimality_ThermalNumber}.

Conversely, in the sub-diffraction regime, $(\Delta \ll 1)$, the gap between separable and joint measurement strategies becomes pronounced for $K>3$. For example, in the pentagonal geometry $K=5$, the single-copy NH bound $c_{\bm{\theta},1}^{\rm NH}$ is shown to be $3\times$ greater than the Helstrom bound when $\Delta = 10^{-2}$. The primary limitation preventing numerical interrogation of object sizes smaller than $\Delta < 10^{-2}$ arises from the explosion of the condition number for the Gram matrix at such scales. This introduces numerical instability in the interior point optimizer used to compute the NH bound. In our simulations, the NH bound was computed using the MOSEK solver in Matlab's \texttt{Yalmip} toolbox \cite{Yalmip}. The SPADE optimization over the manifold $\mathbb{O}(K)$ was performed using the trust-regions solver in Matlab's \texttt{manopt} package~\cite{manopt}.

{\em Adaptive SPADE}---In practice, finding the optimal SPADE measurement for brightness estimation poses a fundamental challenge because $R^{\star}_{\bm{\theta}}$ depends on the brightness parameters which are unknown at the outset. A viable workaround is to invoke an adaptive sequential estimation strategy that iteratively updates the SPADE basis using progressively refined parameter estimates. Letting $\check{\bm{\theta}}^{[t]}\in\Theta$ denote the parameter estimate at the $t^{\rm th}$ iteration (initialized arbitrarily at $t=1$). A simple adaptive SPADE measurement (Ada-SPADE) cycles through three update steps:
\begin{align*}
    R^{[t]} &\leftarrow \argmin_{R\in \mathbb{O}(K)} c^{\rm SP}_{\check{\bm{\theta}}^{[t]}}(R) \qquad \qquad \, \text{Update SPADE Basis}\\
    \mbf{x}^{[t]} &\leftarrow \text{Mu}(\mbf{x}|n^{[t]},P^{[t]}\bm{\theta}) \qquad \quad \,\,\,\text{Collect $n^{[t]}$ Photons}\\
    \check{\bm{\theta}}^{[t+1]}&\leftarrow \max_{\bm{\theta}\in\Theta} p_{\bm{\theta}}(\mbf{x}^{[1:t]}|P^{[1:t]}) \qquad \text{Update ML Estimate}
\end{align*}
The algorithm terminates once the total number of photons collected reaches $n=\sum_{t}n^{[t]}$. This adaptive estimation protocol is asymptotically efficient \cite{Fujiwara:2006} and parallels the receiver defined in \cite{Matlin_Zipp:2022_IncoherentSourceDistr_QuantumLimitedImaging} with the added benefit of directly optimizing the CRB. Figure~\ref{fig: setup}(c) shows simulated reconstructions of a collection of sub-diffraction incoherent pixelated images composed of $8\times8$ unequally-bright incoherent point emitters that fit within the diffraction-limited PSF spot of a circular aperture. For comparison, we also illustrate the maximum likelihood (ML) estimate of the relative brightness parameters obtained from direct imaging measurements. Qualitatively, the ML estimate obtained from adaptive SPADE bears much greater perceptual resemblance to the true scene compared to the ML estimate obtained from direct imaging measurements using the same number of photons. This points to the improved information efficiency afforded by measurements that are optimally adapted to the specific scene being observed.

{\em Optimal Joint Detection Receivers}--- The veritable gap between the Helstrom bound and the NH bound in the sub-diffraction regime incentivizes us to consider implementations of optimal joint measurements. These protocols may be particularly valuable for extreme imaging problems operating in photon-starved settings, wherein extracting maximal information from all state copies remains imperative. Here we describe two possible protocols for realizing joint detection measurements acting on $\hat{\rho}_{\bm{\theta}}^{\otimes n}$ that asymptotically converge to the Helstrom bound -- both invoke a two-stage  approach. The first stage, which is common to both protocols, involves allocating some asymptotically vanishing proportion of copies $\tilde{n} =\sqrt{n}$ towards generating a coarse pre-estimate $\tilde{\bm{\theta}} = \bm{\theta}+\bm{\epsilon}$ of the parameters using sub-optimal measurements (e.g. direct imaging or SPADE). The remaining majority of state copies $n\approx n-\sqrt{n}$ undergo a unitary transformation that leaves them approximately diagonalized. In the case of brightness estimation, this corresponds to decomposing the optical field into an approximation of the Williamson modes using a reconfigurable spatial mode sorter (see Fig.~\ref{fig: setup}(a)) such that $\rho_{\bm{\theta}}\leftarrow \tilde{O}^{\intercal}\rho_{\bm{\theta}}\tilde{O}$ where $\tilde{O}$ is obtained through the spectral decomposition $\rho_{\tilde{\bm{\theta}}}=\tilde{O}\tilde{\Lambda}\tilde{O}^{\intercal}$ as in Eq. \ref{eq:spectral_decomposition}. The second stage involves acting a joint measurement on the remaining state copies in order to determine the residual correction $\bm{\epsilon}$ -- it is here in the second stage that the two protocols differ. 

{\em (1) Transduction into Quantum Memories}---Recent works in optical imaging and laser communications~\cite{Mokeev:2026,Gardner:2026,Da_Silva2013,Werker-Smith2026,Almeida:2025_SuperresCollective} have shown that joint measurements across a multi-mode optical signal can be implemented by transducing each copy of the photonic state into qubit-based quantum memories. Afterwards, realizing a desired receiver becomes a circuit compilation problem on a fault-tolerant quantum computer. Here we envision utilizing a generalization of the Duan-Kimble protocol~\cite{DuanKimble:2004} proposed in Ref.~\cite{Richardson:2026} that deterministically maps a unary-encoded photonic qudit state distributed over $K$ bosonic modes into $\lceil \log_{2}K \rceil$ qubits (also see~\cite{Khabiboulline2019}). Storing the state across $n$ temporal modes each with an independent copy of $\hat{\rho}_{\bm{\theta}}$---which in turn is a state of $1$ photon spread across the span of $K$ linearly-independent spatial modes---thus requires $n\log_{2}K$ qubits. This protocol enjoys the added benefit of heralding the arrival of a photon in a given temporal mode, eliminating the vacuum state in the weak thermal source model and leading to efficient usage of quantum memories.

Assuming this $n$-copy state transduction is performed, the optimal joint measurement follows readily from two seminal papers \cite{Guta:2009,Demkowicz-Dobrzanski:2020} articulating the ideas of quantum local asymptotic normality (QLAN). The central tenet of QLAN is that an $n$-copy quantum state acting on a $K$-dimensional Hilbert space is asymptotically isometric to a classical-quantum Gaussian shift model (GSM) involving $K(K-1)/2$ bosonic modes in the limit $n\rightarrow \infty$. In this GSM, the parameters of interest appear as phase-space displacements of uncorrelated thermal states for which dual homodyne detection is known to be maximally sensitive. 

The second stage involves performing a ``which block" measurement on the Schur-Weyl decomposition of $(\mathbb{C}^{K})^{\otimes n}$ via the quantum Schur transform defined in \cite{Krovi:2019,Bacon:2006}. The resulting state is subsequently measured using a discrete approximation of dual-homodyne detection mapped back to the qudit space in order to extract the residual phase-space displacements. The precise compilation of the joint measurement as a quantum circuit acting on finite $n$ copies, after the generalized Duan-Kimble transduction onto logic-capable atomic quantum memories, is left for future work.

{\em (2) Interaction with Bosonic Ancillas}--- An alternative strategy that does not require state transduction into quantum memories was recently proposed by Tsang~\cite{Tsang:2026_HolevoBound_ManyBody} (see yet another proposal~\cite{zhou:2026_MixedStatePurification}). Here, $K$ bosonic ancillas with quadratures $\hat{q}_{\ell}',\hat{p}_{\ell}'$ are configured to evolve under an interaction Hamiltonian built from the efficient influence operators of the estimation task~\cite{Tsang:2020_SemiParametricEstimation} (see Supplement Sec. \ref{ss: Joint Measurements}). We find that the interaction that must be realized between each temporal mode of the received optical field and the $\ell^{\rm th}$ ancilla involves three-wave mixing terms with specific cooperativity $\beta_{jk}^{\ell}$:
\begin{subequations}
\begin{align}
\hat{H}^{(\ell)}_{\rm int} &= \sum_{j,k}^{K}\beta_{jk}^{(\ell)}(\hat{c}_{j}^{\dagger}\hat{c}_{k} +\hat{c}_{k}^{\dagger}\hat{c}_{j})\hat{p}_{\ell}'\\
\beta_{jk}^{(\ell)} &\equiv \sum_{i=1}^{K} [Q^{+}_{\tilde{\bm{\theta}}}]_{\ell i}\bigg(\frac{\Psi_{ji}\Psi_{ki}}{M_{jk}}\bigg).
\end{align}
\label{eq:interaction_hamiltonian}
\end{subequations}
The ancillas are initialized in vacuum and left to interact with all of the remaining state copies. Finally, the $\hat{q}'_{\ell}$ quadratures of the ancillas are measured using homodyne detection. The measurement outcome is proportional to the correction $\bm{\epsilon}$. Here we envision coupling Williamson modes $o_{k}(u)$ of the optical field to an optomechanical membrane with natural mechanical modes $\phi_{\ell}(u)$ as shown in the latter half of Fig. \ref{fig: setup}(a). The cooperativity between two optical modes mediated by the membrane is given by the spatial overlap integral $\alpha_{jk}^{\ell} = \int o_{j}(u)\phi_{\ell}(u)o_{k}(u)du$. To realize the desired interaction, the natural modes of the oscillator can be designed such that $\alpha_{jk}^{(\ell)} = \beta_{jk}^{(\ell)}$. Additionally, since each mechanical mode $\ell=1,\ldots,K$ oscillates a distinct frequency $\omega_\ell$, one may exploit the sideband frequencies $\omega_{0}\pm \omega_{\ell}$ of the reflected homodyne beam to individually read out each mechanical mode supported on a single membrane.

{\em Conclusion}--- Our work resolves an outstanding question in the domain of quantum-inspired super-resolution: namely, whether separable measurements are sufficient to attain the Helstrom bound for general incoherent imaging. We showed that the gap between the NH bound and the Helstrom bound appears to grow with the number of parameters $K\gg 1$ and with the compactness $\Delta \ll 1$ of the scene. On the way, we derived a closed-form expression for the QFIM for the brightness parameters of any incoherent scene composed of thermal point emitters as well as the CFIM for SPADE and Direct imaging. We found that a numerically-optimized SPADE basis often attains the 1-copy NH bound and consistently outperforms direct imaging in the sub-diffraction regime. 

Several open questions remain. First, while we articulate two protocols for realizing optimal joint measurements predicated on the use of quantum resources (i.e. atomic quantum memories or optomechanical membranes cooled to the motional ground state), it is unknown whether an equivalent measurement can be implemented simply using a spatio-temporal mode sorter that groups individual photons spread across multiple temporal modes into a single temporal mode \cite{Cui:2025}. If possible, this approach would dramatically improve the experimental feasibility of joint measurements. Second, while we employ a discrete approximation of general scenes to guarantee the finiteness of the Hilbert space (rendering the NH bound computable) it would be valuable to expand our analysis to faithful extended objects parameterized by coefficients to an encoding basis (e.g. wavelets, discrete-cosine transforms, etc.). In this context, analyzing the Helstrom bound for sparsely-encoded objects could bridge the insights from quantum estimation theory and compressed sensing theory used throughout computational imaging. Third, the condition numbers of the Gram matrix constitute the primary limiting factor prohibiting us from numerically probing the NH-Helstrom gap under extreme imaging scenarios (e.g. high-contrast with small extents) where joint measurements may hope to offer order-of-magnitude improvements over separable ones. Looking forward, we hope this challenge motivates theoretical development towards making concrete analytical claims about the rate that the NH bound diverges from the Helstrom bound in sub-diffraction multiparameter estimation.

{\em Acknowledgments}--- ND is supported by the National Science Foundation (NSF) Graduate Research Fellowship Program (DGE-2137419). SG acknowledges support from the Department of Energy (DoE) project ``Quantum-aided Label-free In-situ Super-resolution Bioimaging" funded by DoE's Biological and Environmental Research (BER) program office under a prime award to the University of Arizona (grant number DE-SC0025910), and subaward to the University of Maryland (subaward ID 815375). ND, AA, and SG acknowledge insightful discussions with Alex Lvovsky's group. This manuscript was prepared in parallel with their work on simultaneous estimation of Fourier amplitudes for subdiffraction imaging and both groups agreed to submit for publication on the same day. ND, AA, and SG also acknowledge valuable conversations with Johannes Borregard, Maxim Sirotin, Aleksandr Mokeev, and Anthony Brady on exciting complementary work in the domain of subdiffraction multiparameter estimation. Finally, ND is grateful to Lorcan Conlon for bringing recent works \cite{Tsang:2026_HolevoBound_ManyBody,zhou:2026_MixedStatePurification} on asymptotically optimal joint measurements to his attention.

\bibliography{references}

@article{Nair:2015_PhotonCountingOptimality_ThermalNumber,
doi = {10.1088/0004-637X/808/2/125},
url = {https://doi.org/10.1088/0004-637X/808/2/125},
year = {2015},
publisher = {The American Astronomical Society},
volume = {808},
number = {2},
pages = {125},
author = {Nair, Ranjith and Tsang, Mankei},
title = {Quantum Optimality of Photon Counting for Temperature Measurement of Thermal Astronomical Sources},
journal = {Astrophys. Jour.}
}

@ARTICLE{Lee:2023,
  title     = "Quantum-inspired multi-parameter adaptive Bayesian estimation for
               sensing and imaging",
  author    = "Lee, Kwan Kit and Gagatsos, Christos N and Guha, Saikat and
               Ashok, Amit",
  journal   = "IEEE J. Sel. Top. Signal Process.",
  publisher = "Institute of Electrical and Electronics Engineers (IEEE)",
  volume    =  17,
  number    =  2,
  pages     = "491--501",
  month     =  mar,
  year      =  2023
}

@ARTICLE{Khabiboulline2019,
  title     = "Quantum-assisted telescope arrays",
  author    = "Khabiboulline, E T and Borregaard, J and De Greve, K and Lukin, M
               D",
  journal   = "Phys. Rev. A",
  publisher = "American Physical Society",
  volume    =  100,
  number    =  2,
  pages     =  022316,
  month     =  aug,
  year      =  2019
}

@ARTICLE{Werker-Smith2026,
  title     = "Quantum-processing-assisted classical communication",
  author    = "Werker Smith, Kelly and Boroson, Don and Guha, Saikat and
               Borregaard, Johannes",
  journal   = "Phys. Rev. Appl.",
  publisher = "American Physical Society (APS)",
  volume    =  25,
  number    =  4,
  pages     =  044037,
  month     =  apr,
  year      =  2026,
}

@ARTICLE{Da_Silva2013,
  title     = "Achieving minimum-error discrimination of an arbitrary set of
               laser-light pulses",
  author    = "da Silva, Marcus P and Guha, Saikat and Dutton, Zachary",
  journal   = "Phys. Rev. A",
  publisher = "American Physical Society",
  volume    =  87,
  number    =  5,
  pages     =  052320,
  month     =  may,
  year      =  2013
}

@article{Tsang:2016_TwoSource,
  title = {Quantum Theory of Superresolution for Two Incoherent Optical Point Sources},
  author = {Tsang, Mankei and Nair, Ranjith and Lu, Xiao-Ming},
  journal = {Phys. Rev. X},
  volume = {6},
  issue = {3},
  pages = {031033},
  numpages = {17},
  year = {2016},
  month = {Aug},
  publisher = {American Physical Society},
  doi = {10.1103/PhysRevX.6.031033},
  url = {https://link.aps.org/doi/10.1103/PhysRevX.6.031033}
}

@article{Tsang:2019_Qlimits_SubdiffractionImaging_Part1,
  title = {Quantum limit to subdiffraction incoherent optical imaging},
  author = {Tsang, Mankei},
  journal = {Phys. Rev. A},
  volume = {99},
  issue = {1},
  pages = {012305},
  numpages = {12},
  year = {2019},
  month = {Jan},
  publisher = {American Physical Society},
  doi = {10.1103/PhysRevA.99.012305},
  url = {https://link.aps.org/doi/10.1103/PhysRevA.99.012305}
}

@article{Tsang:2021_Qlimits_SubdiffractionImaging_Part2,
  title = {Quantum limit to subdiffraction incoherent optical imaging. II. A parametric-submodel approach},
  author = {Tsang, Mankei},
  journal = {Phys. Rev. A},
  volume = {104},
  issue = {5},
  pages = {052411},
  numpages = {16},
  year = {2021},
  month = {Nov},
  publisher = {American Physical Society},
  doi = {10.1103/PhysRevA.104.052411},
  url = {https://link.aps.org/doi/10.1103/PhysRevA.104.052411}
}

@article{Matlin_Zipp:2022_IncoherentSourceDistr_QuantumLimitedImaging,
   title={Imaging arbitrary incoherent source distributions with near quantum-limited resolution},
   volume={12},
   ISSN={2045-2322},
   url={http://dx.doi.org/10.1038/s41598-022-06644-3},
   DOI={10.1038/s41598-022-06644-3},
   number={1},
   journal={Scientific Reports},
   publisher={Springer Science and Business Media LLC},
   author={Matlin, Erik F. and Zipp, Lucas J.},
   year={2022}}

@article{Weedbrook:2012,
  title = {Gaussian quantum information},
  author = {Weedbrook, Christian and Pirandola, Stefano and Garc\'{\i}a-Patr\'on, Ra\'ul and Cerf, Nicolas J. and Ralph, Timothy C. and Shapiro, Jeffrey H. and Lloyd, Seth},
  journal = {Rev. Mod. Phys.},
  volume = {84},
  issue = {2},
  pages = {621--669},
  numpages = {0},
  year = {2012},
  month = {May},
  publisher = {American Physical Society},
  doi = {10.1103/RevModPhys.84.621},
  url = {https://link.aps.org/doi/10.1103/RevModPhys.84.621},
}

@article{Safranek:2019,
title = {Estimation of Gaussian quantum states},
doi = {10.1088/1751-8121/aaf068},
url = {https://doi.org/10.1088/1751-8121/aaf068},
year = {2018},
month = {dec},
publisher = {IOP Publishing},
volume = {52},
number = {3},
pages = {035304},
author = {Šafránek, Dominik},
journal = {Journal of Physics A: Mathematical and Theoretical}
}

@article{Albarelli:2026_Efficient_SLD_CRB_GaussianStates,
title = {Efficiently evaluating Holevo, RLD and SLD Cramér-Rao bounds for multiparameter quantum estimation with Gaussian states},
author = {Shoukang, Chang and Genoni, Marco G. and Albarelli, Francesco},
year = {2026},
journal = {Commun Phys},
publisher = {Springer},
doi = {10.1038/s42005-026-02550-6},
url = {https://doi.org/10.1038/s42005-026-02550-6},
}

@article{Sorelli:2024,
doi = {10.1088/1367-2630/ad5eb2},
url = {https://doi.org/10.1088/1367-2630/ad5eb2},
year = {2024},
month = {jul},
publisher = {IOP Publishing},
volume = {26},
number = {7},
pages = {073022},
author = {Sorelli, Giacomo and Gessner, Manuel and Treps, Nicolas and Walschaers, Mattia},
title = {Gaussian quantum metrology for mode-encoded parameters},
journal = {New Journal of Physics}
}

@article{Houde:2024,
   title={Matrix decompositions in quantum optics: Takagi/Autonne, Bloch–Messiah/Euler, Iwasawa, and Williamson},
   volume={102},
   ISSN={1208-6045},
   url={http://dx.doi.org/10.1139/cjp-2024-0070},
   DOI={10.1139/cjp-2024-0070},
   number={10},
   journal={Canadian Journal of Physics},
   publisher={Canadian Science Publishing},
   author={Houde, Martin and McCutcheon, Will and Quesada, Nicolás},
   year={2024},
   month=oct, pages={497–507} }

@article{Liu:2020_QFIM,
doi = {10.1088/1751-8121/ab5d4d},
url = {https://doi.org/10.1088/1751-8121/ab5d4d},
year = {2019},
month = {dec},
publisher = {IOP Publishing},
volume = {53},
number = {2},
pages = {023001},
author = {Liu, Jing and Yuan, Haidong and Lu, Xiao-Ming and Wang, Xiaoguang},
title = {Quantum Fisher information matrix and multiparameter estimation},
journal = {Journal of Physics A: Mathematical and Theoretical}
}

@article{Demkowicz-Dobrzanski:2020,
doi = {10.1088/1751-8121/ab8ef3},
url = {https://doi.org/10.1088/1751-8121/ab8ef3},
year = {2020},
month = {aug},
publisher = {IOP Publishing},
volume = {53},
number = {36},
pages = {363001},
author = {Demkowicz-Dobrzański, Rafał and Górecki, Wojciech and Guţă, Mădălin},
title = {Multi-parameter estimation beyond quantum Fisher information},
journal = {Journal of Physics A: Mathematical and Theoretical}
}

@article{Conlon:2021_NHBound,
   title={Efficient computation of the Nagaoka–Hayashi bound for multiparameter estimation with separable measurements},
   volume={7},
   ISSN={2056-6387},
   url={http://dx.doi.org/10.1038/s41534-021-00414-1},
   DOI={10.1038/s41534-021-00414-1},
   number={1},
   journal={npj Quantum Information},
   publisher={Springer Science and Business Media LLC},
   author={Conlon, Lorcán O. and Suzuki, Jun and Lam, Ping Koy and Assad, Syed M.},
   year={2021}
   }

@article{Tsang:2020_SemiParametricEstimation,
  title = {Quantum Semiparametric Estimation},
  author = {Tsang, Mankei and Albarelli, Francesco and Datta, Animesh},
  journal = {Phys. Rev. X},
  volume = {10},
  issue = {3},
  pages = {031023},
  numpages = {28},
  year = {2020},
  month = {Jul},
  publisher = {American Physical Society},
  doi = {10.1103/PhysRevX.10.031023},
  url = {https://link.aps.org/doi/10.1103/PhysRevX.10.031023}
}

@misc{Tsang:2026_HolevoBound_ManyBody,
      title={Approaching the ultimate limit of quantum multiparameter estimation by many-body physics}, 
      author={Mankei Tsang},
      year={2026},
      eprint={2603.17955},
      archivePrefix={arXiv},
      primaryClass={quant-ph},
      url={https://arxiv.org/abs/2603.17955}, 
}

@misc{zhou:2026_MixedStatePurification,
      title={Quantum metrology of mixed states via purification}, 
      author={Sisi Zhou},
      year={2026},
      eprint={2605.03975},
      archivePrefix={arXiv},
      primaryClass={quant-ph},
      url={https://arxiv.org/abs/2605.03975}, 
}

@article{Guta:2009,
author = {Jonas Kahn and Mădălin Guţă},
year = {2009},
title = {Local Asymptotic Normality for Finite Dimensional Quantum Systems},
journal   = {Commun. Math. Phys.},
publisher = {Springer},
volume = {289},
issue = {2},
url = {https://doi.org/10.1007/s00220-009-0787-3},
doi = {10.1007/s00220-009-0787-3}
}

@article{Guta:2007,
  title     = {Local asymptotic normality in quantum statistics},
  author    = {Mădălin Guţă and Anna Jenčová},
  journal   = {Commun. Math. Phys.},
  publisher = {Springer},
  volume    = {276},
  number    = {2},
  pages     = {341-379},
  year      =  {2007}
  }

@ARTICLE{Nurdin:2024,
  author={Nurdin, Hendra I.},
  journal={IEEE Control Systems Letters}, 
  title={Corrections to “Saturability of the Quantum Cramér-Rao Bound in Multiparameter Quantum Estimation at the Single-Copy Level”}, 
  year={2024},
  volume={8},
  number={},
  pages={2111-2113},
  doi={10.1109/LCSYS.2024.3451468}
  }

@article{Hayashi:2023_tightCRB,
  doi = {10.22331/q-2023-08-29-1094},
  url = {https://doi.org/10.22331/q-2023-08-29-1094},
  title = {Tight {C}ram{\'{e}}r-{R}ao type bounds for multiparameter quantum metrology through conic programming},
  author = {Hayashi, Masahito and Ouyang, Yingkai},
  journal = {{Quantum}},
  issn = {2521-327X},
  publisher = {{Verein zur F{\"{o}}rderung des Open Access Publizierens in den Quantenwissenschaften}},
  volume = {7},
  pages = {1094},
  month = aug,
  year = {2023}
}

@misc{Imai:2026_boundhierarchy,
      title={Hierarchy of saturation conditions for multiparameter quantum metrology bounds}, 
      author={Satoya Imai and Jing Yang and Luca Pezzè},
      year={2026},
      eprint={2602.12097},
      archivePrefix={arXiv},
      primaryClass={quant-ph},
      url={https://arxiv.org/abs/2602.12097}, 
}

@article{Matsumoto:2002,
doi = {10.1088/0305-4470/35/13/307},
url = {https://doi.org/10.1088/0305-4470/35/13/307},
year = {2002},
month = {mar},
publisher = {},
volume = {35},
number = {13},
pages = {3111},
author = {K Matsumoto},
title = {A new approach to the
Cramér-Rao-type bound of the pure-state model},
journal = {Journal of Physics A: Mathematical and General},
}

@article{Ragy:2016,
  title = {Compatibility in multiparameter quantum metrology},
  author = {Ragy, Sammy and Jarzyna, Marcin and Demkowicz-Dobrza\ifmmode \acute{n}\else \'{n}\fi{}ski, Rafa\l{}},
  journal = {Phys. Rev. A},
  volume = {94},
  issue = {5},
  pages = {052108},
  numpages = {11},
  year = {2016},
  month = {Nov},
  publisher = {American Physical Society},
  doi = {10.1103/PhysRevA.94.052108},
  url = {https://link.aps.org/doi/10.1103/PhysRevA.94.052108}
}

@article{Yang:2019,
  title = {Optimal measurements for quantum multiparameter estimation with general states},
  author = {Yang, Jing and Pang, Shengshi and Zhou, Yiyu and Jordan, Andrew N.},
  journal = {Phys. Rev. A},
  volume = {100},
  issue = {3},
  pages = {032104},
  numpages = {14},
  year = {2019},
  month = {Sep},
  publisher = {American Physical Society},
  doi = {10.1103/PhysRevA.100.032104},
  url = {https://link.aps.org/doi/10.1103/PhysRevA.100.032104}
}

@article{Albarelli:2020,
title = {A perspective on multiparameter quantum metrology: From theoretical tools to applications in quantum imaging},
journal = {Physics Letters A},
volume = {384},
number = {12},
pages = {126311},
year = {2020},
issn = {0375-9601},
doi = {https://doi.org/10.1016/j.physleta.2020.126311},
url = {https://www.sciencedirect.com/science/article/pii/S0375960120301109},
author = {F. Albarelli and M. Barbieri and M.G. Genoni and I. Gianani}
}

@misc{Conlon:2024_gap_persistence_theorem,
      title={The gap persistence theorem for quantum multiparameter estimation}, 
      author={Lorcán O. Conlon and Jun Suzuki and Ping Koy Lam and Syed M. Assad},
      year={2024},
      eprint={2208.07386},
      archivePrefix={arXiv},
      primaryClass={quant-ph},
      url={https://arxiv.org/abs/2208.07386}, 
}

@article{Albarelli:2019,
  title = {Evaluating the Holevo Cram\'er-Rao Bound for Multiparameter Quantum Metrology},
  author = {Albarelli, Francesco and Friel, Jamie F. and Datta, Animesh},
  journal = {Phys. Rev. Lett.},
  volume = {123},
  issue = {20},
  pages = {200503},
  numpages = {7},
  year = {2019},
  month = {Nov},
  publisher = {American Physical Society},
  doi = {10.1103/PhysRevLett.123.200503},
  url = {https://link.aps.org/doi/10.1103/PhysRevLett.123.200503}
}

@Article{manopt,
    author  = {Boumal, N. and Mishra, B. and Absil, P.-A. and Sepulchre, R.},
    journal = {Journal of Machine Learning Research},
    title   = {{M}anopt, a {M}atlab Toolbox for Optimization on Manifolds},
    year    = {2014},
    number  = {42},
    pages   = {1455--1459},
    volume  = {15},
    url     = {https://www.manopt.org}
}

@INPROCEEDINGS{Yalmip,
  author={Lofberg, J.},
  booktitle={2004 IEEE International Conference on Robotics and Automation (IEEE Cat. No.04CH37508)}, 
  title={YALMIP : a toolbox for modeling and optimization in MATLAB}, 
  year={2004},
  volume={},
  number={},
  pages={284-289},
  doi={10.1109/CACSD.2004.1393890}}

@article{Fujiwara:2006,
doi = {10.1088/0305-4470/39/40/014},
url = {https://doi.org/10.1088/0305-4470/39/40/014},
year = {2006},
month = {sep},
publisher = {},
volume = {39},
number = {40},
pages = {12489},
author = {Fujiwara, Akio},
title = {Strong consistency and asymptotic efficiency for adaptive quantum estimation problems},
journal = {Journal of Physics A: Mathematical and General}
}

@article{Mokeev:2026,
  title = {Enhancing Optical Imaging via Quantum Computation},
  author = {Mokeev, Aleksandr and Saif, Babak and Lukin, Mikhail D. and Borregaard, Johannes},
  journal = {PRX Quantum},
  volume = {7},
  issue = {1},
  pages = {010318},
  numpages = {18},
  year = {2026},
  month = {Jan},
  publisher = {American Physical Society},
  doi = {10.1103/s94k-929p},
  url = {https://link.aps.org/doi/10.1103/s94k-929p}
}

@article{DuanKimble:2004,
  title = {Scalable Photonic Quantum Computation through Cavity-Assisted Interactions},
  author = {Duan, L.-M. and Kimble, H. J.},
  journal = {Phys. Rev. Lett.},
  volume = {92},
  issue = {12},
  pages = {127902},
  numpages = {4},
  year = {2004},
  month = {Mar},
  publisher = {American Physical Society},
  doi = {10.1103/PhysRevLett.92.127902},
  url = {https://link.aps.org/doi/10.1103/PhysRevLett.92.127902}
}

@article{Fiderer:2021,
  title = {General Expressions for the Quantum Fisher Information Matrix with Applications to Discrete Quantum Imaging},
  author = {Fiderer, Lukas J. and Tufarelli, Tommaso and Piano, Samanta and Adesso, Gerardo},
  journal = {PRX Quantum},
  volume = {2},
  issue = {2},
  pages = {020308},
  numpages = {15},
  year = {2021},
  month = {Apr},
  publisher = {American Physical Society},
  doi = {10.1103/PRXQuantum.2.020308},
  url = {https://link.aps.org/doi/10.1103/PRXQuantum.2.020308}
}

@article{Tan:2023,
  title = {Quantum limit to subdiffraction incoherent optical imaging. III. Numerical analysis},
  author = {Tan, Xiao-Jie and Tsang, Mankei},
  journal = {Phys. Rev. A},
  volume = {108},
  issue = {5},
  pages = {052416},
  numpages = {8},
  year = {2023},
  month = {Nov},
  publisher = {American Physical Society},
  doi = {10.1103/PhysRevA.108.052416},
  url = {https://link.aps.org/doi/10.1103/PhysRevA.108.052416}
}

@article{Duplinskiy:2025,
  title={Tsang’s resolution enhancement method for imaging with focused illumination},
  author={Duplinskiy, Alexander and Frank, Jernej and Bearne, Kaden and Lvovsky, AI},
  journal={Light: Science \& Applications},
  volume={14},
  number={1},
  pages={159},
  year={2025},
  publisher={Nature Publishing Group UK London}
}

@misc{lvovsky:2026,
      title={Passive optical superresolution at the quantum limit}, 
      author={A. I. Lvovsky and Michael R. Grace and Saikat Guha and Mankei Tsang and Gerardo Adesso and Nicolas Treps},
      year={2026},
      eprint={2605.10767},
      archivePrefix={arXiv},
      primaryClass={quant-ph},
      url={https://arxiv.org/abs/2605.10767}, 
}

@article{Pushkina:2021,
  title = {Superresolution Linear Optical Imaging in the Far Field},
  author = {Pushkina, A. A. and Maltese, G. and Costa-Filho, J. I. and Patel, P. and Lvovsky, A. I.},
  journal = {Phys. Rev. Lett.},
  volume = {127},
  issue = {25},
  pages = {253602},
  numpages = {6},
  year = {2021},
  month = {Dec},
  publisher = {American Physical Society},
  doi = {10.1103/PhysRevLett.127.253602},
  url = {https://link.aps.org/doi/10.1103/PhysRevLett.127.253602}
}

@article{Tsang:2017,
doi = {10.1088/1367-2630/aa60ee},
url = {https://doi.org/10.1088/1367-2630/aa60ee},
year = {2017},
month = {feb},
publisher = {IOP Publishing},
volume = {19},
number = {2},
pages = {023054},
author = {Tsang, Mankei},
title = {Subdiffraction incoherent optical imaging via spatial-mode demultiplexing},
journal = {New Journal of Physics}
}

@article{Bearne:2021,
author = {Katherine K. M. Bearne and Yiyu Zhou and Boris Braverman and Jing Yang and S. A. Wadood and Andrew N. Jordan and A. N. Vamivakas and Zhimin Shi and Robert W. Boyd},
journal = {Opt. Express},
number = {8},
pages = {11784--11792},
publisher = {Optica Publishing Group},
title = {Confocal super-resolution microscopy based on a spatial mode sorter},
volume = {29},
month = {Apr},
year = {2021},
url = {https://opg.optica.org/oe/abstract.cfm?URI=oe-29-8-11784},
doi = {10.1364/OE.419493}
}

@article{Tan:2023_experiment,
author = {Xiao-Jie Tan and Luo Qi and Lianwei Chen and Aaron J. Danner and Pakorn Kanchanawong and Mankei Tsang},
journal = {Optica},
number = {9},
pages = {1189--1194},
publisher = {Optica Publishing Group},
title = {Quantum-inspired superresolution for incoherent imaging},
volume = {10},
month = {Sep},
year = {2023},
url = {https://opg.optica.org/optica/abstract.cfm?URI=optica-10-9-1189},
doi = {10.1364/OPTICA.493227}
}

@article{Lian:2026,
author = {Qiushuang Lian and Cilong Zhang and Qiaofeng Tan and Jun Zhu and Liangcai Cao},
journal = {Optica},
number = {5},
pages = {876--883},
publisher = {Optica Publishing Group},
title = {Incoherent superresolution via diffraction-based Hermite-Gaussian imaging},
volume = {13},
month = {May},
year = {2026},
url = {https://opg.optica.org/optica/abstract.cfm?URI=optica-13-5-876},
doi = {10.1364/OPTICA.584741}
}

@article{Tsang:2018_SPADE_semiclassical,
  title = {Subdiffraction incoherent optical imaging via spatial-mode demultiplexing: Semiclassical treatment},
  author = {Tsang, Mankei},
  journal = {Phys. Rev. A},
  volume = {97},
  issue = {2},
  pages = {023830},
  numpages = {18},
  year = {2018},
  month = {Feb},
  publisher = {American Physical Society},
  doi = {10.1103/PhysRevA.97.023830},
  url = {https://link.aps.org/doi/10.1103/PhysRevA.97.023830}
}

@article{Tsang:2019_ClassicalSemiparametric,
  title = {Semiparametric estimation for incoherent optical imaging},
  author = {Tsang, Mankei},
  journal = {Phys. Rev. Res.},
  volume = {1},
  issue = {3},
  pages = {033006},
  numpages = {14},
  year = {2019},
  month = {Oct},
  publisher = {American Physical Society},
  doi = {10.1103/PhysRevResearch.1.033006},
  url = {https://link.aps.org/doi/10.1103/PhysRevResearch.1.033006}
}

@article{Bisketzi:2019,
doi = {10.1088/1367-2630/ab58a0},
url = {https://doi.org/10.1088/1367-2630/ab58a0},
year = {2019},
month = {dec},
publisher = {IOP Publishing},
volume = {21},
number = {12},
pages = {123032},
author = {Bisketzi, Evangelia and Branford, Dominic and Datta, Animesh},
title = {Quantum limits of localisation microscopy},
journal = {New Journal of Physics}
}

@article{Tsang:2021_PoissonStates,
  doi = {10.22331/q-2021-08-19-527},
  url = {https://doi.org/10.22331/q-2021-08-19-527},
  title = {Poisson {Q}uantum {I}nformation},
  author = {Tsang, Mankei},
  journal = {{Quantum}},
  issn = {2521-327X},
  publisher = {{Verein zur F{\"{o}}rderung des Open Access Publizierens in den Quantenwissenschaften}},
  volume = {5},
  pages = {527},
  month = aug,
  year = {2021}
}

@misc{Gardner:2026,
      title={Quantum superresolution and noise spectroscopy with quantum computing}, 
      author={James W. Gardner and Federico Belliardo and Gideon Lee and Tuvia Gefen and Liang Jiang},
      year={2026},
      eprint={2602.17862},
      archivePrefix={arXiv},
      primaryClass={quant-ph},
      url={https://arxiv.org/abs/2602.17862}, 
}

@inproceedings{Richardson:2026,
author = {J. Gabriel Richardson and Prajit Dhara and Aqil Sajjad and Saikat Guha},
booktitle = {CLEO 2026},
journal = {CLEO 2026},
pages = {JW1.117},
publisher = {Optica Publishing Group},
title = {Memory Efficient Atom-Photon Interface for Unary Encoded Photonic Qudits},
year = {2026},
url = {https://opg.optica.org/abstract.cfm?URI=CLEO_AT-2026-JW1.117},
doi = {10.1364/CLEO_AT.2026.JW1.117},
}

@ARTICLE{Ben-Haim:2009,
  author={Ben-Haim, Zvika and Eldar, Yonina C.},
  journal={IEEE Signal Processing Letters}, 
  title={On the Constrained CramÉr–Rao Bound With a Singular Fisher Information Matrix}, 
  year={2009},
  volume={16},
  number={6},
  pages={453-456},
  doi={10.1109/LSP.2009.2016831}}

@article{Krovi:2019,
  doi = {10.22331/q-2019-02-14-122},
  url = {https://doi.org/10.22331/q-2019-02-14-122},
  title = {An efficient high dimensional quantum {S}chur transform},
  author = {Krovi, Hari},
  journal = {{Quantum}},
  issn = {2521-327X},
  publisher = {{Verein zur F{\"{o}}rderung des Open Access Publizierens in den Quantenwissenschaften}},
  volume = {3},
  pages = {122},
  month = feb,
  year = {2019}
}

@article{Bacon:2006,
  title = {Efficient Quantum Circuits for Schur and Clebsch-Gordan Transforms},
  author = {Bacon, Dave and Chuang, Isaac L. and Harrow, Aram W.},
  journal = {Phys. Rev. Lett.},
  volume = {97},
  issue = {17},
  pages = {170502},
  numpages = {4},
  year = {2006},
  month = {Oct},
  publisher = {American Physical Society},
  doi = {10.1103/PhysRevLett.97.170502},
  url = {https://link.aps.org/doi/10.1103/PhysRevLett.97.170502}
}

@article{Cui:2025,
  title={Superadditive communication with the green machine as a practical demonstration of nonlocality without entanglement},
  author={Cui, Chaohan and Postlewaite, Jack and Saif, Babak N and Fan, Linran and Guha, Saikat},
  journal={Nature Communications},
  volume={16},
  number={1},
  pages={3760},
  year={2025},
  publisher={Nature Publishing Group UK London}
}

@misc{Tsang:2026_NH,
  author       = {Tsang, Mankei},
  title        = {Cram\'er–Rao-type bounds for quantum multiparameter estimation with separable measurements},
  howpublished = {\url{https://docs.google.com/viewerng/viewer?url=https://blog.nus.edu.sg/mankei/files/2026/08/hayashi-matsumoto2.pdf&hl}},
  year         = {2026}
}

@article{Almeida:2025_SuperresCollective,
  title = {Superresolving collective quantum measurements},
  author = {de Almeida, J. O. and Lewenstein, M. and Skotiniotis, M.},
  journal = {Phys. Rev. A},
  volume = {112},
  issue = {5},
  pages = {052605},
  numpages = {14},
  year = {2025},
  month = {Nov},
  publisher = {American Physical Society},
  doi = {10.1103/9w4p-5d9j},
  url = {https://link.aps.org/doi/10.1103/9w4p-5d9j}
}

\clearpage
\onecolumngrid

\setcounter{equation}{0}
\setcounter{figure}{0}
\setcounter{table}{0}
\setcounter{section}{0}
\renewcommand{\theequation}{S\arabic{equation}}
\renewcommand{\thefigure}{S\arabic{figure}}
\renewcommand{\thetable}{S\arabic{table}}
\renewcommand{\thesection}{S\arabic{section}}
\renewcommand{\thesubsection}{\thesection.\arabic{subsection}}
\makeatletter
\renewcommand{\p@subsection}{}
\makeatother

\begin{center}
\textbf{\large Supplemental Material}
\end{center}

\noindent This supplement contains derivations, accompanied by interpretations, of the primary equations in the Letter \textit{Attaining Fundamental Limits of Multiparameter Incoherent Optical Imaging Using Joint Quantum Measurements}. Specifically, we derive (1) the input-output relationships between populated spatial modes before and after transmitting through an arbitrary aperture, (2) the quantum Fisher information matrix (QFIM) and SLDs for brightness parameters of a $K$-emitter ensemble, (3) the classical Cram\'er-Rao bounds (CRB) for brightness estimation using SPADE and Direct Imaging, (4) the multi-copy interaction Hamiltonian for executing an asymptotically optimal joint measurement on multiple bosonic ancillas. Additionally, we provide definitions for the quantum CRBs highlighted in the main text and prove that they are invariant under an appropriate coordinate definition over the simplex.

\tableofcontents

\let\oldsection\section
\renewcommand\section[1]{\refstepcounter{section}\oldsection{\thesection.\ #1}}
\let\oldsubsection\subsection
\renewcommand\subsection[1]{\refstepcounter{subsection}\oldsubsection{\thesubsection.\ #1}}
\newpage

\begin{figure}
    \centering
    \includegraphics[width=.8\linewidth]{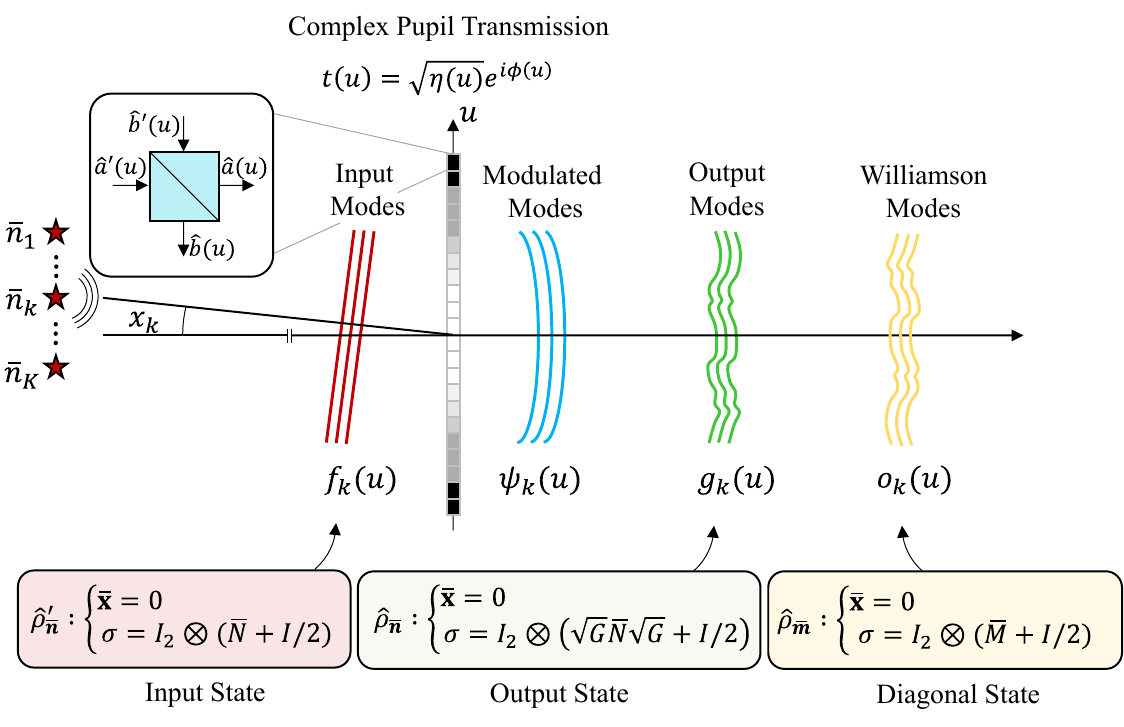}
    \caption{Illustration of the physical model for optical modes passing through an aperture with complex transmission $t(u)$. Each point $u$ on the mask is modeled as an infinitesimal splitter with transmissivity $t(u)$ that mixes the input modes $\hat{a}'(u)$ and ancillas $\hat{b}'(u)$, yielding the output modes $\hat{a}(u)$ and ancillas $\hat{b}(u)$. Additionally, assuming a thermal state input, each of colored block shows the Gaussian moments of the state when defined over each mode basis.}
    \label{fig: Transmission Mask Model}
\end{figure}

\section{Gaussian State Model}
\label{ss: Gaussian State Model}
In this section we consider the evolution of a set of orthonormal input modes through a passive complex transmission mask $t(u)=\sqrt{\eta(u)}e^{i\phi(u)}$ defined at the pupil plane of an imaging system. Our approach involves treating pupil plane as an array of infinitesimally small beam splitters characterized \textit{locally} at the point $u$ by the transmissivity function $t(u)$ as depicted in Fig. \ref{fig: Transmission Mask Model}. Assuming a quasi-monochromatic paraxial field, let $\{f_{k}(u)\}_{k=1}^{K}$ be a finite set of square-integrable orthonormal spatial modes incident on the pupil and satisfying, 
$$
(f_j|f_k) = \int du f_{j}^{*}(u)f_{k}(u) = \delta_{jk}.
$$
We refer to this set of modes as the {\em input mode} basis. Directly after the pupil the input modes become,
\begin{equation}
\psi_{k}(u) = t(u)f_{k}(u)
\label{eq:modulated_modes}
\end{equation}
which we refer to as the {\em modulated modes}. For any realistic pupil transmissivity function $t(u)$ that attenuates the field (i.e. there exist regions of the pupil with non-zero measure for which $\eta(u)<1$) the propagation of the input modes through the pupil destroys their original orthogonality. We collect the overlap between the modulated modes into a Gram matrix,
\begin{equation}
G_{jk} = (\psi_j|\psi_{k}) = \int du \, \eta(u) f_{j}^{*}(u)f_k(u),
\label{eq:gram_matrix}
\end{equation}
which is a positive-semidefinite matrix. Since transmission through the aperture is purely a lossy process, the eigenvalues of $G$ are necessarily less than or equal to one such that $\det G\leq 1$. The diagonal elements $G_{kk}\in[0,1]$ thus correspond to the mode-dependent transmissivity of $f_k(u)$ through the aperture. In the special case where the incident modes are plane waves $f_{k}(u) = \frac{1}{\sqrt{2\pi}}e^{ix_{k}u}$, it is standard to calculate the Gram matrix from the normalized version of the modulated modes $\psi_{k}(u)\rightarrow \psi_{k}(u)/\sqrt{(\psi_k|\psi_k)}$  since plane-waves are not square-integrable and the attenuation happens identically for all incident angles. While we make this simplification in the main text, here our derivation makes no such approximation.

Let $\{g_k(u)\}_{k=1}^{K}$ be an orthonormal mode basis over $\overline{\text{span}}\{\psi_{k}(u)\}_{k=1}^{K}$ satisfying $(g_j|g_k)=\delta_{jk}$. We refer to this as the {\em output mode} basis. Expanding each modulated mode in the output mode basis gives,
$$
\psi_k(u) = \sum_{j=1}^{K}T_{jk}g_j(u),\qquad T_{jk} = (g_j|\psi_k)  = \sqrt{\eta_{jk}}e^{i\phi_{jk}}.
$$
Note that the matrix $T$ may also be interpreted as the aperture-mediated coupling between input and output modes via $T_{jk} = (g_j|\psi_k) = (g_j|T|f_k)$. We emphasize that, throughout this section, $T$ is {\em not} the same as the tangent space projector introduced in the main text. Using the closure identity $I =\sum_{k=1}^{K}|g_k)(g_k|$ on $\overline{\text{span}}\{\psi_{k}(u)\}_{k=1}^{K}$, we find,
\begin{equation}
G = TT^{\dagger}.
\end{equation}
Importantly, even though $\overline{\text{span}}\{f_k(u)\}\neq \overline{\text{span}}\{g_k(u)\}$, we only need to keep track of the output modes since photons in the populated modes $\{f_{k}\}$ are guaranteed to end up in the span of $\{g_k\}$ modes after modulation by the pupil transmissivity function. The transmission matrix $T_{jk}$ models the interaction between these two mode spaces.

We now perform a Stinespring dilation of the modal Hilbert space and introduce ancillary vacuum modes to make the interaction between mode spaces unitary. The ancillary vacuum modes serve as a reservoir to account for the energy lost during the interaction with the pupil transmission. The annihilation operators of the input modes $\hat{a}_{k}'\leftrightarrow f_k(u)$ are given by
\begin{equation}
\hat{a}_k' = \int du f^{*}_{k}(u)\hat{a}'(u),
\label{eq:input_mode_field_op}
\end{equation}
where $\hat{a}'(u)$ are local annihilation operators over the pupil satisfying canonical commutation relation $[\hat{a}'(u),\hat{a}'^{\dagger}(u')]=\delta(u-u')$. As shown in Fig. \ref{fig: Transmission Mask Model}, the operators $\hat{a}'(u)$ are understood to evolve under a \textit{local} beam splitter interaction with vacuum ancilla $\hat{b}'(u)$ as,
$$
\begin{bmatrix}
    \hat{a}(u) \\
    \hat{b}(u)
\end{bmatrix} =\underbrace{
\begin{bmatrix}
    t(u) & r(u) \\
    -r^{*}(u) & t^{*}(u)
\end{bmatrix}}_{B(u)}
\begin{bmatrix}
    \hat{a}'(u) \\
    \hat{b}'(u)
\end{bmatrix},
$$
where $r(u) = \sqrt{1-\eta(u)}e^{i\phi(u)}$. The local beamsplitter interaction matrix $B(u)$ is unitary $B(u)B(u)^\dagger = B(u)^{\dagger}B(u) = I_{2}$. Our goal is to determine how the interaction matrix $T$ between input and output modes appears in the evolution of the optical field. First, we complete the input mode basis $\{f_{1},\ldots,f_{K},\ldots,f_{\infty}\}$ such that it admits a closure relation on the full Hilbert space of square-integrable spatial modes
$$
\delta(u-u') = \sum_{k=1}^{\infty}f_{k}(u)f_k^{*}(u') 
$$
Then we have the identity,
\begin{align*}
    \hat{a}'(u) &= \int du' \delta(u-u')\hat{a}'(u')\\
    &= \int du'\bigg(\sum_{k=1}^{\infty}f_{k}(u)f_{k}^{*}(u')\bigg) \hat{a}'(u') \\    &=\sum_{k=1}^{\infty}f_k(u)\hat{a}'_k  = \sum_{k=1}^{K}f_k(u)\hat{a}'_k + \hat{a}'_{\perp}(u)
\end{align*}
After the evolution, the output modes are given by,
\begin{align*}
    \hat{a}_j &= \int\, du\, g_{j}^{*}(u)\hat{a}(u) \\
    &=\int du\, g_{j}^{*}(u)[t(u)\hat{a}'(u) + r(u)\hat{b}'(u)] \\
    &= \int du\, g_{j}^{*}(u)\bigg[ t(u) \bigg(\sum_{k=1}^{K}f_{k}(u)\hat{a}'
    _k + \hat{a}_{\perp}'(u)\bigg) + r(u)\hat{b}'(u) \bigg] \\
    &= \sum_{k=1}^{K}T_{jk}\hat{a}'_k + \int du \,g^{*}_{j}(u)[t(u)\hat{a}'_{\perp}(u)+ r(u)\hat{b}'(u)] \\
    &= \sum_{k=1}^{K} T_{jk}\hat{a}'_k + r_j \hat{v}'_j,
\end{align*}
where $\hat{v}'_j$ is composed entirely of vacuum modes and has the property that $[\hat{v}'_j,\hat{a}_{k}'^{\dagger}]=0$ for $k=1,\ldots,K$. We massage this expression into a canonical beam splitter interaction by introducing the ancillary vacuum modes $\hat{b}_{k}'$ over the span of modes $\hat{v}_{j}'$ with canonical commutation relation $[\hat{b}_{j}',\hat{b}_{k}'^{\dagger}]=\delta_{jk}$ and orthogonality $[\hat{a}'_j,\hat{b}_{k}'^{\dagger}]=0$. Then we may rewrite $r_j \hat{v}'_j = \sum_{k}R_{jk}\hat{b}'_{k}$. The output mode operators are thus related to the input mode operators via,
\begin{subequations}
\begin{align}
\hat{a}_{j} &= \sum_{k=1}^{K} \bigg(T_{jk}\hat{a}_{k}' + R_{jk}\hat{b}_{k}'\bigg), \\
\hat{b}_{j} &= \sum_{k=1}\bigg(- R_{jk}^{*}\hat{a}_{k}' + T_{jk}^{*}\hat{b}_{k}'\bigg),
\end{align}
\end{subequations}
which may be expressed in a block matrix equation
\begin{equation}
\begin{bmatrix}
    \hat{\mathbf{a}} \\
    \hat{\mathbf{b}}
\end{bmatrix}  = 
\underbrace{\begin{bmatrix}
     T & R \\
    -R^{*} & T^{*}
\end{bmatrix}}_{B}
\begin{bmatrix}
    \hat{\mathbf{a}}' \\
    \hat{\mathbf{b}}'
\end{bmatrix} .
\label{eq:block_beamsplitter}
\end{equation}
Imposing the canonical commutation relation on the output mode annihilation operators $[\hat{a}_{j},{\hat{a}}_{k}^{\dagger}] = [\hat{b}_{j},{\hat{b}}_{k}^{\dagger}]= \delta_{jk}$ and $[\hat{a}_{j},{\hat{b}}_{k}^{\dagger}] = 0$ gives rise to the constraints,
\begin{equation}
TT^\dagger + RR^\dagger = T^\dagger T+ R^\dagger R = I,\qquad TR^{\intercal}=RT^{\intercal}. 
\label{eq:transmission_reflection_constraints}
\end{equation}
In turn, these imply that  $B$ is unitary matrix: $BB^{\dagger} = B^\dagger B = I$. At this point, we can make a convenient choice for the output mode basis $\{g_k\}$ that gives an explicit form for $T$ and $R$ with respect to quantities defined so far. In particular, we take the transmission matrix $T$ to be the positive square root of the Gram matrix $T = \sqrt{G}$. Specifically, since $G$ is a positive semidefinite Hermitian matrix, it admits a spectral decomposition $G = UYU^\dagger$ where the eigenvalues of the diagonal matrix $Y$ are real and non-negative. We therefore pick $T = \sqrt{G} = U\sqrt{Y}U^\dagger$ to be the {\em positive semidefinite square root} of $G$ (i.e. we take the positive square roots of the eigenvalues). Moreover, since the eigenvalues of $G$ are less than or equal to one, it is readily shown that the matrix $I-G$ is also positive semidefinite: 
$$
I-G = I-UYU^\dagger = U(I-Y)U^\dagger,
$$
where $I-Y\geq 0$. Therefore, we can similarly define the reflection matrix $R = \sqrt{I-G}$ as the positive semidefinite square root of the matrix $I-G$. We will henceforth use,
\begin{equation}
T = \sqrt{G}, \qquad R = \sqrt{I-G},
\end{equation}
which is consistent with all the constraints of Eq. \ref{eq:transmission_reflection_constraints}. With this choice, the unitary matrix relating input and output modes assumes the form,
$$
\begin{bmatrix}
    \hat{\mbf{a}}\\
    \hat{\mbf{b}}
\end{bmatrix} = 
\underbrace{\begin{bmatrix}
    \,\,\,\sqrt{G}& \sqrt{I-G} \\
    -\sqrt{I-G} & \sqrt{G}
\end{bmatrix}}_{B} \begin{bmatrix}
    \hat{\mbf{a}}'\\
    \hat{\mbf{b}}'
\end{bmatrix}.
$$
Moreover, since $T$ and $R$ are chosen to be the positive semidefinite square roots, they are also Hermitian matrices $T=T^\dagger$ and $R=R^\dagger$.
Given this simplified beam-splitter relation, our goal will be to evolve a Gaussian state defined on the input modes through the unitary channel given by the beam-splitter matrix $B$ and subsequently trace over the ancillary mode space. To that end, we will now invoke the Gaussian state formalism on the field quadratures defined as
\begin{equation}
\hat{q}_j = (\hat{a}_j+ \hat{a}_j^\dagger)/\sqrt{2} ,\qquad \hat{p}_j = -i(\hat{a}_j-\hat{a}_j^\dagger)/\sqrt{2}
\label{eq:quadrature_convention}
\end{equation}
which satisfy canonical commutation relation $[\hat{q}_j,\hat{p}_k] = i \delta_{jk}$. We will use the `$qqpp$' ordering with separation between physical modes and ancillary modes such that the quadrature vector is given by,
\begin{equation}
\hat{\mbf{x}} = 
\begin{bmatrix}
\hat{\mbf{q}}_{a}\\
\hat{\mbf{p}}_{a}\\
\hat{\mbf{q}}_{b}\\
\hat{\mbf{p}}_{b}
\end{bmatrix} = \underbrace{\frac{1}{\sqrt{2}}
\begin{bmatrix}
I & I & 0 & 0 \\
-iI & iI & 0 & 0\\
0 & 0 & I & I  \\
0 & 0 &-iI & iI 
\end{bmatrix}}_{U}
\begin{bmatrix}
    &\hat{\mbf{a}}\\
    &\hat{\mbf{a}}^\dagger\\
    &\hat{\mbf{b}}\\
    &\hat{\mbf{b}}^\dagger\\
\end{bmatrix}
\label{eq:field_operator_to_quadrature}
\end{equation}
The displacement vector and the covariance matrix of a Gaussian state are defined as
\begin{equation}
\bar{\mbf{x}} = \langle \hat{\mbf{x}}\rangle \qquad \sigma_{jk} = \frac{1}{2}\langle \{ \hat{x}_{j}-\bar{x}_{j},\hat{x}_{k}-\bar{x}_j \}\rangle
\label{eq:displacement_covariance_convention}
\end{equation}
where $\{\cdot,\cdot\}$ is the anti-commutator.
Extending the field operator relations of Eq. \ref{eq:block_beamsplitter} to include the creation operators, we get
\begin{equation}
\begin{bmatrix}
    \hat{\mbf{a}}\\
    \hat{\mbf{a}}^\dagger\\
    \hat{\mbf{b}}\\
    \hat{\mbf{b}}^\dagger
\end{bmatrix} =\underbrace{ 
\begin{bmatrix}
    T & 0 &R & 0 \\
    0 & T^{*} & 0 & R^{*} \\
    -R & 0 & T & 0 \\
    0 & -R^{*} & 0 & T^{*}    
\end{bmatrix}}_{\tilde{B}}
\begin{bmatrix}
    &\hat{\mbf{a}}'\\
    &\hat{\mbf{a}}'^\dagger\\
    &\hat{\mbf{b}}'\\
    &\hat{\mbf{b}}'^\dagger\\
\end{bmatrix}.
\label{eq:field_operator_evolution}
\end{equation}
Combining Eqs. \ref{eq:field_operator_to_quadrature} and \ref{eq:field_operator_evolution}, the conversion between canonical quadratures in the input and output mode bases is given by a symplectic transformation,
$$
\hat{\mbf{x}} = S \hat{\mbf{x}}' = U\tilde{B}U^{-1}\hat{\mbf{x}}',
$$
where $U^{-1} = U^{\dagger}$ is the inverse of the linear map from field operators to quadratures and the symplectic matrix is given by the block form,
\begin{equation}
S = 
\begin{bmatrix}
 \Re{T} & -\Im{T} &| & \Re{R} & -\Im{R} \\
 \Im{T} &  \Re{T} &| &  \Im{R} &  \Re{R} \\
 \hline
-\Re{R} & \Im{R} &| &  \Re{T} &  -\Im{T} \\
 -\Im{R} & -\Re{R} &| & \Im{T} &  \Re{T}
\end{bmatrix}.
\label{eq:symplectic_transform}
\end{equation}
Assuming the input modes $\hat{a}_{k}'$ are excited in a Gaussian state and the ancilla modes $\hat{b}_{k}'$ are in vacuum such that the input density operator (the state entering the pupil) may be written as $\hat{\rho}_{ab}' = \hat{\rho}_{a}'\otimes\dyad{\mathbf{0}}$, the field is initially in a Gaussian state with displacement vector and covariance
$$
\bar{\mbf{x}}' = 
\begin{bmatrix}
\bar{\mbf{x}}_{a}' \\
\mbf{0}
\end{bmatrix},\qquad 
\sigma =
\begin{bmatrix}
\sigma_{a}' & 0 \\
0 & \tfrac{1}{2}I
\end{bmatrix}
$$
The beam splitter transformation constitutes a Gaussian channel. Therefore, the state of the field after passing through the pupil is also Gaussian with mean and covariance
$$
\bar{\mbf{x}} = S\bar{\mbf{x}}',\qquad \sigma = S\sigma' S^{\intercal}.
$$
To make the partition of the populated and ancillary vacuum modes in the input mode space more explicit, we note that the mean displacement vector transforms to,
$$
\bar{\mbf{x}} =
\begin{bmatrix}
    \bar{\mbf{x}}_{a}\\
    \bar{\mbf{x}}_{b}
\end{bmatrix} = \underbrace{
\begin{bmatrix}
    S_{a} & S_{ab}\\
    S_{ba} & S_{b}
\end{bmatrix}}_{S}
\begin{bmatrix}
    \bar{\mbf{x}}_{a}'\\
    \mbf{0}
\end{bmatrix} =
\begin{bmatrix}
    S_{a}\,\bar{\mbf{x}}_{a}'\\
    S_{ba}\bar{\mbf{x}}_{a}'
\end{bmatrix},
$$
while the covariance matrix transforms to,

\begin{align*}
\sigma &= 
\begin{bmatrix}
\sigma_{a} &  \sigma_{ab}\\
\sigma_{ba} & \sigma_{b}
\end{bmatrix} = 
\begin{bmatrix}
S_{a} & S_{ab} \\
S_{ba} & S_{b}
\end{bmatrix}
\begin{bmatrix}
\sigma_{a}' & 0 \\
0 & \tfrac{1}{2}I
\end{bmatrix}
\begin{bmatrix}
S_{a}^{\intercal} & S_{ba}^{\intercal} \\
S_{ab}^{\intercal} & S_{b}^{\intercal}
\end{bmatrix} \\
&=
\begin{bmatrix}
\big(S_{a}\sigma_{a}'S_{a}^{\intercal} + \tfrac{1}{2} S_{ab} S_{ab}^{\intercal}\big) & \big(S_{a}\sigma_{a}'S_{ba}^{\intercal} + \tfrac{1}{2} S_{ab}S_{b}^{\intercal}\big)\\
\big(S_{ba}\sigma_{a}'S_{a}^{\intercal} + \tfrac{1}{2}S_{b}S_{ab}^{\intercal}\big) & \big(S_{ba}\sigma_{a}'S_{ba}^{\intercal} + \tfrac{1}{2}S_{b}S_{b}^{\intercal}\big)
\end{bmatrix}
\end{align*}
Since the ancillary modes $\hat{b}_{j}$ are a mathematical convenience, not a physically available resource, we trace over these modes leading to a Gaussian state on the output modes $\hat{a}_j$ with reduced Gaussian parameters,
\begin{subequations}
\begin{align}
\bar{\mbf{x}} & \leftarrow \bar{\mbf{x}}_{a} = S_{a}\bar{\mbf{x}}_a'\\ 
\sigma & \leftarrow \sigma_{a} = S_{a}\sigma_{a}'S_{a}^{\intercal} + \tfrac{1}{2}S_{ab} S_{ab}^{\intercal},
\end{align}
\label{eqn: Evolved Gaussian Moments}
\end{subequations}
where,
\begin{subequations}
\begin{align}
    S_{a} &= 
    \begin{bmatrix}
     \Re\sqrt{G} & -\Im\sqrt{G} \\
     \Im\sqrt{G} & \Re\sqrt{G}
    \end{bmatrix}\\
        S_{ab} &= 
    \begin{bmatrix}
     \Re\sqrt{I-G} & -\Im\sqrt{I-G} \\
     \Im\sqrt{I-G} & \Re\sqrt{I-G}
    \end{bmatrix}.
\end{align}
\label{eq:symplectic sub-blocks}
\end{subequations}
In the evolved covariance matrix $\sigma_{a}$, the term $S_{ab}S_{ab}^{\intercal}$ results from the noise added from mixing with the ancillary vacuum modes. Physically, this accounts for the photons in the populated modes $\{f_k\}$ which are lost to the aperture. We have modeled the aperture as a pure-loss quantum channel by introducing a beam splitter interaction with ancillary modes and then `forgetting' the state of the ancillas.

\subsection{Thermal States}
In the special case where the Gram matrix is real such that $G_{jk} = G_{kj}$, the symplectic transformation simplifies dramatically. The Gram matrix is real (and symmetric) if the amplitude transmissivity function is inversion symmetric $\eta(-u) = \eta(u)$ and the input modes $f_k(u)$ are Hermitian $f_k(-u)=f_k(u)^{*}$. The plane waves $f_{k}(u)=\frac{1}{\sqrt{2\pi}}e^{ir_{k}u}$ used in the main text are an example of a set of Hermitian input modes. Under these conditions, it is easily verified that $G$ is real symmetric,
\begin{align*}
G_{jk} &= \int_{-\infty}^{\infty} du \, \eta(u)f^{*}_{j}(u)f_{k}(u) = \int_{-\infty}^{\infty} du\,\eta(-u) f_{j}(-u)f_{k}(-u)^{*} \\
&= - \int_{\infty}^{-\infty} du' \, \eta(u') f_{j}(u')f_{k}(u')^{*} = \int_{-\infty}^{\infty}du \, \eta(u)f_{j}(u)f_{k}^{*}(u) =G_{kj} 
\end{align*}
 In this case, the relevant blocks of the symplectic matrix $S$ in Eq. \ref{eq:symplectic sub-blocks} reduce to,
$$
S_{a} = I_2\otimes \sqrt{G}
,\qquad S_{ab} = I_2\otimes \sqrt{I-G}.
$$
Invoking these in Eq. \ref{eqn: Evolved Gaussian Moments} gives,
\begin{subequations}
\begin{align}
\bar{\mbf{x}} &= (I_{2}\otimes\sqrt{G})\bar{\mbf{x}}'\\
\sigma &= (I_{2}\otimes \sqrt{G})\sigma' (I_2 \otimes \sqrt{G}) + I_2\otimes \tfrac{1}{2}(I-G).
\end{align}
\label{eq: General Output Moments}
\end{subequations}
Let us further suppose that the input state is a tensor product thermal state,
$$
\hat{\rho}_{\bar{\mbf{n}}}' = \bigotimes_{k=1}^{K}\frac{1}{(\bar{n}_{k}+1)}\bigg(\frac{\bar{n}_{k}}{\bar{n}_{k}+1}\bigg)^{\hat{n}_{k}'}
$$
where $\hat{n}_{k}'\equiv\hat{a}_{k}'^{\dagger}\hat{a}_{k}'$ are the number operators for the input modes. This state has Gaussian parameters,
\begin{align*}
\bar{\mbf{x}}' &= \mbf{0}\\
\sigma' &= I_2\otimes(\bar{N} + \tfrac{1}{2}I),
\end{align*}
where $\bar{N} = \text{Diag}(\bar{n}_1,\ldots,\bar{n}_{K})$ from the main text. Inserting these definitions into Eq. \ref{eq: General Output Moments}, we find that the Gaussian parameters over the output modes are given by
\begin{subequations}
\begin{align}
\bar{\mbf{x}} &= \mbf{0},\label{eq:gauss_params_maintext_xbar}\\
\sigma &=I_{2}\otimes(\sqrt{G}\bar{N}\sqrt{G} + \tfrac{1}{2}I), \label{eq:gauss_params_maintext_sigma}
\end{align}
\end{subequations}
which is a primary result reported in the main text.
Comparing the input and output covariance matrices, $\sigma' = I_2 \otimes(\bar{N}+\tfrac{1}{2}I)$ and $\sigma=I_{2}\otimes(\sqrt{G}\bar{N}\sqrt{G} + \tfrac{1}{2}I)$, the effect of the aperture is easily interpretable. The aperture attenuates and redistributes the photons in the orthonormal input modes $f_k(u)$ to the orthonormal output modes $g_j(u)$ in a correlated manner (off-diagonal terms in the covariance matrix are non-zero). These correlations are directly related to the overlaps of the modulated modes $\psi_k(u)$ captured by the Gram matrix. 

\subsection{Williamson Decomposition}
\noindent In this section we derive the Williamson decomposition of the post-aperture Gaussian state defined by the displacement vector and covariance matrix of Eqs. \ref{eq:gauss_params_maintext_xbar} and \ref{eq:gauss_params_maintext_sigma}. Any positive definite matrix $\sigma$ admits a symplectic diagonalization,
$$
\sigma = SDS^\intercal,
$$
where $S$ is symplectic (i.e. $S\Omega S^\intercal = \Omega$) and $D = I_{2}\otimes \mathcal{V}$ is a diagonal matrix of symplectic eigenvalues $\{\nu_{k}\}_{k=1}^{K}$ satisfying $\nu_k\geq \tfrac{1}{2}$. Additionally, under the `$qqpp$' ordering the primitive symplectic representation matrix is,
$$
\Omega = \omega\otimes I_{K},\qquad \omega\equiv
\begin{bmatrix}
 0 & 1\\
 -1 & 0
\end{bmatrix}
$$
Applying Williamson's theorem to the covariance matrix of Eq. \ref{eq:gauss_params_maintext_sigma}, it is straightforward to verify that,
$$
S = I_{2}\otimes O,\qquad D = I_{2}\otimes \underbrace{(\bar{M}+\tfrac{1}{2}I)}_{\mathcal{V}}
$$
where $O$ is an orthogonal matrix and $\bar{M}=\text{Diag}(\bar{m}_1,\ldots,\bar{m}_{K})$ is a diagonal matrix obtained from the spectral decomposition.
$$
\sqrt{G}\bar{N}\sqrt{G} = O\bar{M}O^{\intercal}.
$$
First, we show that the hypothesized Williamson decomposition recovers the original covariance matrix:
\begin{align*}
SDS^{\intercal} &= I_{2}\otimes (O(\bar{M}+\tfrac{1}{2}I)O^{\intercal})\\
&=I_{2}\otimes( O\bar{M}O^{\intercal} + \tfrac{1}{2}I)\\
&= I_{2}\otimes(\sqrt{G}\bar{N}\sqrt{G}+\tfrac{1}{2}I)\\
&= \sigma.
\end{align*}
Next, we show that $S$ is indeed symplectic:
\begin{align*}
S\Omega S^{\intercal} &= (I_{2}\otimes O)(\omega\otimes I)(I_{2}\otimes O^{\intercal}) \\
&=\omega \otimes(OO^{\intercal}) = \omega\otimes I_{K}\\
&= \Omega.
\end{align*}
Finally, we show that the symplectic eigenvalues satisfy the uncertainty limit $\nu_{k} =\bar{m}_{k}+\tfrac{1}{2} \geq \frac{1}{2}$. This follows directly from the fact that $\sqrt{G}\bar{N}\sqrt{G}$ is positive semidefinite:
\begin{align*}
    0&\leq \sqrt{G}\bar{N}\sqrt{G} = O\bar{M}O^{\intercal}\\
    \implies 0&\leq \bar{m}_{k} \\
    \implies \tfrac{1}{2} &\leq \bar{m}_k + \tfrac{1}{2}=\nu_{k}.
\end{align*}
The physical meaning of $\bar{\mbf{m}}$ is now abundantly clear. It represents the mean photon numbers of the uncorrelated tensor product thermal states defined on Williamson modes. Furthermore, the mean photon number before and after the aperture is 
\begin{subequations}
    \begin{align}
    \bar{n}_0 &= \tr \bar{N} = \sum_{k=1}^{K}\bar{n}_k\\
 \bar{m}_0 &=\tr \bar{M} = \sum_{k=1}^{K}G_{kk}\bar{n}_k
    \end{align}
    \label{eq:mean_photons}
\end{subequations}
The total mean photon number in the output modes $\bar{m}_0$ corresponds to the sum of mean photons $\bar{n}_k$ supplied by the input state weighted by the mode-dependent transmissivity $G_{kk}$. After this diagonalization, we may write the output state once again as a tensor product of thermal states
\begin{equation}
    \hat{\rho}_{\bar{\mbf{n}}} \equiv  \bigotimes_{k=1}^{K}\frac{1}{(\bar{m}_{k}+1)}\bigg(\frac{\bar{m}_{k}}{\bar{m}_{k}+1}\bigg)^{\hat{m}_{k}}
\end{equation}
where $\hat{m}_{k} = \hat{c}_{k}^{\dagger}\hat{c}_{k}$ are the number operators for the Williamson modes with annihilation $\hat{c}_{k}\leftrightarrow o_{k}(u)$. We now seek to determine the explicit form of $o_{k}(u)$ and $\hat{c}_{k}$ in terms of the output mode basis $g_{k}(u)$ and $\hat{a}_{k}$. The electric field over the pupil plane may be expressed as,
$$
\hat{E}^{(+)}(u) = \sum_{k}o_{k}(u)\hat{c}_{k}
$$
This quantity should be invariant under a unitary change of mode basis which is commensurately accounted for by a unitary transformation of the field operators. Let the spatial mode basis $\{o_{k}(u)\}$ be expressable in terms of the $\{g_{k}(u)\}$ basis via,
$$
o_{k}(u) = \sum_{j}(g_j|o_{k})g_j(u) = \sum_{j}O_{jk}g_{j}(u)
$$
So the columns of the unitary matrix $O_{jk} = (g_j|o_k)$ are the $o_{k}(u)$ modes represented in the $g_k(u)$ basis with the property that $O^\dagger O = OO^\dagger = I$. We now look to express the electric field operator in a new basis

\begin{align*}
\hat{E}^{(+)}(u) &= \sum_{k}\underbrace{\bigg[\sum_{i}O_{ik}g_{i}(u)\bigg]}_{o_{k}(u)}\underbrace{\bigg[\sum_{j} O_{k j}^\dagger\hat{a}_{j}\bigg]}_{\hat{c}_{k}} \\
&=\sum_{ij}\underbrace{\bigg[\sum_{k}O_{ik}O_{kj}^\dagger\bigg]}_{\delta_{ij}}g_{i}(u)\hat{a}
_{j}\\
&=\sum_{k}g_{k}(u)\hat{a}_{k}
\end{align*}
Therefore the modes and their field operators transform inversely. To summarize,
$$
o_{k}(u) = \sum_{j}O_{jk}
g_{j}(u),
\qquad \hat{c}_k = \sum_{j}O^{*}_{jk}\hat{a}_{j}
$$
Applied to the Williamson modes (where $O$ is a real orthogonal matrix), we get
$$
\boxed{
o_{k}(u) = \sum_{j}O_{jk}g_{j}(u),\qquad \hat{c}_{k} =\sum_{j}O_{jk}\hat{a}_{j}\iff \hat{\mbf{c}} = O^{\intercal}\hat{\mbf{a}}
}
$$

\subsection{QFIM for Brightness Parameters}
\label{ss: QFIM}
We derive the QFIM $Q_{\bar{\mbf{n}}}$ by applying its definition under Gaussian states given by Liu {\em et al.} in Theorem 2.10 of Ref. \cite{Liu:2020_QFIM}. An equivalent expression is used in Ref. \cite{Sorelli:2024}, while alternative expressions for computing the QFIM of a multimode Gaussian state are provided in Refs. \cite{Safranek:2019,Albarelli:2026_Efficient_SLD_CRB_GaussianStates}. The QFIM expression from Theorems 2.9 and 2.10 are reported using $'qpqp'$ operator ordering convention. Therefore, for this section, we assume $\hat{\mbf{x}} = [\hat{q}_1,\hat{p}_1,\ldots,\hat{q}_K,\hat{p}_K]^{\intercal}$ which will amount to a simple permutation of the phase-insensitive covariance matrix derived in Eq. \ref{eq:gauss_params_maintext_sigma}. In particular, transforming from `$qqpp$' to `$qpqp$' conventions just rearranges the order of Kronecker product, 
\begin{align*}
\text{`$qqpp$' ordering} &\qquad |  \qquad \text{`$qpqp$' ordering}\\
\sigma' = I_{2}\otimes (\sqrt{G}\bar{N}\sqrt{G}+\tfrac{1}{2}I)&\longrightarrow\sigma' = (\sqrt{G}\bar{N}\sqrt{G}+\tfrac{1}{2}I)\otimes I_{2}\\
S = I_{2}\otimes O &\longrightarrow S = O \otimes I_2\\
D = I_{2}\otimes \mathcal{V} &\longrightarrow D = \mathcal{V} \otimes I_{2}
\end{align*} 

\rule{\linewidth}{.5pt}
\noindent \textbf{Ref. \cite{Liu:2020_QFIM} Theorem 2.10}: \\
\\
For a continuous variable bosonic $K$-mode Gaussian state with quadrature operator ordering $\hat{\mbf{x}} = [\hat{q}_1,\hat{p}_{1},\ldots,\hat{q}_{K},\hat{p}_{K}]$, displacement vector $\bar{\mbf{x}} = \langle\hat{\mbf{x}}\rangle$, and covariance matrix $\sigma_{jk} = \frac{1}{2}\langle\{\hat{x}_j-\bar{x}_j,\hat{x}_k-\bar{x}_j\} \rangle$ having Williamson decomposition $\sigma = S(\mathcal{V} \otimes I_{2}) S^{\intercal}$ where $S$ is symplectic and $\mathcal{V} = \diag(\nu_{1},\ldots,\nu_{K})$ is the matrix of symplectic eigenvalues, the entries of the QFIM are given by
\begin{equation}
Q_{ab} = \tr[H_{a}\partial_{b}\sigma] + (\partial_{a}\bar{\mbf{x}}^{\intercal})\sigma^{-1}(\partial_{b}\bar{\mbf{x}})
\label{eq:Gaussian_qfim_closed_form}
\end{equation}
where 
\begin{align*}
H_{a} &= \sum_{j,k=1}^{K}\sum_{\ell=0}^{3} \frac{h_{\ell,a}^{(jk)}}{4\nu_{j}\nu_k + (-1)^{\ell + 1}} (S^{\intercal})^{-1}A^{(jk)}_{\ell}S^{-1}\\
h_{\ell,a}^{(jk)} &= \tr[S^{-1}(\partial_{a}\sigma)(S^{\intercal})^{-1}A_{\ell}^{(jk)}]
\end{align*}
and
$$
A_{\ell}^{(jk)} = i\sigma_{y}^{(jk)},\,\,\sigma_{z}^{(jk)},\,\,I_{2}^{(jk)}, \,\, \sigma_{x}^{(jk)}
$$
for $\ell= 0,1,2,3$. The matrices $\sigma_{\ell}^{(jk)}$ are $2K$-dimensional matrices with all the entries zero except a $2\times2$ sub-block wherein we insert a Pauli matrix (or $I_2$ identity) into the $(j,k)$ sub-block,
$$
\sigma_{\ell}^{(jk)} = (\mbf{1}_{j}\mbf{1}_{k}^{\intercal})\otimes\sigma_{\ell} 
$$
where $\mbf{1}_j = \underbrace{[0,\ldots,1,\ldots,0]^{\intercal}}_{1\text{ in $j^{\rm th}$ index}}$ is a one-hot column vector consisting of a $1$ in the $j^{\rm th}$ entry and zeroes in all other entries.

\rule{\linewidth}{.5pt}
We now look to apply this expression to our Gaussian state model after the aperture. Since the state is zero-mean we may immediately discard the second term in Eq. \ref{eq:Gaussian_qfim_closed_form}. Therefore, $
Q_{ab} = \tr [H_{a}\partial_{b}\sigma]$. For the remainder of the derivation, it will be convenient to define the matrix,
$$
\Psi_{jk} \equiv [O^{\intercal}\sqrt{G}]_{jk} = \mbf{o}_{j}^{\intercal}\mbf{g}_k 
$$
where $\mbf{o}_{j}$ and $\mbf{g}_j$ are the $j^{\rm th}$ column of matrices $O$ and $\sqrt{G}$ respectively. The columns of $\Psi$ are therefore the representation of the modulated modes in the Williamson mode basis. The derivative of the covariance matrix with respect to $\bar{n}_{a}$ is given by,
$$
\partial_{a}\sigma' = \mbf{g}_{a}\mbf{g}_{a}^{\intercal} \otimes I_{2}
$$
where we use the shorthand $\partial_{a} = \partial/\partial \bar{n}_{a}$. Computing the weights,
\begin{align*}
h_{\ell,a}^{(jk)} &= \tr[S^{-1}(\partial_{a}\sigma')(S^{\intercal})^{-1}A_{\ell}^{(jk)}]\\
&=\tr[(O^{\intercal}\otimes I_{2})(\mbf{g}_{a}\mbf{g}_{a}^{\intercal}\otimes I_2)(O\otimes I_{2})(\mbf{1}_{j}\mbf{1}_k^{\intercal}\otimes i^{\delta_{0\ell}} \sigma_{\ell})]\\
&=\tr[(O^{\intercal}\mbf{g}_{a}\mbf{g}_{a}^{\intercal}O\mbf{1}_{j}\mbf{1}_{k}^{\intercal})\otimes (i^{\delta_{0\ell}} \sigma_{\ell})]\\
&= \tr[(O^{\intercal}\mbf{g}_{a}\mbf{g}_{a}^{\intercal}O\mbf{1}_{j}\mbf{1}_{k}^{\intercal})]\cdot \tr[i^{\delta_{0\ell}} \sigma_{\ell}]\\
&= (\mbf{o}_{k}^{\intercal}\mbf{g}_{a}\mbf{g}_{a}^{\intercal}\mbf{o}_{j})\cdot \delta_{2\ell}\tr[I_2] \\
&=2(\Psi_{ja}\Psi_{ka})\cdot \delta_{2\ell}
\end{align*}
We observe that only the $\ell = 2$ Pauli terms survive.  This simplifies the sum over $\ell$ such that $H_a$ is
\begin{align*}
H_{a} &= \sum_{j,k=1}^{K} \frac{h_{2,a}^{(jk)}}{4\nu_{j}\nu_{k}+(-1)^{2+1}}(S^{\intercal})^{-1}A_{2}^{(jk)}S^{-1} \\
&= \sum_{j,k=1}^{K}\frac{h_{2,a}^{(jk)}}{4\nu_{j}\nu_{k}-1}\bigg[(O\otimes I_{2})(\mbf{1}_{j}\mbf{1}_k^{\intercal}\otimes I_{2})(O^{\intercal}\otimes I_{2})\bigg]\\
&=\sum_{j,k=1}^{K}\frac{h_{2,a}^{(jk)}}{4\nu_{j}\nu_{k}-1}\bigg[(O\mbf{1}_{j}\mbf{1}_{k}^{\intercal}O^{\intercal})\otimes(I_{2})\bigg]\\
&=\sum_{j,k=1}^{K}\frac{h_{2,a}^{(jk)}}{4\nu_{j}\nu_{k}-1}(\mbf{o}_{j}\mbf{o}_{k}^{\intercal})\otimes I_{2}\\
&=\frac{1}{2}\sum_{j,k=1}^{K} \frac{\Psi_{ja}\Psi_{ka}}{M_{jk}}(\mbf{o}_{j}\mbf{o}_{k}^{\intercal})\otimes I_{2}
\end{align*}
Finally, inserting both expression into the definition of the QFIM we have,
\begin{align*}
Q_{ab} &= \tr[H_{a}\partial_{b}\sigma] \\
&= \frac{1}{2}\sum_{j,k=1}^{K}\frac{\Psi_{ja}\Psi_{ka}}{M_{jk}}\tr[(\mbf{o}_{j}\mbf{o}_{k}^{\intercal}\otimes I_2)(\mbf{g}_{b}\mbf{g}_{b}^{\intercal} \otimes I_{2})]\\
&= \frac{1}{2}\sum_{j,k=1}^{K}\frac{\Psi_{ja}\Psi_{ka}}{M_{jk}}\tr[\mbf{o}_{j}\mbf{o}_{k}^{\intercal} \mbf{g}_{b}\mbf{g}_{b}^{\intercal}] \cdot \tr [I_2]\\
&=\sum_{jk} \frac{\Psi_{ja}\Psi_{ka}\Psi_{jb}\Psi_{kb}}{M_{jk}}
\end{align*}
In summary,
\begin{equation}
\boxed{
[Q_{\bar{\mbf{n}}}]_{ab} = \sum_{j,k=1}^{K}  \frac{\Psi_{ja}\Psi_{ka}\Psi_{jb}\Psi_{kb}}{M_{jk}},\qquad \Psi = O^{\intercal}\sqrt{G},\qquad M_{jk} = \bar{m}_{j}\bar{m}_{k} +\frac{1}{2}(\bar{m}_{j}+\bar{m}_{k})
}
\end{equation}

\subsection{Symmetric Logarithmic Derivatives}

\rule{\linewidth}{.5pt}
\noindent\textbf{Ref. \cite{Liu:2020_QFIM} Theorem 2.9:}

Under `qpqp' ordering of the quadrature operator $\hat{\mbf{x}} = [\hat{q}_{1},\hat{p}_{1},\ldots,\hat{q}_{K},\hat{p}_{k}]^{\intercal}$, the SLDs for a gaussian state defined by displacement $\bar{\mbf{x}} = \langle\hat{\mbf{x}}\rangle$ and covariance matrix $\sigma = \frac{1}{2}\langle \{(\hat{\mbf{x}}-\bar{\mbf{x}}),(\hat{\mbf{x}}-\bar{\mbf{x}})\}\rangle$, the SLDs are given by,
$$
\hat{L}_{a} = L_{a}^{(0)} I_{2K} + (\mbf{L}_{a}^{(1)})^{\intercal}\hat{\mbf{x}} + \hat{\mbf{x}}^{\intercal}H_{a}\hat{\mbf{x}}
$$
where,
\begin{align*}
    L_{a}^{(0)} &= \bar{\mbf{x}}^{\intercal}H_{a}\bar{\mbf{x}} - (\partial_{a}\bar{\mbf{x}})^{\intercal}\sigma^{-1}\bar{\mbf{x}} - \tr[H_{a}\sigma]\\
\mbf{L}_{a}^{(1)} &= \sigma^{-1}(\partial_{a}\bar{\mbf{x}}) - 2 H_{a}\bar{\mbf{x}}
\end{align*}

\rule{\linewidth}{.5pt}
Applying this theorem to our problem, 
we immediately see that $\bar{\mbf{x}}=0$ simplifies $L_{a}^{(0)} = -\tr[H_{a}\sigma]$ and $\mbf{L}_{a}^{(1)} = 0$ such that,
$$
\hat{L}_{a} = \hat{\mbf{x}}^{\intercal}H_{a}\hat{\mbf{x}} - \tr[H_{a}\sigma]
$$
In the following derivation we will make use of the fact that $\hat{c}_{k} =\mbf{o}_{k}^{\intercal}\hat{\mbf{a}}$. Calculating the first term in the SLD equation gives,
\begin{align*}
\hat{\mbf{x}}^{\intercal}H_{a}\hat{\mbf{x}} &= \frac{1}{2}\sum_{jk} \frac{\Psi_{ka}\Psi_{ja}}{M_{jk}} \hat{\mbf{x}}^{\intercal}[(\mbf{o}_{j}\mbf{o}_{k}^{\intercal})\otimes I_{2}]\hat{\mbf{x}}\\
    &=\frac{1}{2}\sum_{jk} \frac{\Psi_{ka}\Psi_{ja}}{M_{jk}}[\hat{\mbf{q}}^{\intercal}(\mbf{o}_{j}\mbf{o}_{k}^{\intercal})\hat{\mbf{q}} + \hat{\mbf{p}}^{\intercal}(\mbf{o}_{j}\mbf{o}_{k}^{\intercal})\hat{\mbf{p}}]\\
    &=\frac{1}{2}\sum_{jk} \frac{\Psi_{ka}\Psi_{ja}}{M_{jk}} \frac{1}{2}\big[(\hat{\mbf{a}}+\hat{\mbf{a}}^{\dagger})^{\intercal}(\mbf{o}_{j}\mbf{o}_{k}^{\intercal})(\hat{\mbf{a}}+\hat{\mbf{a}}^{\dagger}) - (\hat{\mbf{a}}-\hat{\mbf{a}}^{\dagger})^{\intercal}(\mbf{o}_{j}\mbf{o}_{k}^{\intercal})(\hat{\mbf{a}}-\hat{\mbf{a}}^{\dagger})\big]\\
    &=\frac{1}{2}\sum_{jk} \frac{\Psi_{ka}\Psi_{ja}}{M_{jk}}(\hat{c}_{j}\hat{c}_{k}^\dagger + \hat{c}_{j}^{\dagger}\hat{c}_k)\\
    &= \frac{1}{2}\sum_{jk}\frac{\Psi_{ka}\Psi_{ja}}{M_{jk}}(\hat{c}_{k}^{\dagger}\hat{c}_{j}+c_{j}^{\dagger}\hat{c}_{k} + \delta_{jk})\\
    &=\sum_{jk}\frac{\Psi_{ja}\Psi_{ka}}{M_{jk}}\hat{c}_{j}^{\dagger}\hat{c}_{k} + \frac{1}{2}\sum_{k}\frac{\Psi_{ka}^2}{M_{kk}}
\end{align*}
Calculating the second term in the SLD equation gives,
\begin{align*}
    \tr[H_{a}\sigma] &= \frac{1}{2} \sum_{jk} \frac{\Psi_{ja}\Psi_{ka}}{M_{jk}} \tr[(\mbf{o}_{j}\mbf{o}_{k}^{\intercal}\otimes I_{2} )(\sqrt{G}\bar{N}\sqrt{G}+\tfrac{1}{2}I)\otimes I_{2})] \\
    &=\frac{1}{2} \sum_{jk} \frac{\Psi_{ja}\Psi_{ka}}{M_{jk}} \mbf{o}_{k}^{\intercal}(O\bar{M}O^{\intercal}+\tfrac{1}{2}I)\mbf{o}_{j}\cdot \tr[I_2]\\
    &=\sum_{jk}\frac{\Psi_{ja}\Psi_{ka}}{M_{jk}}(\bar{m}_{j}+\tfrac{1}{2})\delta_{jk} \\
    &=\sum_{k}\frac{\Psi_{ka}^2(\bar{m}_{k}+\tfrac{1}{2})}{M_{kk}}\\
    &= \sum_{k}\frac{\Psi_{ka}^{2}\bar{m}_{k}}{M_{kk}} + \frac{1}{2}\sum_{k}\frac{\Psi_{ka}^2}{M_{kk}}
\end{align*}
Combining both terms gives,
\begin{equation}
\boxed{
\hat{L}_{a} = \underbrace{\sum_{j k}\frac{\Psi_{ja}\Psi_{k a}}{M_{jk}}\hat{c}_{j}^{\dagger}\hat{c}_{k}}_{\hat{E}_{a}}  - \underbrace{\sum_{k}\frac{\Psi_{ka}^2}{M_{kk}}\bar{m}_{k}}_{\langle \hat{E}_{a}\rangle }
}
\label{eqn:sld_eqn}
\end{equation}
As a sanity check, we verify that these SLDs recover the QFIM, whose entries are given by given by,
\begin{align*}
    [Q_{\bar{\mbf{n}}}]_{ab} &=\langle \hat{L}_{a}\circ \hat{L}_{b} \rangle = \langle \hat{E}_{a}\circ \hat{E}_{b}\rangle - \langle \hat{E}_{a}\rangle\langle \hat{E}_{b}\rangle 
\end{align*}
where $A\circ B = \frac{1}{2}(AB+BA)$ is the Jordan product. The operator correlation,
\begin{align*}
\langle \hat{E}_{a}\circ \hat{E}_{b}\rangle &= \sum_{ijk\ell}\frac{\Psi_{ia}\Psi_{ja}\Psi_{kb}\Psi_{\ell b}}{M_{ij}M_{k\ell}} \langle \hat{c}_{i}^{\dagger}\hat{c}_{j}\circ\hat{c}_{k}^{\dagger}\hat{c}_{\ell}\rangle 
\end{align*}
involves terms of the form
\begin{align*}
\langle \hat{X}_{ij}\circ \hat{X}_{k\ell} \rangle =  \langle \hat{c}_{i}^{\dagger}\hat{c}_{j}\circ\hat{c}_{k}^{\dagger}\hat{c}_{\ell}\rangle &=\sum_{\mbf{m}}p_{\mbf{m}}\frac{1}{2}\mel{\mbf{m}}{\hat{c}_{i}^{\dagger}\hat{c}_{j}\hat{c}_{k}^\dagger \hat{c}_{\ell }+ \hat{c}_{k}^\dagger \hat{c}_{\ell }\hat{c}_{i}^{\dagger}\hat{c}_{j}}{\mbf{m}}
\end{align*}
where we have introduced the cross-mode operators $\hat{X}_{ij} = \hat{c}_{i}^{\dagger}\hat{c}_{j}$ for notational brevity. The only non-zero elements of the Fock-state expectations are given by indices $i,j,k,\ell$ that give rise to an equal number of raising and lowering operators of the same mode:
$$
\mel{\mbf{m}}{\hat{c}_{i}^{\dagger}\hat{c}_{j}\hat{c}_{k}^{\dagger}\hat{c}_{\ell } + \hat{c}_{k}^{\dagger}\hat{c}_{\ell }\hat{c}_{i}^{\dagger}\hat{c}_{j} }{\mbf{m}} =
\begin{cases}
2m_{i}^{2} & i=j,\, k=\ell,\, i=k \\
2m_{i}m_{k} & i=j,\, k=\ell,\, i\neq k\\
m_{i}(m_{j}+1) + m_{j}(m_{i}+1)& i=\ell,\, j=k, i\neq j\\
0 & {\rm otherwise}
\end{cases}\\
$$
Hence,
\begin{align*}
\langle \hat{X}_{ij}\circ \hat{X}_{k\ell} \rangle&=\frac{1}{2}\sum_{\mbf{m}}p_{\mbf{m}} \bigg[2m_{i}^2\delta_{ij}\delta_{k\ell}\delta_{ik} + 2m_{i}m_{k}\delta_{ij}\delta_{k\ell}\delta_{i\neq k} + [m_{i}(m_j + 1) + m_j(m_i+1)]\delta_{i\ell}\delta_{jk}\delta_{i\neq j} \bigg]\\
&=\delta_{ij}\delta_{k\ell}\bigg( \E[m_i^2]\delta_{ik} + \E[m_i m_{k}]\delta_{i\neq k}\bigg) + \E[m_{i}m_{j}+\frac{1}{2}(m_{i} +m_{j})]\delta_{i\ell}\delta_{jk}\delta_{i\neq j} \\
&=\delta_{ij}\delta_{k\ell}\bigg(\bar{m}_{i}\bar{m}_{k} + \delta_{ik}M_{ii}\bigg) + \delta_{i\ell}\delta_{jk}\delta_{i\neq j}M_{ij}
\end{align*}
where we used the following properties of the joint geometric distribution to simplify: $
\E[m_i] = \bar{m}_i$ and $\E[m_{i}m_{j}]=\bar{m}_{i}\bar{m}_{j} +\delta_{ij}M_{ii}$
Plugging into the first expectation term in the QFIM, we get
\begin{align*}
\langle \hat{E}_{a}\circ \hat{E}_{b}\rangle &= \sum_{ijk\ell}\frac{\Psi_{ia}\Psi_{ja}\Psi_{kb}\Psi_{\ell b}}{M_{ij}M_{k\ell}}\bigg[ \delta_{ij}\delta_{k\ell}\bigg(\bar{m}_{i}\bar{m}_{k} +\delta_{ik}M_{ii}\bigg) + \delta_{i\ell}\delta_{jk}\delta_{i\neq j}\, M_{ij} \bigg]\\
&=\sum_{ik}\frac{\Psi_{ia}^{2}\Psi_{kb}^{2}}{M_{ii}M_{kk}}(\bar{m}_{i}\bar{m}_{k} +\delta_{ik}M_{ii}) + \sum_{i\neq j}\frac{\Psi_{ia}\Psi_{ja}\Psi_{ib}\Psi_{jb}}{M_{ij}^2}M_{ij} \\
&=\underbrace{\sum_{ij}\frac{\Psi_{ia}^2 \Psi_{jb}^2}{M_{ii}M_{jj}}\bar{m}_{i}\bar{m}_{j}}_{\langle \hat{E}_{a}\rangle \langle \hat{E}_{b}\rangle} + \underbrace{\sum_{ij}\frac{\Psi_{ia}\Psi_{ja}\Psi_{ib}\Psi_{jb}}{M_{ij}}}_{[Q_{\bar{\mbf{n}}}]_{ab}} \\
&= \langle \hat{E}_{a}\rangle \langle\hat{E}_{b}\rangle + [Q_{\bar{\mbf{n}}}]_{ab}
\end{align*}
The QFIM is therefore recovered.

\subsection{SLD Commutativity}
The commutators of the SLDs are given by
\begin{align*}
[\hat{L}_{a},\hat{L}_{b}] &= [\hat{E}_{a},\hat{E}_{b}]
=\sum_{ijk\ell} \frac{\Psi_{ia}\Psi_{ja}\Psi_{kb}\Psi_{\ell b}}{M_{ij}M_{k\ell}} [\hat{c}_{i}^{\dagger}\hat{c}_{j},\hat{c}_{k}^{\dagger}\hat{c}_{\ell}]\\
&=\sum_{ijk\ell}\frac{\Psi_{ia}\Psi_{ja}\Psi_{kb}\Psi_{\ell b}}{M_{ij}M_{k\ell}}\bigg(\delta_{jk}\hat{X}_{i\ell} - \delta_{i\ell}\hat{X}_{kj}\bigg)\\
&= \sum_{i\ell}\bigg(\sum_{j}\frac{\Psi_{ja}\Psi_{jb}}{M_{ij}M_{j\ell}}\bigg) \Psi_{ia} \hat{X}_{i\ell} \Psi_{\ell b} - \sum_{jk}\bigg(\sum_{i}\frac{\Psi_{ia}\Psi_{ib}}{M_{ij}M_{ki}}\bigg) \Psi_{ja}\hat{X}_{kj}\Psi_{kb}\\
&=\sum_{ik}\bigg(\sum_{j}\frac{\Psi_{ja}\Psi_{jb}}{M_{ij}M_{jk}}\bigg)\Psi_{ia}\hat{X}_{ik}\Psi_{kb} - \sum_{ik}\bigg(\sum_{j} \frac{\Psi_{ja}\Psi_{jb}}{M_{ji}M_{kj}}\bigg)\Psi_{ia}\hat{X}_{ki}\Psi_{kb}\\
&=\sum_{ik}\bigg(\sum_{j}\frac{\Psi_{ja}\Psi_{jb}}{M_{ij}M_{jk}}\bigg)\Psi_{ia}\bigg(\hat{X}_{ik}-\hat{X}_{ki}\bigg)\Psi_{kb}\\
&=\sum_{i\neq k}\bigg(\sum_{j}\frac{\Psi_{ja}\Psi_{jb}}{M_{ij}M_{jk}}\bigg)\Psi_{ia}\bigg(\hat{X}_{ik}-\hat{X}_{ki}\bigg)\Psi_{kb}\\
&=\sum_{i<k}\bigg(\sum_{j}\frac{\Psi_{ja}\Psi_{jb}}{M_{ij}M_{jk}}\bigg)\bigg[ (\hat{X}_{ik} - \hat{X}_{ki})\Psi_{ia}\Psi_{kb} + (\hat{X}_{ki}-\hat{X}_{ik})\Psi_{ka}\Psi_{ib}\bigg]\\
&=\sum_{i<k}\bigg(\sum_{j}\frac{\Psi_{ja}\Psi_{jb}}{M_{ij}M_{jk}}\bigg) \bigg[ (\Psi_{ia}\Psi_{kb}-\Psi_{ka}\Psi_{ib})\hat{X}_{ik} + (\Psi_{ka}\Psi_{ib} - \Psi_{ia}\Psi_{kb})\hat{X}_{ki}\bigg]\\
&=\sum_{i\neq k}\underbrace{\bigg(\sum_{j}\frac{\Psi_{ja}\Psi_{jb}}{M_{ij}M_{jk}}\bigg)  (\Psi_{ia}\Psi_{kb}-\Psi_{ka}\Psi_{ib})}_{C_{ik}}\hat{X}_{ik} \\
&= \sum_{ik}C_{ik}\hat{X}_{ik}
\end{align*}
The matrix of coefficients,
$$
C_{ik}=\bigg(\sum_{j}\frac{\Psi_{ja}\Psi_{jb}}{M_{ij}M_{jk}}\bigg)  (\Psi_{ia}\Psi_{kb}-\Psi_{ka}\Psi_{ib})
$$ 
must be zero everywhere for the SLDs to commute. Overall, it is difficult to ascertain a collection of necessary and sufficient conditions under which this condition is satisfied. As a trivial example, the SLDs commute when the Gram matrix is diagonal such that the modes remain orthogonal after propagating through the aperture. In this situation,
$$
G = \bar{M} \implies O = I \implies \Psi = O^{\intercal}\sqrt{G} = \sqrt{\bar{M}} \implies C = 0
$$

Physically, such a situation arises when the emitters in an ensemble are infinitely far away from each other, or when the input modes $f_{k}(u)$ just so happen to be eigenmodes of the pupil transmission mask $t(u)f_{k}(u) \propto f_{k}(u)$ such that they remain orthogonal after propagation through the pupil. 

Finally, given the commutator it is straightforward to prove that the SLDs weakly commute. 
$$
\langle [\hat{L}_{a},\hat{L}_{b}]\rangle = \sum_{jk}C_{jk}\langle \hat{X}_{jk}\rangle = \sum_{jk}C_{jk}\delta_{jk}\bar{m}_{k} \\
=\sum_{k}\underbrace{C_{kk}}_{0}\bar{m}_{k} = 0
$$
where in the last equality we use the fact that the diagonal entries $C_{kk}=0$.

\subsection{Weak Thermal Source Approximation}
For completeness, we briefly derive the weak thermal source model in the Williamson basis. The weak thermal model assumes the total mean photon across all spatial modes is small $\bar{m}_0 = \tr[\bar{m}] \ll 1 \implies \bar{m}_{k}\ll 1$. Expressing the post-aperture thermal state in the Fock basis of the Williamson modes we have,
$$
\hat{\rho}_{\bar{\mbf{n}}} = \sum_{\mbf{m}\in\mathbb{Z}_{+}^{K}}p_{\bar{\mbf{m}}}(\mbf{m})\dyad{\mbf{m}},\qquad  p_{\bar{\mbf{m}}}(\mbf{m}) = \prod_{k=1}^{K} \frac{1}{1+\bar{m}_{k}}\bigg(\frac{\bar{m}_{k}}{1+\bar{m}_{k}}\bigg)^{m_{k}}
$$
It is clear that $p_{\bar{\mbf{m}}}(\mbf{m}) \sim \bar{m}_{1}^{m_1}\cdots \bar{m}_{K}^{m_K}\leq \bar{m}_0^{m_{1}+\cdots+m_{K}}$ which becomes vanishingly small for multi-photon states (i.e. Fock states $\ket{\mbf{m}}$ such that $\sum_{k=1}^{K}m_k >1$). Thus the dominant terms in the Fock state expansion are the vacuum state along with single-photon excitations of the Williamson modes:
\begin{equation}
\hat{\rho}_{\bar{\mbf{n}}}' \approx  (1-\epsilon)\dyad{0} + \epsilon \sum_{k=1}\lambda_{k}\dyad{o_k}
\label{eq: WeakThermalModel}
\end{equation}
where we have introduced the shorthand $\ket{o_{k}} = \hat{c}_{k}^{\dagger} \ket{0}$ to represent the single-photon state of the $k^{\rm th}$ Williamson mode $\braket{o_{j}}{o_{k}} = \delta_{jk}$. Moreover, we have introduced $\epsilon \equiv \frac{\bar{m}_0}{1+\bar{m}_0}\ll 1$ and the probability weights $\lambda_k = \frac{\bar{m}_{k}}{\bar{m}_{0}}$. For $K$ emitters with non-zero brightness that are not coincident in space, the support of this state is $(K+1)$-dimensional.

Assuming the loss through the aperture for all modes $f_k(u)$ is uniform, post-selecting on the single-photon component of the weak-thermal source model gives,
$$
\hat{\rho}_{{\bm{\theta}}} = \sum_{k=1}^{K} \lambda_k \dyad{o_k} = \sum_{k}\theta_{k} \dyad{\psi_k}
$$
where $\theta = \frac{\bar{n}_k}{\bar{n}_0}$ where $\bm{\theta}\in\Theta$ is a discrete probability distribution living in the $K$-simplex $\Theta = \{\bm{\theta}\in\mathbb{R}^{K}:\mbf{1}^{\intercal}\bm{\theta}=1,\theta_k>0\}$.

\subsection{Helstrom Bound for Relative Brightness Parameters}
\label{sub:HelstromBound_RelativeBrightnessParams}

Assume we are working with the normalized Gram matrix of the main text so that $\bar{n}_0 =\bar{m}_0$. The QFIM $Q_{\bm{\theta}}$ for the relative brightness parameters is ostensibly obtained from a Jacobian transformation $\frac{\partial \bar{n}_{j}}{\partial \theta_{k}} = \bar{n}_{0}\delta_{jk}$ of the mean-photon number QFIM $Q_{\bar{\mbf{n}}}$. However, here we emphasize that naive application of the Jacobian transformation leads one astray. Instead, we will show that one must invoke the projector $\Gamma  = I -\frac{1}{K}\mbf{1}\mbf{1}^{\intercal}$ onto the tangent space of the simplex. 

\begin{figure}
    \centering
    \includegraphics[width=\linewidth]{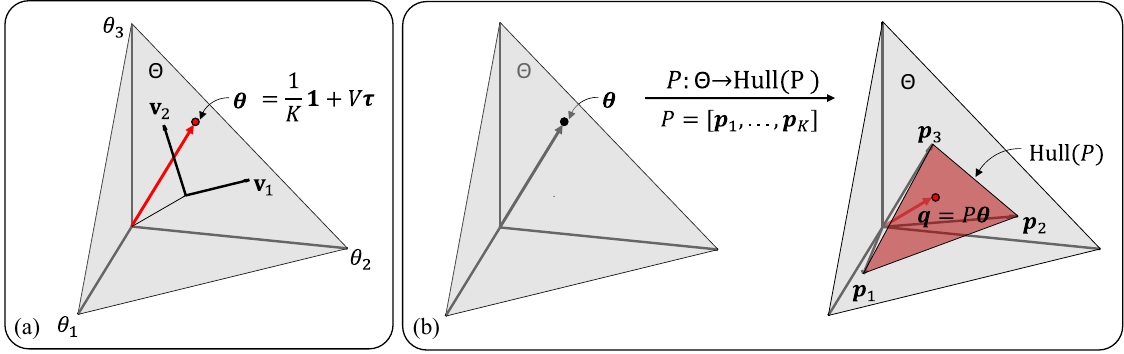}
    \caption{(a) Visualization of the probability simplex for $K=3$ embedded in $\mathbb{R}^{K}$. The coordinate vectors $\mbf{v}_1$ and $\mbf{v}_2$ form an orthonormal basis for the simplex tangent space $\mathcal{T}_{\bm{\theta}}\Theta$, giving rise to the re-parameterization $\bm{\theta}=\frac{1}{K}\mbf{1} + V\bm{\tau}$ in terms of the unconstrained variables $\bm{\tau}\in\mathbb{R}^{K-1}$. (b) Geometric description of how SPADE maps the relative brightness parameters $\bm{\theta}$ into the convex subset in $\Theta$. A SPADE measurement defined through the orthogonal matrix $R\in\mathbb{O}(K)$ produces a probability transfer matrix $P_{jk} = [R^{\intercal}\sqrt{G}]_{jk}^{2}$ with columns $\mbf{p}_{k}\in\Theta$. The action of SPADE on the parameter space is to compress the simplex into $\text{Hull}(P)\subseteq \Theta$ such that relative brightness parameters map to the mode detection probabilities $P:\bm{\theta}\mapsto \mbf{q}$. Different choices of the modes $R$ produce measurements with varying sensitivity to changes in $\bm{\theta}$.}
    \label{fig: simplex_parameterization}
\end{figure}

For now, suppose the parameters $\bm{\theta}$ were unconstrained. Under this relaxed assumption, the Jacobian transformation $\frac{\partial \bar{n}_{j}}{\partial \theta_{k}} = \bar{n}_{0}\delta_{jk}$ is valid but we must further normalize by the single photon probability $\epsilon \approx \bar{n}_0\ll 1$ since we consider post-selecting on the single-photon component. After doing so, we get the unconstrained per-photons QFIM $\frac{1}{\bar{n}_{0}}(\bar{n}_0 I)Q_{\bar{\mbf{n}}}(\bar{n}_0 I) = \bar{n}_0 Q_{\bar{\mbf{n}}}$ given by,
$$
[\bar{n}_0 Q_{\bar{\mbf{n}}}]_{ab} = \sum_{jk}\frac{\Psi_{ja}\Psi_{ka}\Psi_{jb}\Psi_{kb}}{\Lambda_{jk}},\qquad [\Lambda]_{jk} = [M]_{jk}/\bar{n}_0 = \bar{n}_0\lambda_{j}\lambda_{k} + \frac{1}{2}(\lambda_{j}+\lambda_{k})
$$
Here $\lambda_{j} \equiv \frac{\bar{m}_j}{\bar{n}_0}$ are the eigenvalues of $\hat{\rho}_{\bm{\theta}}$. However, the denominator still has dependence on $\bar{n}_0$ which may be neglected in the weak source limit. Hence we take $\Lambda_{jk} \approx \frac{1}{2}(\lambda_{j}+\lambda_{k})$ and define the unconstrained per-photon QFIM,
$$
[\tilde{Q}_{\bm{\theta}}]_{ab} \equiv 2\sum_{jk}\frac{\Psi_{ja}\Psi_{ka}\Psi_{jb}\Psi_{kb}}{\lambda_{j}+\lambda_{k}}
$$
such that $\tilde{Q}_{\bm{\theta}} \approx \bar{n}_0 Q_{\bar{\mbf{n}}}$. While $\tilde{Q}_{\bm{\theta}}$ is valid for the single-photon model assuming unconstrained brightness parameters, it ignores the fact that we have implicitly reduced a degree of freedom (namely the total mean photon number $\bar{n}_0$) by post-selecting the single-photon component. Indeed, the only information carried by the vacuum term in the weak source approximation is about the relative prominence of the single photon state (i.e. flux). The vacuum carries no information about the relative brightness of each sources - this information is entirely contained in the single-photon component. This means that $\tilde{Q}_{\bm{\theta}}$ does not take into account the fact that $\bm{\theta}$ is actually required to satisfy
$$
\gamma(\bm{\theta}) = \mbf{1}^{\intercal}\bm{\theta}-1 = 0.
$$
The constraint function $\gamma$ induces a tangent space for the probability simplex at the point $\bm{\theta}$. This space is defined as the set of directions along which $\bm{\theta}$ is allowed to be perturbed such that the constraint is locally preserved. Formally, the tangent space can be identified by finding the directional derivative of the constraint,
$$
\mathcal{D}_{\mbf{v}}[\gamma(\bm{\theta})] = \lim_{t\rightarrow 0} \frac{\gamma(\bm{\theta}+ t\mbf{v})-\gamma(\bm{\theta})}{t} =  \mbf{1}^{\intercal}\mbf{v}
$$
and considering the span of vectors for which it is zero,
$$
\mathcal{T}_{\bm{\theta}}\Theta = \{\mbf{v}\in\mathbb{R}^{K} : \mbf{1}^{\intercal}\mbf{v} =0\}
$$
This tangent space has a very clear interpretation -- it represents the collection of vectors where the entries sum to zero (see Fig. \ref{fig: simplex_parameterization}(a)). Ref. \cite{Ben-Haim:2009} showed how Cram\'er-Rao bounds and Fisher information matrices are modified in the context of constrained parameterizations. We briefly review the main conclusions from that paper assuming unbiased estimators.

\noindent\rule{\textwidth}{0.4pt}
\textbf{Ref~\cite{Ben-Haim:2009} - Theorem 1:}\\
Let $\bm{\theta}\in\mathbb{R}^{K}$ be a set of $K$ parameters and $\tilde{F}_{\bm{\theta}}\in\mathbb{R}^{K\times K}$ be a Fisher information matrix (quantum or classical) associated with an unconstrained treatment of the parameters. Now, suppose it is known that $\bm{\theta}$ satisfies a collection of $D<K$ constraint equalities captured by the vector-valued constraint function $\bm{\gamma}:\mathbb{R}^{K}\rightarrow\mathbb{R}^{D}$ such that the feasible parameter space is,
$$
\Theta = \{\bm{\theta}\in\mathbb{R}^{K}:\bm{\gamma}(\bm{\theta}) = \mbf{0}\}
$$
Assuming the matrix $H = \frac{\partial \bm{\gamma}}{\partial \bm{\theta}} \in \mathbb{R}^{K\times D}$ has full row rank (i.e. no redundant constraints) then there exists a matrix $V\in\mathbb{R}^{K\times(K-D)}$ such that $HV = 0$ and $V^{\intercal}V = I_{K-D}$. Defining the projection onto the tangent space induced by the constraints $\Gamma = VV^{\intercal}\in\mathbb{R}^{K\times K}$, the Fisher information matrix transforms to,
$$
F_{\bm{\theta}} = \Gamma\tilde{F}_{\bm{\theta}}\Gamma 
$$
while the Cram\'er-Rao bound for any unbiased estimator is beholden to,
$$
\mathbb{V}(\check{\bm{\theta}}|\bm{\theta}) \geq F_{\bm{\theta}}^{+} = V(V^{\intercal}\tilde{F}_{\bm{\theta}}V)^{-1}V^{\intercal}
$$
where $A^{+}$ denotes the Moore-Penrose pseudo-inverse of the matrix $A$.

\noindent\rule{\textwidth}{0.4pt}

Turning back to our brightness estimation task we explore some of the intricacies associated with defining a coordinate system $V$ on the tangent space $\mathcal{T}_{\bm{\theta}}\Theta$, This coordinate system will be equivalent to picking a $(K-1)$-dimensional parameterization for $\bm{\theta}$. To this end, let $V \in \mathbb{R}^{K\times(K-1)}$ be a matrix whose columns $\mbf{v}_{i}\in\mathbb{R}^{K}$ are orthonormal and span the tangent space $\mathcal{T}_{\bm{\theta}}\Theta$. In particular, this matrix satisfies $V^{\intercal}V = I_{K-1}$ and $\mbf{1}^{\intercal}V=\mbf{0}$, thereby constituting an orthogonal basis for the tangent space. As an example, consider the Helmert basis given by the vectors,
$$
\mbf{v}_{k} = \frac{1}{\sqrt{k(k+1)}}[\underbrace{1,\ldots,1}_{k},-k,0,\ldots,0]^{\intercal}\in\mathbb{R}^{K},\qquad k=1,\ldots,K-1
$$
which satisfy $\mbf{v}_{j}^{\intercal}\mbf{v}_{k}=\delta_{jk}$ and $\mbf{1}^{\intercal}\mbf{v}_{k}=0$ as desired. Any relative brightness vector can thus be expressed in terms of some unconstrained variables $\bm{\tau} = [\tau_{1},\ldots,\tau_{K-1}]^{\intercal}\in\mathbb{R}^{K}$ via,
$$
\bm{\theta} = \frac{1}{K}\mbf{1} + V\bm{\tau}
$$
Next, consider some other orthogonal coordinate system $V' = VR$ given by rotating the original coordinate systems with a matrix from the orthogonal group $R\in\mathbb{O}(K-1)$. We now show that the Helstrom bound is invariant under any such rotation. Projecting the QFIM onto the coordinates $V$ and computing its inverse trace gives,
\begin{equation}
c_{\bm{\theta}}^{\rm SLD}\equiv \tr [(V'^{\intercal}Q_{\bm{\theta}}V')^{-1}] = \tr [(R^{\intercal}V^{\intercal}Q_{\bm{\theta}}VR)]^{-1} = \tr[(V^{\intercal}Q_{\bm{\theta}}V)(RR^{\intercal})^{-1}] = \tr[(V^{\intercal}Q_{\bm{\theta}}V)^{-1}]
\label{eq:Helstrom_Bound_RelativeBrightness}
\end{equation}
The right-most equality is the same Helstrom bound recovered by projecting onto $V$ instead of $V'$. Hence, this quantity is invariant under rotations of orthogonal coordinate systems on the simplex and is reported in the main text Eq. \ref{eq:helstrom_bound}.

As a cautionary tale, we note that the naive "leave-one-out" strategy, wherein one arbitrarily delegates a single relative brightness parameter to be fixed $\theta_{j}=1-\sum_{k\neq j}\theta_{k}$ and the rest $\theta_{k\neq j}$ to be free variables, is an invalid means of computing the Helstrom bound as it produces a quantum bound that depends on the choice of $j$. This approach induces a set of {\em non-orthogonal} coordinate vectors on $\mathcal{T}_{\bm{\theta}}\Theta$ given by the columns of the matrix $T_{j}\in\mathbb{R}^{K\times(K-1)}$,

$$
T_{j} = P_{j}T_{K},\qquad T_{K}\equiv
\begin{bmatrix}
I_{(K-1)} \\
  \hline
-1,\ldots,-1
\end{bmatrix}
$$
where $T_{K}\in\mathbb{R}^{K\times(K-1)}$ is a block matrix where the top of the block is identity and the bottom row consists of all $-1$'s while $P_j$ is a permutation matrix exchanging rows $j\leftrightarrow K$ when multiplying on the left. We note that each $T_{j}$ is really a Jacobian for an unconstrained re-parameterization in terms of the variables $\bm{\theta}_{|j} = [\theta_{1},\ldots,\theta_{k\neq j},\ldots,\theta_{K}]\in\mathbb{R}^{K-1}$ given by,
$$
\bm{\theta} = \mbf{1}_{j} + T_{j}\bm{\theta}_{|j}
$$
For clarity, we show these matrices for when $K=4$, each corresponding to a different choice of the fixed parameter $\theta_{j}$:
$$
T_{1} = 
\begin{bmatrix}
-1 & -1 & -1 \\
1 & 0 & 0 \\
0 & 1 & 0 \\
0 & 0 & 1
\end{bmatrix},
\qquad
T_{2} = 
\begin{bmatrix}
1 & 0 & 0 \\
-1 & -1 & -1 \\
0 & 1 & 0 \\
0 & 0 & 1
\end{bmatrix},
\qquad
T_{3} = 
\begin{bmatrix}
1 & 0 & 0 \\
0 & 1 & 0 \\
-1 & -1 & -1 \\
0 & 0 & 1
\end{bmatrix},
\qquad
T_{4} = 
\begin{bmatrix}
1 & 0 & 0 \\
0 & 1 & 0 \\
0 & 0 & 1 \\
-1 & -1 & -1
\end{bmatrix}
$$
Note that while the columns of $T_j$ are linearly independent and satisfy $\mbf{1}^{\intercal}T_{j}=\mbf{0}$ such that they form a valid coordinate system for the simplex tangent space, this coordinate system is not orthogonal $T_{j}^{\intercal}T_{j}\neq I_{K-1}$. Nevertheless, it is instructive to see what happens if we choose to proceed computing the Helstrom bound for this parameterization. Specifically, we will show that the Helstrom bound induced by projecting onto $T_{j}$ is {\em not} invariant under the choice of $j$. To see this, note that the span of the columns of $V$ and $T_{j}$ are the same. Thus there exists an invertible matrix $R_j \in\mathbb{R}^{(K-1)\times(K-1)}$ such that $T_{j} = V R_j$. Then,
\begin{equation}
c_{\bm{\theta}|j}^{\rm SLD}\equiv \tr [(T_j^{\intercal}Q_{\bm{\theta}}T_{j})^{-1}] = \tr [(R_j^{\intercal}V^{\intercal}Q_{\bm{\theta}}V R_{j})^{-1}]  = \tr[(V^{\intercal}Q_{\bm{\theta}}V)^{-1}(R_jR_j^{\intercal})^{-1}]
\label{eq:leave_one_out_Helstrom}
\end{equation}
Since $R_{j}R_{j}^{\intercal}\neq I_{K-1}$, we see that the right-most term does not disappear as it did in the case of an orthogonal system - instead it varies for different choices of $j$. In fact, we can explicitly compute $(R_{j}R_{j}^{\intercal})^{-1}$ to see how $c_{\bm{\theta}|j}^{\rm SLD}$ compares to $c_{\bm{\theta}}^{\rm SLD}$. First, we note that the product of the tangent space matrices reduces to a convenient form,
\begin{align*}
    T_{j}T_{j}^{\intercal} &= P_{j}T_{K}T_{K} P_{j}^{\intercal} = I_{K}-\mbf{1}\mbf{1}_{j}^{\intercal}-\mbf{1}_{j}\mbf{1}^{\intercal} + K \mbf{1}_{j}\mbf{1}_{j}^{\intercal}
\end{align*}
Using the fact that $R_{j}=V^{\intercal}T_{j}$ we then find,
\begin{align*}
R_{j}R_{j}^{\intercal} &= V^{\intercal}T_{j}T_{j}^{\intercal}V = I_{K-1} + K \bar{\mbf{v}}_{j}\bar{\mbf{v}}_{j}^{\intercal}
\end{align*}
where $(V^{\intercal}\mbf{1}_{j}) = \bar{\mbf{v}}_{j}\in\mathbb{R}^{K-1}$ denotes the column vector composed from the $j^{\rm th}$ {\em row} of $V$ (not to be confused with $\mbf{v}_{j}$ which denotes the column vector composed from the $j^{\rm th}$ {\em column} of $V$). Realizing that $R_{j}R_{j}^{\intercal}$ is an invertible matrix plus a rank-1 projector we use the Sherman-Morrison equation to compute it's inverse:
\begin{align*}
(R_{j}R_{j}^{\intercal})^{-1} = (I_{K-1} + K \bar{\mbf{v}}_{j}\bar{\mbf{v}}_{j}^{\intercal} )^{-1} = I_{K-1} - \frac{K\bar{\mbf{v}}_{j}\bar{\mbf{v}}_{j}^{\intercal}}{1+ K\bar{\mbf{v}}_{j}^{\intercal}\bar{\mbf{v}}_{j}}
\end{align*}
To calculate $\bar{\mbf{v}}_{j}^{\intercal}\bar{\mbf{v}}_{j}$ in the denominator, we recall that $\Gamma = VV^{\intercal} = I - \frac{1}{K}\mbf{1}\mbf{1}^{\intercal}$ is a projector onto the tangent space. Therefore, we have $\bar{\mbf{v}}_{j}^{\intercal}\bar{\mbf{v}}_{j}=  \mbf{1}_{j}^{\intercal}VV^{\intercal}\mbf{1}_{j} = \frac{K-1}{K}$. Finally, we find that
$$
(R_{j}R_{j}^{\intercal})^{-1} = I_{K-1} - \bar{\mbf{v}}_{j}\bar{\mbf{v}}_{j}^{\intercal}
$$
Applying this identity to the erroneous "leave-$j$-out" Helstrom bound of Eq. \ref{eq:leave_one_out_Helstrom} we find,
\begin{subequations}
\begin{align}
c_{\bm{\theta}|j}^{\rm SLD} &= c_{\bm{\theta}}^{\rm SLD} - \bar{\mbf{v}}_{j}^{\intercal}(V^{\intercal}Q_{\bm{\theta}}V)^{-1}\bar{\mbf{v}}_{j}\\
c_{\bm{\theta}|j}^{\rm SLD} &\leq c_{\bm{\theta}}^{\rm SLD} 
\end{align}
\label{eq:Naive_v_Real_Helstrom}
\end{subequations}
where the inequality comes from the fact that $(V^{\intercal}Q_{\bm{\theta}}V)^{-1}\geq 0$. We see from Eq. \ref{eq:Naive_v_Real_Helstrom} that the Helstrom bound $c_{\bm{\theta}|j}^{\rm SLD}$, computed naively assuming $\theta_{j}$ is fixed by the other brightness values, not only varies for different choices of the index $j$ but is also, in fact, too optimistic.

\section{Summary of Quantum Cram\'er-Rao Bounds}
\label{ss: CRBs}

The landscape of multiparameter quantum estimation is a vast and ever-growing area of research. Unlike the single-parameter setting, the presence of multiple unknown parameters often faces parameter incompatibility - a situation where optimal observables for individual parameters do not generally commute, and therefore cannot be simultaneously measured. In the presence of this complexity, researchers in multiparameter estimation theory have ascertained a series of operationally useful uncertainty bounds for unbiased estimators that impose fundamental limits on the collective precision with which different incompatible parameters may be jointly estimated~\cite{Nurdin:2024,Albarelli:2020,Tsang:2020_SemiParametricEstimation,Matsumoto:2002,Ragy:2016,Hayashi:2023_tightCRB,Imai:2026_boundhierarchy,Demkowicz-Dobrzanski:2020}. For context, we review some of the most ubiquitous multiparameter matrix and scalar estimation bounds. We will first introduce the {\em relationships} between the bounds before providing their explicit mathematical definitions. For the duration of this section, the presentation holds generally for any parameter estimation problem and is not specific to the brightness estimation task explored in the main text.

In general, our task at hand is to estimate free parameters $\bm{\theta}=[\theta_{1},\ldots,
\theta_{d}]^{\intercal}\in\Theta \subseteq  \mathbb{R}^{d}$ from measurements on $n$ identical copies of the state $\hat{\rho}^{\otimes n}_{\bm{\theta}}$. The parameter space $\Theta$ is assumed to be an open subset of $\mathbb{R}^{d}$. Let $\Pi = \{\hat{\Pi}_{x}\}_{x\in\mathcal{X}}$ be a POVM on the joint Hilbert space for all copies $\mathcal{H}^{\otimes n}$. The probability of observing an event $x$ is given by the Born rule, and induces a classical statistical model,
$$
p_{\bm{\theta}}(x) = \Tr[\hat{\Pi}_{x}\hat{\rho}_{\bm{\theta}}^{\otimes n}]
$$ 
Let $\check{\bm{\theta}}(x)$ be a locally unbiased (l.u.) estimator satisfying,
$$
\sum_{x\in\mathcal{X}}p_{\bm{\theta}}(x) \check{\bm{\theta}}(x) = \bm{\theta},\qquad\qquad \bm{\partial}\sum_{x\in\mathcal{X}}p_{\bm{\theta}}(x)\check{\bm{\theta}}^{\intercal}(x) = I_{d}
$$
where we have defined the gradient $\bm{\partial}\equiv  [\partial_{\theta_1},\ldots,\partial_{\theta_d}]^{\intercal}$. The mean-squared error (MSE) matrix (which in the case of unbiased estimators is equivalent to the covariance matrix) is given by
\begin{equation}
{\rm V}_{\bm{\theta}}(\check{\bm{\theta}},\Pi) = \sum_{x\in\mathcal{X}} p_{\bm{\theta}}(x)(\check{\bm{\theta}}(x)-\bm{\theta})(\check{\bm{\theta}}(x)-\bm{\theta})^{\intercal}.
\end{equation}
In general, the MSE matrix for l.u. estimators is subject to the chain of matrix inequalities,
\begin{equation}
{\rm V}_{\bm{\theta}}(\check{\bm{\theta}},\Pi)\geq F_{\bm{\theta}}^{-1}(\Pi) \geq \frac{1}{n} Q^{-1}_{\bm{\theta}}
\label{eq:matrix_ineqs}
\end{equation}
where $F_{\bm{\theta}}(\Pi)$ is the classical Fisher information matrix (CFIM) of the statistical model $p_{\bm{\theta}}(x)$ and $Q_{\bm{\theta}}$ is the QFIM. The matrix inequality $A\geq B$ is understood in a positive semidefinite sense $A-B\geq 0$. Determining necessary and sufficient conditions for the attainability of QFIM ({\em i.e.} whether or not there exists a POVM such that $F_{\bm{\theta}}(\Pi) = nQ_{\bm{\theta}}$) has received considerable theoretical attention due to parameter incompatibility. Unlike in the single-parameter setting, direct optimization of the classical Fisher information matrix in Eq. \ref{eq:matrix_ineqs} with respect to the measurement $\Pi$ is generally not possible because the space of positive semidefinite matrices admits only a partial order. Explicitly, this means that there may exist measurements $\Pi_{A}$ and $\Pi_{B}$ such that neither $F_{\bm{\theta}}(\Pi_{A})-F_{\bm{\theta}}(\Pi_{B})$ nor $F_{\bm{\theta}}(\Pi_{B})-F_{\bm{\theta}}(\Pi_{A})$ are positive semidefinite. This motivates introducing an operationally relevant optimization objective which admits total ordering. 

To evaluate the performance an estimator, it is thus standard to use the (weighted) MSE $c^{\rm MSE}_{\bm{\theta}}(\check{\bm{\theta}},\Pi) \equiv \tr[W{\rm V}_{\bm{\theta}}(\check{\bm{\theta}},\Pi)]$ where $W>0$ is a parameter-independent positive-definite weight matrix. From the matrix inequality of Eq. \ref{eq:matrix_ineqs}, it follows that the MSE is lower bounded by the classical scalar CRB 
$$
c^{\rm MSE}_{\bm{\theta}}(\check{\bm{\theta}},\Pi) \geq c_{\bm{\theta}}^{\rm CR}(\Pi) \equiv \tr[WF_{\bm{\theta}}^{-1}(\Pi)].
$$ 
Quantum information theory then tells us that for any measurement $\Pi$ on $\mathcal{H}^{\otimes n}$, the following chain of inequalities holds: 
\begin{equation}
c_{\bm{\theta}}^{\rm CR}(\Pi) \geq c_{\bm{\theta},n}^{\rm NH} \geq \tfrac{1}{n} c_{\bm{\theta}}^{\rm H} \geq  \tfrac{1}{n}c_{\bm{\theta}}^{\rm SLD}
\end{equation}
where $c_{\bm{\theta},n}^{\rm NH}$ is the NH bound \cite{Conlon:2021_NHBound}, $c_{\bm{\theta}}^{\rm H}$ is the Holevo bound, and $c^{\rm SLD}_{\bm{\theta}}$ is the Helstrom bound. The Holevo and Helstrom bounds are {\em additive} $c(\hat{\rho}^{\otimes n}_{\bm{\theta}})= \frac{1}{n}c(\hat{\rho}_{\bm{\theta}})$ while the NH bound is {\em subadditive} $c(\hat{\rho}^{\otimes n}_{\bm{\theta}})\leq \frac{1}{n}c(\hat{\rho}_{\bm{\theta}})$ and has thus been adorned with an additional copy number subscript `$n$' as a reminder $c_{\bm{\theta},n}^{\rm NH}$. Importantly, the subadditivity of the NH bound provides an avenue for exploring how separable measurements compare to joint measurements. The operational significance of each quantity in inequality chain is as follows: the NH bound furnishes a bound on the MSE for measurements acting on finite numbers of state copies $n<\infty$. This bound can be formulated as a semidefinite program \cite{Conlon:2021_NHBound} that is efficiently computable for states on finite-dimensional Hilbert spaces. Moreover, the NH bound converges to the Holevo bound asymptotically $nc^{\rm NH}_{\bm{\theta},n} \xrightarrow[]{n\rightarrow\infty} c^{\rm H}_{\bm{\theta}}$. In turn, the Holevo bound furnishes an asymptotically tight quantum bound that accounts for parameter incompatibility. That is, in the limit  of infinite state copies, there exists an optimal (possibly joint) measurement $\Pi^{\star}_{n}$ such that $nc^{\rm CR}(\Pi^{\star}_{n})\xrightarrow[]{n\rightarrow\infty}c^{\rm H}_{\bm{\theta}}$~\cite{Guta:2007,Guta:2009,Demkowicz-Dobrzanski:2020}. The Holevo bound may also be posed as a semidefinite program making it computationally tractable~\cite{Albarelli:2019}. Finally, we arrive at the Helstrom bound which enjoys the benefit of having an analytical closed-form. Critically, the Helstrom bound is equal to the Holevo bound if and only if the SLDs {\em weakly} commute.
$$
\Tr[\hat{\rho}_{\bm{\theta}}[\hat{L}_{a}\hat{L}_{b}]] = 0 \qquad \forall\,\,a,b\in\{1,\ldots,d\} \iff c_{\bm{\theta}}^{\rm H} = c_{\bm{\theta}}^{\rm SLD}
$$
In the finite-copy setting $n<\infty$ the Helstrom bound is achievable on a multicopy quantum state if and only if it is achievable at the single-copy level, a property known as the {\em persistence gap} theorem~\cite{Conlon:2024_gap_persistence_theorem}. While joint measurements on progressively larger numbers of copies may bring one closer to the Helstrom bound in the presence of the persistence gap, the gap only shrinks to zero in the infinite copy limit. Finally, in a short yet surprising proof, Ref. \cite{Yang:2019} showed that attaining the Helstrom bound is equivalent to attaining the matrix quantum CRB.  That is, if there exists a single-copy measurement $\Pi^{\star}$ on $\mathcal{H}$ that attains the Helstrom bound, then $c^{\rm CR}_{\bm{\theta}}(\Pi^{\star}) = c^{\rm SLD}_{\bm{\theta}} \iff F_{\bm{\theta}}(\Pi^{\star})=Q_{\bm{\theta}}$. With this backdrop in place, we now move to the definitions.

\subsection{Definitions of Scalar Quantum Bounds}
\label{sub: CRB Definitions}
Here we borrow the notation of Tsang, Albarelli, and Datta's semiparametric estimation framework \cite{Tsang:2020_SemiParametricEstimation} to concisely  define all of the bounds under one umbrella. Let $\mathcal{H}$ be a $q$-dimensional Hilbert space and let $\mathcal{Y}$ be the set of bounded self-adjoint operators on $\mathcal{H}$. For notational convenience, it will be valuable to define the bilinear forms local to the point $\hat{\rho}_{\bm{\theta}}$,
$$
\langle \hat{A},\hat{B} \rangle \equiv \frac{1}{2}\Tr[\hat{\rho}_{\bm{\theta}}(\hat{A}\hat{B}+\hat{B}\hat{A})],\qquad\qquad
( \hat{A},\hat{B} ) \equiv \frac{1}{2i}\Tr[\hat{\rho}_{\bm{\theta}}(\hat{A}\hat{B}-\hat{B}\hat{A})]
$$
which give the real and imaginary part of the expectation of a matrix product $\langle \hat{A} \hat{B}\rangle = \langle \hat{A},\hat{B}\rangle + i(\hat{A},\hat{B})$. Additonally, we will denote column vectors of operators with boldface $\mbf{A}=[\hat{A}_{1},\ldots,\hat{A}_{K}]^{\intercal}\in\mathcal{Y}^{\oplus K}$. The bilinear forms acting to two such vectoral operators is taken to apply element-wise to yield a matrix with entries. For example,
$$
[\langle \mbf{A},\mbf{B}\rangle ]_{jk} = \langle \hat{A}_{j},\hat{B}_{k}\rangle.
$$
Let the space of zero-mean operators be denoted by,
$$
\mathcal{Z} \equiv \{\hat{h} \in \mathcal{Y}: \Tr \hat{\rho}_{\bm{\theta}} \hat{h}  = \langle \hat{h},\hat{I}\rangle = 0 \}.
$$
Let the vector of symmetric logarithmic derivatives for the parameters be denoted as $\mbf{L} = [L_1,\ldots,L_{K}]^{\intercal} \in \mathcal{Z}^{\oplus K}$. Defining the gradient $\bm{\partial} = [\partial_{\theta_1},\ldots,\partial_{\theta_{K}}]^{\intercal}$, the SLDs satisfy the Lyapunov equation,
$$
\bm{\partial} \hat{\rho}_{\bm{\theta}} = \mbf{L}\circ \hat{\rho}_{\bm{\theta}} 
$$
where above the Jordan product is understood to apply element-wise. The SLDs form a tangent space $\mathcal{T} = \overline{\text{span}}(\hat{L}_{1},\ldots,\hat{L}_{k})\subset \mathcal{Z}$ the point $\hat{\rho}_{\bm{\theta}}$ for the manifold defined by the quantum statistical model. The space of influence operators is given by,
$$
\mathcal{D} \equiv \{\bm{\delta} \in \mathcal{Z}^{\oplus K} : \langle \mbf{L},\bm{\delta} \rangle =  I \}
$$
which comes from the local unbiasedness condition,
$$
\langle \hat{L}_{j},\hat{\delta}_{k}\rangle = \Tr [\hat{\rho}_{\bm{\theta}} (\hat{L}_{j}\circ \hat{\delta}_{k})] = \Tr[ \hat{\delta}_{k} (\hat{\rho}_{\bm{\theta}} \circ \hat{L}_{j})] = \Tr[\hat{\delta}_{k}\partial_{j}\hat{\rho}_{\bm{\theta}}] = \delta_{jk}
$$
With this notation established, the quantum bounds can be defined as,

\begin{subequations}
\begin{empheq}[box=\boxed]{align}
c_{\bm{\theta}}^{\rm SLD} &\equiv \min_{\bm{\delta}\in\mathcal{D}} \tr W \langle\bm{\delta},\bm{\delta}\rangle
\label{eq:SLD_CRB} \\
 c_{\bm{\theta}}^{\rm H} &\equiv \min_{\bm{\delta}\in\mathcal{D}}\tr[W \langle \bm{\delta},\bm{\delta}\rangle  + | \sqrt{W} ( \bm{\delta},\bm{\delta}) \sqrt{W}|]
 \label{eq:Holevo_CRB}\\
 c_{\bm{\theta}}^{\rm NH} &\equiv \min_{\bm{\delta}\in\mathcal{D},\mathbb{L}} \{ \tr W\langle \mathbb{L}\rangle :\quad\mathbb{L} \geq \bm{\delta}\bm{\delta}^{\intercal},\,\, \mathbb{L}_{jk} = \mathbb{L}_{kj}\in\mathcal{Y}\}
 \label{eq:NH_CRB}
\end{empheq}
\end{subequations}
where we introduced $|A| = \sqrt{A^{\dagger}A}$ in the Holevo bound as well as the symmetric operator-valued matrix $\mathbb{L} \in \mathcal{Y}^{d\times d}$ in the NH bound. The expectation of the operator matrix in the NH bound is taken element-wise to produce a matrix $[\langle \mathbb{L}\rangle]_{jk} = \langle \mathbb{L}_{jk}\rangle = \Tr[\mathbb{L}_{jk}\hat{\rho}_{\bm{\theta}}]$.

\subsection{Invariance Under Orthogonal Reparameterization of the Simplex}
\label{sub: CRB Invariance}
In the context of brightness estimation and the previous discussion on the simplex re-parameterization of Supplement Sec. \ref{sub:HelstromBound_RelativeBrightnessParams}, we seek to prove that all of the quantum bounds defined in Eqs.~\ref{eq:SLD_CRB},~\ref{eq:Holevo_CRB}, and~\ref{eq:NH_CRB} are all invariant under a change of orthonormal coordinate system over the simplex tangent space assuming an agnostic weight matrix  $W=I$. Suppose we have two possible $d$-dimensional parameterization of the $K = d+1$ brightness parameters
$$
\bm{\theta} = \frac{1}{K}\mbf{1} + V\bm{\tau} = \frac{1}{K}\mbf{1} + V'\bm{\tau}'
$$
where $V,V'\in\mathbb{R}^{K\times d}$ both define orthonormal coordinate vectors on the simplex such that $V^{\intercal}V = V'^{\intercal}V' =  I_{d}$ and $\mbf{1}^{\intercal}V =\mbf{1}^{\intercal}V' =\mbf{0}$. We may always define $V' = VR$ where $R\in\mathbb{O}(d)$ is an orthonormal rotation matrix and $\bm{\tau}' = R^{\intercal}\bm{\tau}$ define two different parameter vectors. The density operator can then be written in two ways,
$$
\hat{\rho}_{\bm{\theta}} = \hat{\rho}_{0} + \bm{\psi}^{\intercal}V\bm{\tau} = \hat{\rho}_{0} + \bm{\psi}^{\intercal}V'\bm{\tau}' 
$$
where $\hat{\rho}_0 = \frac{1}{K}\bm{\psi}^{\intercal}\mbf{1}= \frac{1}{K}\sum_{k=1}^{K}\dyad{\psi_{k}}$ and $\bm{\psi} = [\hat{\psi}_{1},\ldots,\hat{\psi}_{K}]^{\intercal}$ with $\hat{\psi}_{k}=\dyad{\psi_{k}}$. To prove the invariance of the bounds under a rotation of the simplex coordinates, we first examine the feasible sets $\mathcal{D}$ and $\mathcal{D}'$ under both parameterizations:
\begin{align*}
    \mathcal{D} &= \{\bm{\delta}\in\mathcal{Z}^{\oplus d} : \Tr[(\partial_{\bm{\tau}}\hat{\rho}_{\bm{\theta}})\bm{\delta}^{\intercal}]=I\}\\
    \mathcal{D}' &= \{\bm{\delta}'\in\mathcal{Z}^{\oplus d} : \Tr[(\partial_{\bm{\tau}'}\hat{\rho}_{\bm{\theta}})\bm{\delta}'^{\intercal}]=I\}
\end{align*}
Given that the derivative is $\partial_{\bm{\tau}} \hat{\rho}_{\bm{\theta}} = V^{\intercal}\bm{\psi}$ the constraints for both feasible sets require that,
\begin{align*}
&I = \Tr[\big(V^{\intercal}\bm{\psi}\big)\bm{\delta}^{\intercal}] = V^{\intercal} \Tr[\bm{\psi}\bm{\delta}^{\intercal}] = V'^{\intercal} \Tr[\bm{\psi}\bm{\delta}'^{\intercal}] \\
\implies &V = \Tr[\bm{\psi}\bm{\delta}^{\intercal}]\qquad \text{and} \qquad V'=VR = \Tr[\bm{\psi}\bm{\delta}'^{\intercal}]\\
\implies  & \Tr[\bm{\psi}\bm{\delta}^{\intercal}]R = \Tr[\bm{\psi}\big(\bm{\delta}^{\intercal}R\big)] =  \Tr[\bm{\psi}\bm{\delta}'^{\intercal}]\\
\implies &\bm{\delta}' = R^{\intercal}\bm{\delta}
\end{align*}
Therefore, the feasible sets are an equivalence class up to rotations of influence operators. We can now efficiently prove that the optimization problem associated with each bound is invariant under this coordinate rotation.

\noindent \textbf{Helstrom Bound}
\begin{align*}
    \min_{\bm{\delta}'\in\mathcal{D}'}\tr \langle\bm{\delta}',\bm{\delta}'\rangle = \min_{\bm{\delta}\in\mathcal{D}} \tr\langle R^{\intercal}\bm{\delta},R^{\intercal}\bm{\delta}\rangle = \min_{\bm{\delta}\in\mathcal{D}} \tr R^{\intercal}\langle \bm{\delta},\bm{\delta}\rangle R = \min_{\bm{\delta}\in\mathcal{D}} \tr \langle\bm{\delta},\bm{\delta}\rangle (RR^{\intercal}) = \min_{\bm{\delta}\in\mathcal{D}} \tr \langle\bm{\delta},\bm{\delta}\rangle
\end{align*}
\textbf{Holevo Bound}\\
\begin{align*}
    &\min_{\bm{\delta}'\in\mathcal{D}'}\tr[ \langle\bm{\delta}',\bm{\delta}'\rangle + |(\bm{\delta}',\bm{\delta}')|]
    = \min_{\bm{\delta}\in\mathcal{D}} \tr\langle \bm{\delta},\bm{\delta}\rangle  +\tr |R^{\intercal}(\bm{\delta},\bm{\delta})R^{\intercal}|
\end{align*}
First term is the same as the Helstrom bound, so we focus our attention on the second term:
\begin{align*}
    \tr |R^{\intercal}\underbrace{(\bm{\delta},\bm{\delta})}_{A}R| =\tr \sqrt{R^{\intercal}A^{\intercal}RR^{\intercal}AR} = \tr \sqrt{R^{\intercal}A^{\intercal}A R}
\end{align*}
Next define the spectral decomposition of the real symmetric matrix $A^{\intercal}A=O^{\intercal}DO$ such that,
\begin{align*}
    \tr \sqrt{R^{\intercal}A^{\intercal}A R} = \tr R^{\intercal}O^{\intercal}\sqrt{D}OR = \tr \sqrt{D} = \tr |A| = \tr|(\bm{\delta},\bm{\delta})|
\end{align*}
Therefore, we have shown that, $\tr |R^{\intercal}(\bm{\delta},\bm{\delta})R|= \tr|(\bm{\delta},\bm{\delta})|$. Consequently, the Holevo bound is invariant under rotation of coordinate axes on the simplex
$$
\min_{\bm{\delta}'\in\mathcal{D}'} \tr[\langle\bm{\delta}',\bm{\delta}'\rangle + |(\bm{\delta}',\bm{\delta}')|] = \min_{\bm{\delta}\in\mathcal{D}} \tr[\langle\bm{\delta},\bm{\delta}\rangle + |(\bm{\delta,\bm{\delta}})|]
$$
\noindent \textbf{Nagaoka-Hayashi Bound}\\
\begin{align*}
&\min_{\bm{\delta}'\in\mathcal{D}',\mathbb{L}'} \{\tr\langle \mathbb{L}'\rangle : \mathbb{L}'\geq \bm{\delta}'\bm{\delta}'^{\intercal} ,\mathbb{L}_{jk}' =\mathbb{L}_{kj}' \in\mathcal{Y} \}\\
=&\min_{\bm{\delta}\in\mathcal{D},\mathbb{L}'} \{\tr\langle \mathbb{L}'\rangle : \mathbb{L}'\geq R^{\intercal}\bm{\delta}\bm{\delta}^{\intercal} R ,\mathbb{L}_{jk}' =\mathbb{L}_{kj}' \in\mathcal{Y} \}\\
=&\min_{\bm{\delta}\in\mathcal{D},\mathbb{L}'} \{\tr\langle \mathbb{L}'\rangle : R\mathbb{L}'R^{\intercal}\geq \bm{\delta}\bm{\delta}^{\intercal} ,\mathbb{L}_{jk}' =\mathbb{L}_{kj}' \in\mathcal{Y} \}
\end{align*}
Next we define the transformed operator matrix $\mathbb{L}=R\mathbb{L}'R^{\intercal}$. Note that this matrix is still symmetric under arbitrary rotation since $\mathbb{L}'$ is assumed to be symmetric.
$$
\mathbb{L}_{jk} = \sum_{ab}R_{ja}\mathbb{L}'_{ab}R_{bk}^{\intercal} = \sum_{ab}R_{bk}^{\intercal}\mathbb{L}'_{ba}R_{bk}^{\intercal}R_{ja} =\sum_{ab}R_{kb}\mathbb{L}'_{ba}R_{bk}^{\intercal}R_{aj}^{\intercal} = \mathbb{L}_{kj}    
$$
Hence we may re-write the NH optimization problem as,
$$
\min_{\bm{\delta}\in\mathcal{D},\mathbb{L}} \{\tr\langle R^{\intercal}\mathbb{L}R \rangle : \mathbb{L}\geq \bm{\delta}\bm{\delta}^{\intercal} ,\mathbb{L}_{jk} =\mathbb{L}_{kj} \in\mathcal{Y} \}
$$
Finally, it is straightforward to show that the objective is invariant:
$$
\tr\langle R^{\intercal}\mathbb{L}R \rangle = \tr R^{\intercal}\langle\mathbb{L}\rangle R = \tr R^{\intercal}\langle\mathbb{L}\rangle (RR^{\intercal}) = \tr \langle \mathbb{L}\rangle
$$
Therefore, we have shown that the optimization for the parameters $\bm{\tau}'
$ is identical to that for $\bm{\tau}$: 
$$
\min_{\bm{\delta}\in\mathcal{D},\mathbb{L}} \{\tr\langle \mathbb{L} \rangle : \mathbb{L}\geq \bm{\delta}\bm{\delta}^{\intercal} ,\mathbb{L}_{jk} =\mathbb{L}_{kj} \in\mathcal{Y} \}
$$
All of the quantum bounds are therefore invariant under a reparameterization corresponding to rotations of a coordinate system on the simplex tangent space.

\section{Separable Measurements}
\label{ss: Separable Measurements}
In this section, we derive the Cram\'er-Rao bounds for brightness estimation using SPADE and Direct Imaging measurements on $n$ copies of the single-photon state $\hat{\rho}_{\bm{\theta}}$. We assume inversion-symmetry of the pupil such that $\hat{\rho}_{\bm{\theta}}$ lives on the real span $\ket{\psi_{k}}$. 

\subsection{SPADE}
Suppose a SPADE system is configured to sort a particular mode basis $\{\phi_k(u)\}_{k=1}^{K}$ with $\overline{\rm span}(\phi_k) = \overline{\rm span}(\psi_k)$. We define the probability transfer matrix $P \in \mathbb{R}^{K\times K}$ (i.e. a matrix whose columns sum to one), 
$$
P(R) = (R^{\intercal}\sqrt{G})^{\odot 2}
$$
where $R\in\mathbb{O}(K)$ is an orthogonal matrix denoting a particular representation of the SPADE basis, and ${(\cdot)}^{\odot \mu}$ denotes element-wise raising to the power $\mu$. The matrix element $[R^{\intercal}\sqrt{G}]_{jk}=(\phi_j|\psi_{k})$ is the overlap between a SPADE mode and a modulated mode such that the element $P_{jk}$ is the probability of detecting a photon in the mode $\phi_{j}$ given that a photon was in the mode $\psi_{k}$. Additionally, note that the $P$ induced by any mode basis depends solely on the spatial configuration of the emitters yet remains entirely independent of their brightnesses. If $n$ copies of $\hat{\rho}_{\bm{\theta}}$ are measured with the same SPADE configuration, the outcome is a multinomial random variable $\mbf{x} = [x_1,\ldots,x_K]^{\intercal}$ (not to be confused with the emitter positions!) with distribution
$$
p_{\bm{\theta}}(\mbf{x}) = {\rm Mu}(n,\mbf{q} = P\bm{\theta}) = \frac{n!}{x_1!\cdots x_{K}!}q_{1}^{x_1}\cdots q_{K}^{x_{K}},\qquad \mbf{x}\in\mathcal{X}\equiv\{\mbf{x}\in \mathbb{Z}^{K}_{+} : \mbf{1}^{\intercal}\mbf{x}=n\}
$$
where $\mbf{q} = [q_0,\ldots,q_{K}]^{\intercal}$ is a probability vector over the multinomial categories $\{\ket{\phi_{1}},\ldots,\ket{\phi_{K}}\}$ (see Fig. \ref{fig: simplex_parameterization}(b)). From here, we calculate the classical Fisher information of the measurement, first treating $\bm{\theta}$ as free-parameters and subsequently projecting onto the tangent space of the simplex. Let the log-likelihood be $\ell = \log p_{\bm{\theta}}(\mbf{x})$ and the scores with respect to the category probabilities be $z_{k} = \partial_{q_k} \ell$. These scores are related to the scores of the relative brightness parameters $s_{k} = \partial_{\theta_{k}}\ell$ through the chain rule: 
$$
\partial_{ \theta_{k}}=\sum_{j=1}^{K} \big(\frac{\partial q_j}{\partial \theta_{k}}\big)\partial_{q_{j}} = \sum_{j=1}^{K}P_{jk}\partial_{q_j} \implies s_{k} = \sum_{j}P_{jk}z_{j}
$$
which we may write compactly in matrix form as $\mbf{s} = P^{\intercal}\mbf{z}$. Computing the scores $z_{j}$ explicitly gives,
$$
\ell = \sum_{j=1}^{K}x_j\log q_j + \text{const},\qquad z_{j} = \frac{x_j}{q_j}
$$
The Fisher information matrix of the multinomial probability vector $\mbf{q}$ is given by,
$$
[\tilde{F}_{\mbf{q}}]_{jk} = {\rm Cov}(z_j,z_k) =\frac{1}{q_jq_k}{\rm Cov}(x_j,x_j) = \frac{1}{q_jq_k}\bigg(n q_j(\delta_{jk}  - p_k)\bigg) = n\big(\frac{\delta_{jk}}{q_k} - 1\big)
$$
which we may write in matrix form as, 
$$
\tilde{F}_{\mbf{q}} = n(D_{1/\mbf{q}} -\mbf{1}\mbf{1}^{\intercal})
$$
with $D_{1/\mbf{q}} = \text{Diag}(\frac{1}{q_1},\ldots,\frac{1}{q_{K}})$. As an aside, we note that $\tilde{F}_{\mbf{q}}$ is not invertible. This can quickly be proven by invoking the matrix determinant lemma, 
$$
\det(A + \mbf{u}\mbf{v}^{\intercal}) = \det(A)(1 + \mbf{v}^{\intercal}A^{-1}\mbf{u}).
$$
In our case, we find that the term $(1-\mbf{1}^{\intercal}D_{1/\mbf{q}}^{-1}\mbf{1})= (1-\mbf{q}^{\intercal}\mbf{1}) =0$ so that the Fisher information matrix has determinant zero $\det \tilde{F}_{\mbf{q}}=0$ and hence cannot be inverted. Carrying forward with our analysis and applying the relation between the scores $\mbf{s} = P^{\intercal}\mbf{z}$, the Fisher information matrix for the (unconstrained) relative brightness parameters $\bm{\theta}$ is obtained from the Jacobian transformation,
$$
\tilde{F}_{\bm{\theta}}^{\rm SP} = P^{\intercal}\tilde{F}_{\mbf{q}}P = n(P^{\intercal}D_{1/\mbf{q}}P - P^{\intercal}\mbf{1}\mbf{1}^{\intercal}P) = n(P^{\intercal}D_{1/\mbf{q}}P - \mbf{1}\mbf{1}^{\intercal})  
$$
where in the second term of the second equality we used the fact that $\mbf{1}^{\intercal}P=1$ since the columns of $P$ sum to 1. Our final step is to project this matrix onto the probability simplex such that
$$
F_{\bm{\theta}}^{\rm SP} = \Gamma \tilde{F}_{\bm{\theta}}\Gamma = n \Gamma P^{\intercal}D_{1/\mbf{q}}P\Gamma - \Gamma\mbf{1}\mbf{1}^{\intercal}\Gamma =n \Gamma P^{\intercal}D_{1/\mbf{q}}P \Gamma
$$
where in the second equality we used the fact that $\Gamma\mbf{1}=0$ since the vector $\mbf{1}$ is orthogonal to the simplex tangent space by definition. In summary, the CFIM for a particular SPADE measurement defined by $R\in\mathbb{O}(K)$ is,
\begin{equation}
\boxed{
F_{\bm{\theta}}^{\rm SP}(R) = n \Gamma P^{\intercal} D_{1/\mbf{q}}P\Gamma,\qquad [P]_{jk} \equiv [R^{\intercal}\sqrt{G}]_{jk}^{2}
}
\label{eq: SPADE CFIM}
\end{equation}
For a given weight matrix $W>0$, the scalar CRB for SPADE is thus,
\begin{equation}
\boxed{
c_{\bm{\theta}}^{\rm SP}  = \tr[W(F_{\bm{\theta}}^{\rm SP})^{+}]
}
\label{eq: SPADE CRB}
\end{equation}

\subsection{Direct Imaging}
To ascertain the CFIM of direct imaging, we make use of a convenient observation regarding the derivation of the CFIM for SPADE. In particular, note that our derivation made heavy use of the linear map between parameters $\mbf{q} = P\bm{\theta}$ which in turn produced a linear map between scores $\mbf{s} = P^{\intercal}\mbf{z}$. While for the case of SPADE the transition probability matrix was square $P\in\mathbb{R}^{K\times K}$ with columns summing to 1, all of the arguments used previously hold generally for a non-square transition probability matrix $P\in\mathbb{R}^{K'\times K}$~\footnote{For $K'<K$ it is guaranteed that $\tilde{F}_{\bm{\theta}} = P^{\intercal}\tilde{F}_{\mbf{q}} P$ will be singular, but that's overwhelmingly not the situation in direct imaging where $K\gg K'$ since $P = [p_{1}(x),\ldots,p_{K}(x)]^{\intercal}$. Even so, the pseudo-inverse later on in the definition of the scalar CRB handles the singular nature of the CFIM in such cases.}. Given this insight, we take the continuum limit where $P = [p_{1}(x),\ldots,p_{K}(x)]$ is an infinitely tall 'matrix' $(K'\gg K)$ whose columns are the probability distributions associated with the shifted intensity point spread functions produced by each emitter $p_{k}(x) = p(x-x_k)$. By extension, it is straightforward to see that,

$$
\mbf{q} = P\bm{\theta}\qquad \rightarrow \qquad q_{j} = \sum_{k}P_{jk}\theta_k \qquad \rightarrow \qquad q(x) = \sum_{k=1}^{K} p_k(x) \theta_k
$$
such that, the following quantities are similarly mapped as 
$$
\tilde{F}_{\bm{\theta}} = nP^{\intercal}D_{1/\mbf{q}}P \qquad \rightarrow \qquad [\tilde{F}_{\bm{\theta}}]_{ab} =n \sum_{j}\frac{P^{\intercal}_{aj}P_{jb}}{(\sum_{k}P_{jk}\theta_{k})} \qquad \rightarrow \qquad [\tilde{F}_{\bm{\theta}}]_{ab} = n\int_{-\infty}^{\infty} \frac{p_a(x)p_b(x)}{(\sum_{k}p_{k}(x)\theta_{k})}
$$
Therefore the CFIM of direct imaging, when projected back onto the tangent space of the simplex, is
\begin{equation}
    \boxed{
    F_{\bm{\theta}}^{\rm DI} = \Gamma \tilde{F}_{\bm{\theta}}^{\rm DI}\Gamma,\qquad \qquad [\tilde{F}_{\bm{\theta}}^{\rm DI}]_{ab} = n\int_{-\infty}^{\infty} \frac{p_a(x)p_b(x)}{\sum_{k}p_{k}\theta_{k}}
    }
\end{equation}
Thereafter, the scalar CRB is given by,
\begin{equation}
    \boxed{
    c_{\bm{\theta}}^{\rm DI} = \tr[( F_{\bm{\theta}}^{\rm DI})^{+}]
    }
\end{equation}

\section{Joint Measurements}
\label{ss: Joint Measurements}
Recently, Tsang~\cite{Tsang:2026_HolevoBound_ManyBody} and Zhou~\cite{zhou:2026_MixedStatePurification} independently proposed two-stage estimation protocols that realize asymptotically optimal measurements saturating the Holevo bound for multiparameter estimation. In both procedures, the first stage allocates an asymptotically vanishing fraction of the state copies to performing a coarse pre-estimate of the target parameters $\tilde{\bm{\theta}}$ such that the true parameters $\bm{\theta} = \tilde{\bm{\theta}} + \bm{\epsilon}$ lie within a perturbative neighborhood of the pre-estimate. Conditioned on the pre-estimates, the second stage involves constructing a measurement that is locally optimal and may be applied to the remaining state copies. While the existence of an optimal joint measurement strategy was known from the works of Guta and others in their development of quantum local asymptotic normality (QLAN), \cite{Demkowicz-Dobrzanski:2020,Guta:2007,Guta:2009} a concrete implementation of this measurement in general parameter estimation tasks had been elusive before Tsang and Zhou's publications. Both references provide distinct approaches: Zhou's approach uses state purification by way of interaction with environmental degrees of freedom followed by copy-by-copy measurements. On the other hand, Tsang's approach introduces ancillary bosonic systems coupled to the signal-bearing quantum system whose dynamics are governed by a specific interaction Hamiltonian. Subsequently, only the quadratures of the ancillas need be measured. Here we provide further detail about the receiver based on Tsang's general protocol that issues a Helstrom-optimal joint measurement for brightness estimation of an arbitrary ensemble of emitters using an array of optomechanical ancillary systems cross-coupled to the Williamson modes of the field. 

\subsection{Derivation of the Efficient Influence Operator}
Ref. \cite{Tsang:2020_SemiParametricEstimation} Theorem 7 asserts that the Helstrom bound for the parameters of interest is given by,
$$
c^{\rm SLD} = \min_{\bm{\delta} \in \mathcal{D}} \tr[W \langle \bm{\delta},\bm{\delta}\rangle ]= \tr[W \langle \bm{\delta}^{\rm eff},\bm{\delta}^{\rm eff}\rangle ]
$$
where the efficient influence operator $\bm{\delta}^{\rm eff}$ is the {\em unique} projection of $\mathcal{D}$ onto the SLD tangent space
$$
\bm{\delta}^{\rm eff} = \text{proj}(\bm{\delta}|\mathcal{T}^{\oplus d}),\qquad \hat{\delta}^{\rm eff}_{k} = \text{proj}(\hat{\delta}_{k}|\mathcal{T}).
$$
Since the efficient influence operator lives on the SLD tangent space (a real Hilbert space), we know that
$$
\bm{\delta}^{\rm eff} = A \mbf{L},\qquad A \in \mathbb{R}^{d\times d}
$$
where $A$ is a matrix of real expansion coefficients. Applying the unbiasedness constraint of $\mathcal{D}$ we have,
$$
\langle \mbf{L},\bm{\delta}^{\rm eff}\rangle = A \langle \mbf{L},\mbf{L}\rangle = A Q_{\bm{\theta}} = I_{K}
$$
Therefore, assuming the QFIM is invertible, we find that the coefficients are $A=Q_{\bm{\theta}}^{-1}$ such that the vectoral efficient influence operator is given by,
$$
\boxed{
\bm{\delta}^{\rm eff} = Q_{\bm{\theta}}^{-1}\mbf{L} 
}
$$
\subsection{Interaction Hamiltonian for Tsang's Protocol}
With the efficient influence operator in hand, we now turn to implementing the second stage of Tsang's measurement protocol that (in our case) achieves the Helstrom bound. Assuming $\tilde{\bm{\theta}} \in \Theta$ is a close pre-estimate of the target parameters, we define a particular set of joint observables on $n$ state copies (see Ref. \cite{Tsang:2026_HolevoBound_ManyBody} Eq. 13 and Ref. \cite{Tsang:2020_SemiParametricEstimation} Eq. 3.32) given by,

$$
X^{(n)}_{K} = \frac{1}{\sqrt{n}} \sum_{t=1}^{n} I^{\otimes (t-1)}\otimes X_{k} \otimes I^{\otimes (n-t)}
$$
where $I$ is the identity operator on the Hilbert space of the received state $\rho$ and $X_{k} = \delta^{\rm eff}_{k} + \tilde{\theta}_{k}$ for $k=1,\ldots,K$ such that $\mbf{X} = \bm{\delta}^{\rm eff} + \tilde{\bm{\theta}}I$. As Tsang notes ``remarkably, for diffraction-limited incoherent imaging, where $\Tr[\rho [L_{a},L_{b}]]=0$ for all $\bm{\theta}$, all entries of $\mbf{X}^{(n)}$ are asymptotically compatible with one another and one can simply set $\mbf{X}^{(n)} = \mbf{Y}^{(n)} = \mbf{x}^{(n)}$'', where $\mbf{x}^{(n)}$ is a set of observables with conjugates $\mbf{y}^{(n)}$ such that $[x_{j}^{(n)},y_{k}^{(n)}]\rightarrow i \delta_{jk}I$ in the limit of large $n$ such that they approach canonical quadratures of a bosonic mode. Introducing bosonic ancilla's with quadrature operators $q_{k}',p_{k}'$ for $k = 1,\ldots,K$, the interaction Hamiltonian we wish to realize is then given by,
$$
H = \gamma \sum_{k=1}^{K}X_{k}^{(n)}p_{k}'
$$
where $\gamma$ is an arbitrary global coupling coefficient between the quantum system harboring the state copies and the ancillas. According to Tsang's protocol, one initializes the ancillas in a zero-mean low-variance state (e.g. vacuum) and allows them to evolve under $H$ with all $n$ state copies. At the end one performs a homodyne measurement to read out the position quadrature $q_{k}'$ of each ancilla which coincides with the measurement,
$$
q_k' \approx \gamma t X_{k}^{(\infty)}
$$
for large $\gamma t$. The expectation value of the observable converges to the parameters $\langle \mbf{q}' \rangle \rightarrow \gamma t \bm{\theta}$ in the limit $n\gg 1$. Since we found that the SLDs are quadratic in the Williamson mode ladder operators, the single-copy interaction Hamiltonian for a single temporal mode assumes the form,
$$
H = \sum_{j,k,\ell=1}^{K} \beta_{jk}^{(\ell)}\underbrace{(\hat{c}_{j}^{\dagger}\hat{c}_{k}+\hat{c}_{k}^{\dagger}\hat{c}_{j})\hat{p}_{\ell}'}_{\text{3-wave mixing}} + \xi^{(\ell)}\hat{p}_{\ell}'
$$ 
where the coupling coefficients are defined as,
$$
\beta_{jk}^{(\ell)} = \frac{\gamma}{2} \sum_{a=1}^{K} [Q^{-1}_{\tilde{\bm{\theta}}}]_{\ell a}\bigg(\frac{\Psi_{ja}\Psi_{ka}}{M_{jk}}\bigg),\qquad \xi^{(\ell)} =\gamma \sum_{k,a=1}^{K}[Q_{\tilde{\bm{\theta}}}^{-1}]_{\ell a} \bigg( \frac{\Psi_{ka}^2}{M_{kk}}\bar{m}_{k}\bigg)
$$ 
Next, we put the beam-splitter-type terms in a form that is ubiquitous  in quantum optomechanics. In particular, we have
$$
\hat{c}^{\dagger}_{j}c_{k} + c_{k}^{\dagger}c_{j} = \hat{m}_{+}^{(jk)} - \hat{m}_{-}^{(jk)}
$$
where $\hat{m}_{\pm}^{(jk)} = (\hat{c}^{(jk)}_{\pm})^{\dagger}(\hat{c}^{(jk)}_{\pm})$ are photon number operators for the symmetric and antisymmetric combinations of the Williamson modes $\hat{c}^{(jk)}_{\pm} = (\hat{c}_{j}\pm \hat{c}_{k})/\sqrt{2}$. Consequently, the three-wave mixing term in the interaction Hamiltonian may be expressed as,
$$
\sum_{j,k,\ell=1}^{K}\beta_{jk}^{\ell}(\hat{m}_{+}^{(jk)}- \hat{m}_{-}^{(jk)})\hat{p}_{\ell}'
$$
which is a linear combination of multimode optomechanical interaction Hamiltonians (i.e. radiation pressure from multiple optical modes driving the dynamics of a single mechanical oscillator).

\section{Simulation Results (Extended)}

\begin{figure}[h!]
    \centering
    \includegraphics[width=\linewidth]{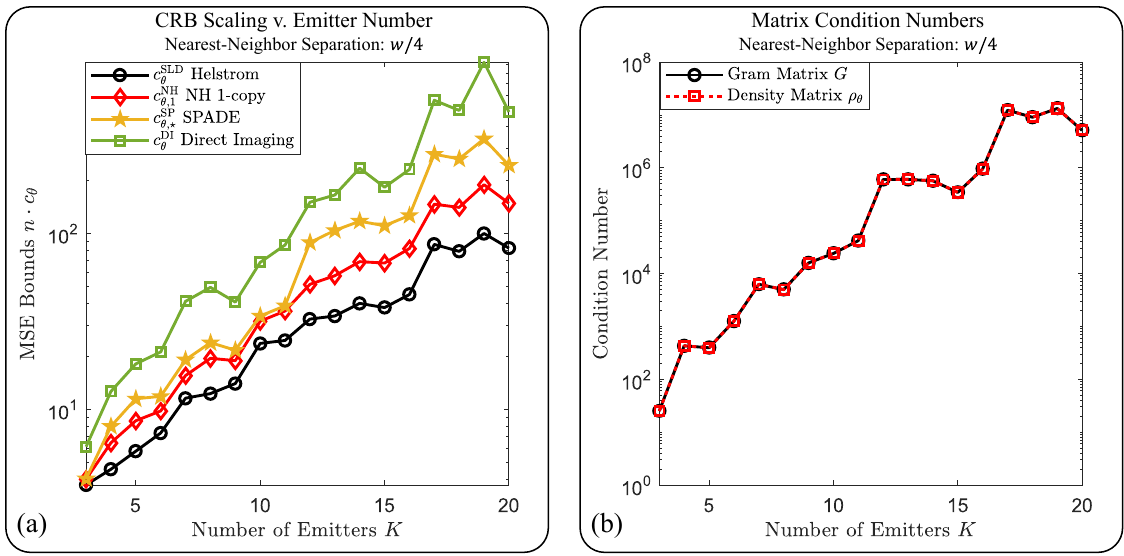}
    \caption{Matrix condition numbers for each emitter geometry in the $K$-sweep of main text. For each geometry, the threshold tolerance for the dual-primal convergence of the SDP MOSEK solver was initialized at $10^{-8}$ and progressively increased by an order of magnitude each time the dual and primal problems failed to converge.}
    \label{fig: extended_1}
\end{figure}

\begin{figure}[h!]
    \centering
    \includegraphics[width=\linewidth]{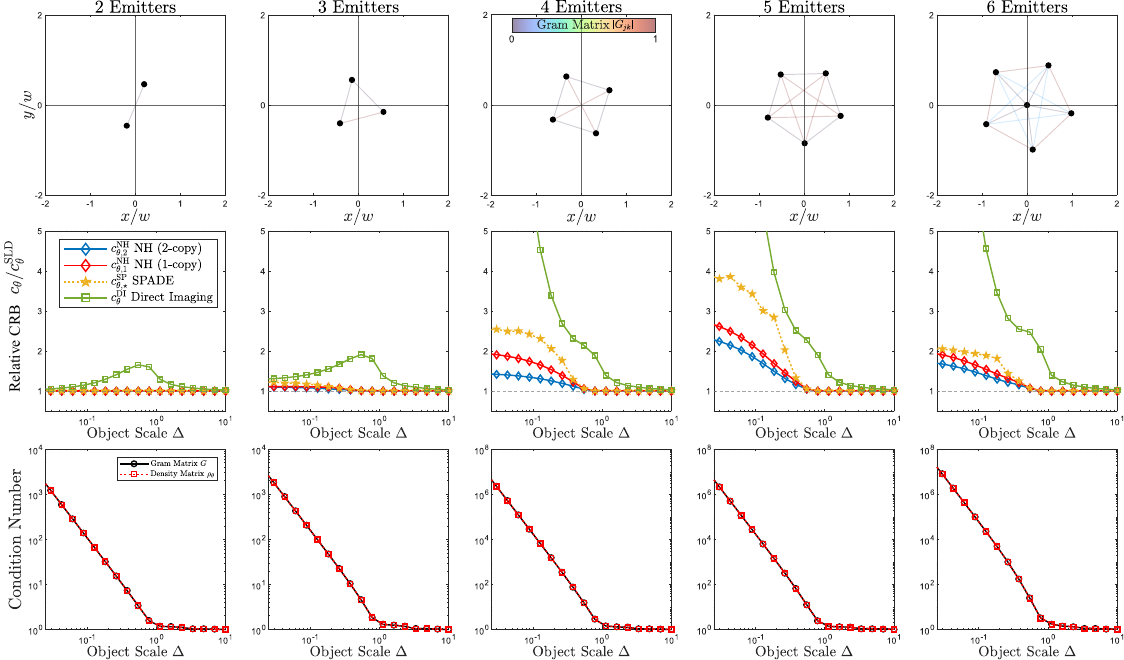}
    \caption{Relative CRBs normalized to the Helstrom bound as a function of scene dilation (scale factor $\Delta$) for different numbers of emitters $K=2,\ldots, 6$.}
    \label{fig: extended_2}
\end{figure}

\begin{figure}[h!]
    \centering
    \includegraphics[width=\linewidth]{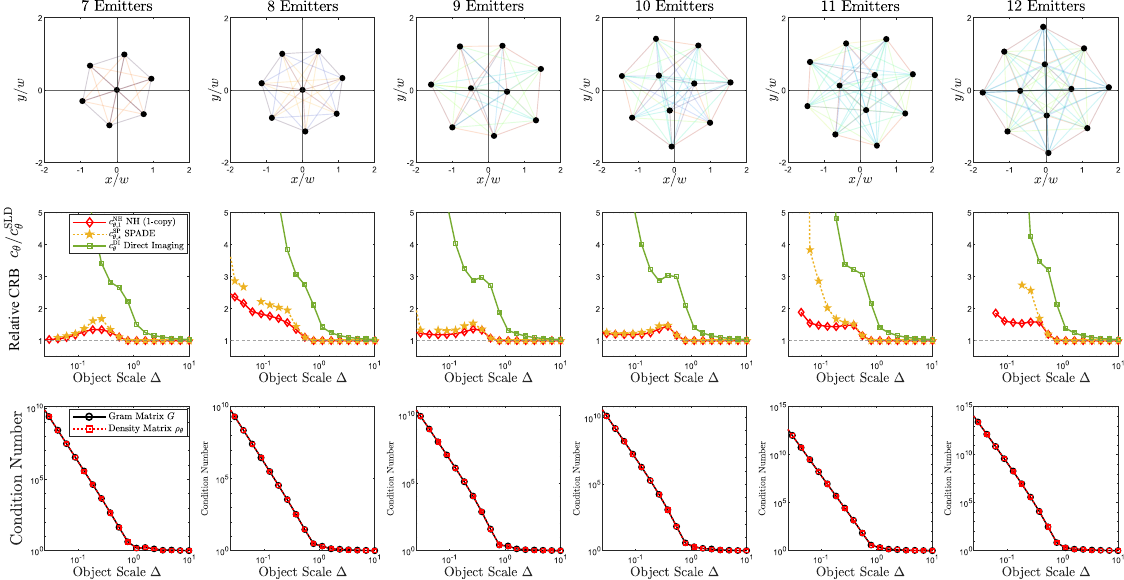}
    \caption{Relative CRBs normalized to the Helstrom bound as a function of scene dilation (scale factor $\Delta$) for different numbers of emitters $K=7,\ldots, 12$.}
    \label{fig: extended_3}
\end{figure}

\end{document}